\documentclass[aps,prd,preprint,showpacs,eqsecnum,nofootinbib]{revtex4}
\usepackage{graphicx,epsf,amssymb}
\newcommand{\be}{\begin{equation}}
\newcommand{\ee}{\end{equation}}

\newcommand{\bea}{\begin{eqnarray}}
\newcommand{\eea}{\end{eqnarray}}
\newcommand{\Tr}{{\rm Tr}}
\newcommand{\nn}{\nonumber}

\newcommand{\I}{{\cal I}}
\newcommand{\dd}{\mbox{d}}

\newcommand{\Gm}{\Gamma}
\newcommand{\lm}{\lambda}

\def\d{{\rm d}}

\newif\ifdraft
\drafttrue
\newif\ifpreprint
\preprinttrue

\def\sect#1{section~{\ref{#1}}}
\def\app#1{appendix~{\ref{#1}}}

\def\fig#1{fig.~{\ref{#1}}}
\def\figs#1#2{figs.~{\ref{#1}} and {\ref{#2}}}
\def\Fig#1{Figure~{\ref{#1}}}

\def\tree{{\rm tree}}

\def\Tr{\, {\rm Tr}}
\def\SYM{MSYM}
\def\hf{{\textstyle{1\over2}}}

\def\NeqFour{{\cal N}=4}
\def\NeqOne{{\cal N}=1}
\def\NeqEight{{\cal N}=8}

\def\Nc{N_c}
\def\spa#1.#2{\left\langle#1\,#2\right\rangle}
\def\spb#1.#2{\left[#1\,#2\right]}
\def\sand#1.#2.#3{%
\left\langle\smash{#1}{\vphantom1}^{-}\right|{#2}%
\left|\smash{#3}{\vphantom1}^{-}\right\rangle}
\def\sandp#1.#2.#3{%
\left\langle\smash{#1}{\vphantom1}^{-}\right|{#2}%
\left|\smash{#3}{\vphantom1}^{+}\right\rangle}
\def\sandpp#1.#2.#3{%
\left\langle\smash{#1}{\vphantom1}^{+}\right|{#2}%
\left|\smash{#3}{\vphantom1}^{+}\right\rangle}
\def\sandpm#1.#2.#3{%
\left\langle\smash{#1}{\vphantom1}^{+}\right|{#2}%
\left|\smash{#3}{\vphantom1}^{-}\right\rangle}
\def\sandmp#1.#2.#3{%
\left\langle\smash{#1}{\vphantom1}^{-}\right|{#2}%
\left|\smash{#3}{\vphantom1}^{+}\right\rangle}
\def\sandmm#1.#2.#3{%
\left\langle\smash{#1}{\vphantom1}^{-}\right|{#2}%
\left|\smash{#3}{\vphantom1}^{-}\right\rangle}
\def\spab#1.#2.#3{\sandmm#1.#2.#3}
\def\spba#1.#2.#3{\sandpp#1.#2.#3}
\def\spaa#1.#2.#3.#4{\sandmp#1.{#2#3}.#4}
\def\spbb#1.#2.#3.#4{\sandpm#1.{#2#3}.#4}
\def\spash#1.#2{\spa{\smash{#1}}.{\smash{#2}}}

\newbox\charbox
\newbox\slabox
\def\s#1{{      
        \setbox\charbox=\hbox{$#1$}
        \setbox\slabox=\hbox{$/$}
        \dimen\charbox=\ht\slabox
        \advance\dimen\charbox by -\dp\slabox
        \advance\dimen\charbox by -\ht\charbox
        \advance\dimen\charbox by \dp\charbox
        \divide\dimen\charbox by 2
        \raise-\dimen\charbox\hbox to \wd\charbox{\hss/\hss}
        \llap{$#1$}
}}

\def\eqn#1{eq.~(\ref{#1})}
\def\Eqn#1{Equation~(\ref{#1})}
\def\eqns#1#2{eqs.~(\ref{#1}) and~(\ref{#2})}
\def\Eqns#1#2{Eqs.~(\ref{#1}) and~(\ref{#2})}
\def\qb{{\overline {\kern-0.7pt q\kern -0.7pt}}}

\def\e{\epsilon}
\def\ep{\epsilon}

\def\lr{\leftrightarrow}

\def\Li{\mathop{\rm Li}\nolimits}

\def\tree{{(0)}}
\def\Lloop{{(L)}}

\def\oneloop{{(1)}}
\def\twoloop{{(2)}}
\def\threeloop{{(3)}}
\def\fourloop{{(4)}}

\def\Ord{{\cal O}}

\def\sandp#1.#2.#3{%
\left\langle\smash{#1}{\vphantom1}^{+}\right|{#2}%
\left|\smash{#3}{\vphantom1}^{+}\right\rangle}

\newbox\ourfigbox
\def\SizedFigureWithCaption#1#2#3{%
\setbox\ourfigbox
  \hbox{\hss\epsfxsize #1 \epsfbox{#2}\hss}
\hbox to \wd\ourfigbox{\vbox{\noindent\copy\ourfigbox\break
\vskip -6mm      \hbox to \wd\ourfigbox{\hss#3\hss}}}
}

\def\spa#1.#2{\left\langle#1\,#2\right\rangle}
\def\spb#1.#2{\left[#1\,#2\right]}
\def\lor#1.#2{\left(#1\,#2\right)}
\def\sand#1.#2.#3{%
\left\langle\smash{#1}{\vphantom1}^{-}\right|{#2}%
\left|\smash{#3}{\vphantom1}^{-}\right\rangle}
\def\sandpp#1.#2.#3{%
\left\langle\smash{#1}{\vphantom1}^{+}\right|{#2}%
\left|\smash{#3}{\vphantom1}^{+}\right\rangle}
\def\sandpm#1.#2.#3{%
\left\langle\smash{#1}{\vphantom1}^{+}\right|{#2}%
\left|\smash{#3}{\vphantom1}^{-}\right\rangle}
\def\sandmp#1.#2.#3{%
\left\langle\smash{#1}{\vphantom1}^{-}\right|{#2}%
\left|\smash{#3}{\vphantom1}^{+}\right\rangle}

\def\ah{\hat{a}}

\begin{document}
\hfuzz 15 pt


\ifpreprint
\noindent
 UCLA/06/TEP/28
\hfill SLAC--PUB--12172
\hfill Saclay/SPhT--T06/135
\fi

\title{The Four-Loop Planar Amplitude and Cusp Anomalous Dimension 
in Maximally Supersymmetric Yang-Mills Theory}%

\author{Zvi Bern}
\affiliation{{} Department of Physics and Astronomy, UCLA\\
\hbox{Los Angeles, CA 90095--1547, USA}
}

\author{Michael Czakon}
\affiliation{ Institut f\"ur Theoretische Physik und Astrophysik,
 Universit\"at W\"urzburg,\\
 Am Hubland, D-97074 W\"urzburg, Germany, \\
 Department of Field Theory and Particle Physics, 
 Institute of Physics,\\
 University of Silesia, Uniwersytecka 4, PL-40007 Katowice, Poland
}

\author{Lance J. Dixon}
\affiliation{ Stanford Linear Accelerator Center \\
              Stanford University\\
             Stanford, CA 94309, USA
}

\author{David A. Kosower}
\affiliation{Service de Physique Th\'eorique\footnote{Laboratory
   of the {\it Direction des Sciences de la Mati\`ere\/}
   of the {\it Commissariat \`a l'Energie Atomique\/} of France},
   CEA--Saclay\\
          F--91191 Gif-sur-Yvette cedex, France
}

\author{Vladimir A. Smirnov}
\affiliation{Nuclear Physics Institute of Moscow State University\\
\hbox{Moscow 119992, Russia}
}

\date{October, 2006}

\begin{abstract}
We present an expression for the leading-color (planar) four-loop
four-point amplitude of $\NeqFour$ supersymmetric Yang-Mills theory in
$4-2\e$ dimensions, in terms of eight separate integrals.  The
expression is based on consistency of unitarity cuts and infrared
divergences.  We expand the integrals around $\e=0$, and obtain
analytic expressions for the poles from $1/\e^8$ through $1/\e^4$.  We
give numerical results for the coefficients of the $1/\e^3$ and
$1/\e^2$ poles.  These results all match the known exponentiated
structure of the infrared divergences, at four separate kinematic
points.  The value of the $1/\e^2$ coefficient allows us to test a
conjecture of Eden and Staudacher for the four-loop cusp (soft)
anomalous dimension.  We find that the conjecture is incorrect,
although our numerical results suggest that a simple modification of
the expression, flipping the sign of the term containing $\zeta_3^2$,
may yield the correct answer.  Our numerical value can be used, in a
scheme proposed by Kotikov, Lipatov and Velizhanin, to estimate the
two constants in the strong-coupling expansion of the cusp anomalous
dimension that are known from string theory.  The estimate works to
2.6\% and 5\% accuracy, providing non-trivial evidence in support of
the AdS/CFT correspondence.  We also use the known constants in the
strong-coupling expansion as additional input to provide
approximations to the cusp anomalous dimension which should be
accurate to under one percent for all values of the coupling.  When
the evaluations of the integrals are completed through the finite
terms, it will be possible to test the iterative, exponentiated
structure of the finite terms in the four-loop four-point amplitude,
which was uncovered earlier at two and three loops.
\end{abstract}

\pacs{11.15.Bt, 11.15.Pg, 11.25.Db, 11.25.Tq, 12.60.Jv  \hspace{1cm}}

\maketitle

\renewcommand{\thefootnote}{\arabic{footnote}}
\setcounter{footnote}{0}

\section{Introduction}
\label{IntroSection}

Maximally supersymmetric $\NeqFour$ Yang-Mills theory (\SYM) has
attracted a great deal of theoretical interest over the years.  It is
widely believed that the 't~Hooft (planar) limit of \SYM, in which the
number of colors $N_c$ is taken to infinity, is dual at strong coupling
($\lambda \equiv g^2 N_c \to \infty$) to weakly-coupled
gravity in five-dimensional anti-de Sitter space~\cite{Maldacena}.
The duality implies that the full quantum anomalous dimensions of
various series of gauge-invariant composite operators are equal to
energies of different gravity modes or configurations of strings in
anti-de Sitter
space~\cite{BPS,OtherAnomalousDim,BMN,Nontrivial,DhokerTasi}.
Heuristically, the Maldacena duality conjecture hints that even
quantities unprotected by supersymmetry should have perturbative
series that can be resummed in closed form.  The strong-coupling limits
of these resummed expressions, possibly supplemented by
non-perturbative contributions, should match results for the
appropriate observables in weakly-coupled supergravity or string theory.

This intuition does appear to apply to the higher-loop on-shell
scattering amplitudes of color-non-singlet gluons, even though the
Maldacena conjecture does not directly address on-shell amplitudes of
massless colored quanta.  There is now significant evidence of a very simple
structure in the planar limit.  In particular, the planar
contributions to the two-loop and three-loop four-gluon amplitudes
have been shown to obey iterative relations~\cite{Iterate2,Iterate3}:
The dimensionally regularized amplitudes ($d=4-2\e$) can be expressed
in terms of lower-loop amplitudes, along with a set of three constants
for each order in the loop expansion.  

An analogous relation is conjectured to hold for generic
maximally-helicity-violating (MHV) $n$-gluon scattering amplitudes, to
all loop orders~\cite{Iterate2,Iterate3}.  The MHV amplitudes are
those for which two of the gluons have negative helicity and the
remaining $(n-2)$ have positive helicity, or the parity conjugate
case.  Indirect evidence for the extension to the $n$-point MHV case
was provided first by studying the consistency of collinear limits at
two loops~\cite{Iterate2}.  Later, the iteration relation was
demonstrated to hold directly for the two-loop five-gluon amplitude,
for the ``even'' terms in the amplitude~\cite{CSV06} and soon thereafter
for the ``odd'' terms as well, {\it i.e.} for the complete
amplitude~\cite{BCKRS}.  (Odd and even refer to the behavior of the
ratio of terms in the loop amplitude to the tree amplitude, under
parity.)  After infrared divergences have been subtracted from the
proposed all-loop iterative relation, the resulting finite remainder
is neatly proportional to an exponential of the product of the
one-loop finite remainder with the so-called cusp anomalous dimension.
Presumably the weak--strong duality between anti-de Sitter space
and conformal field theory (AdS/CFT) plays a role in this simplicity.

The form of the proposed iterative structure of multi-loop planar
\SYM\ is based on the understanding of how soft and collinear infrared
singularities factorize and exponentiate in gauge
theory~\cite{Akhoury,Sudakov,Sen83,SoftGluonSummation,MagneaSterman,%
IROneLoop,CataniIR,TYS}.  For the pole terms in the amplitudes, such
behavior is universal, and holds in any massless gauge theory.  What
is remarkable in planar \SYM\ is that the finite terms in the MHV
scattering amplitudes can be organized into the same exponentiated
form as the divergent terms.  In a non-supersymmetric theory, QCD,
exponentiation of finite terms has also been observed in the context
of threshold resummation of the Drell-Yan cross
section~\cite{EynckLaenenMagnea}.  In the case of four-gluon
amplitudes in \SYM, the behavior is valid for arbitrary values of the
scattering angle (ratio of $t/s$).  For amplitudes with more than
four gluons, there are many kinematical variables,
and so the constraints imposed by the iterative structure are
even stronger.  It is clearly of interest to test whether 
this structure persists beyond three loops.  In this paper we shall 
provide an integral representation for the planar four-loop four-gluon
amplitude.  Our result will enable such a test to be performed at 
four loops, once the relevant integrals have been evaluated to 
sufficiently high order in $\e$.

Another remarkable property of planar \SYM\ is the integrability of the
dilatation operator, interpreted as a Hamiltonian, for many sectors of
the theory.  Integrable structures were identified in 
anomalous dimension matrices in QCD a while ago~\cite{QCDIntegrable}.
In planar \SYM, Minahan and Zarembo~\cite{MinahanZarembo} mapped the 
one-loop dilatation operator for non-derivative single-trace operators
to an integrable spin-chain Hamiltonian, and used a Bethe ansatz to
compute the anomalous dimensions.   Such integrable structures have
since been extended to higher perturbative orders for various sectors 
of the theory~\cite{MoreIntegrable,BeisertDispersion,Integrable}. 
They are also known to be present at strong coupling, from the
form of the classical sigma model on target space 
AdS${}_5\times S^5$~\cite{BPR}.  (Berkovits has given a formal
argument that the integrability extends to the quantum level
on the world sheet~\cite{Berkovits}.)
The iterative structure of \SYM\ amplitudes may somehow be related
to integrability.  If an infinite number of conserved charges
are present, the form of the quantum corrections should be severely
constrained, as it would be by the proposed iterative
structure~\cite{Iterate2,Iterate3}.  An iterative structure may also
arise in correlation functions of gauge-invariant composite operators
in planar \SYM~\cite{Schubert}; but its precise structure, if it 
exists in this context, has not yet been clarified.

Integrability is a powerful computational tool.
Integrability, or in some cases, the assumption of integrability, 
has been employed to compute a variety of one-loop and
multi-loop anomalous dimensions in planar \SYM\ 
from Bethe ans\"atze~\cite{MinahanZarembo,MoreIntegrable,BeisertDispersion}
and from the Baxter equation~\cite{BGK06,Belitsky}.
One of the most interesting developments along these lines has been the
all-orders proposal of Eden and Staudacher (ES), based on an
asymptotic all-loop Bethe ansatz~\cite{AsymptoticBA}, for the
large-spin limit of the anomalous dimensions of leading-twist
operators in \SYM~\cite{EdenStaudacher}.  This quantity,
$\gamma_K(\alpha_s)$, is also referred to as the cusp (or sometimes,
soft) anomalous dimension associated with a Wilson line.
Equivalently, it represents the leading large-$x$ behavior~\cite{KM}
of the DGLAP kernel $P_{ii}(x)$ for parton evolution, $i\to i$,
\be
P_{ii}(x) \to { \gamma_K(\alpha_s) \over 2 \, (1-x)_+ } 
+ B(\alpha_s) \, \delta(1-x) + \ldots, 
\qquad  {\rm as}\ \ x\to 1.
\label{largexP}
\ee
Taking Mellin moments, $\gamma(j) \equiv - \int_0^1 dx \, x^{j-1} P(x)$,
we see that 
the cusp anomalous dimension gives the dominant behavior of the 
leading-twist anomalous dimensions as the spin $j\to\infty$,
\begin{equation}
\gamma(j) = {1\over2} \gamma_K(\alpha_s)\ \ln(j) + \Ord(j^0)\,.
\label{gammajgammaK}
\end{equation}

From the asymptotic all-loop Bethe ansatz, ES derived an integral 
equation for a fluctuation density $\hat\sigma$, from which the 
cusp anomalous dimension can be determined.  The integral equation 
is straightforward to solve perturbatively in $\alpha_s$.  
In terms of the expansion parameter
\be 
\ah \equiv { g^2 N_c\over 8 \pi^2 } = { N_c \alpha_s \over 2 \pi} \,,
\label{ahdef}
\ee
the ES prediction for (one quarter of) the cusp anomalous dimension 
is~\cite{EdenStaudacher}
\bea 
{\gamma_K(\ah)\over4} 
&\equiv& f_0(\ah)
\label{f0def}\\
 &=& \ah - \zeta_2 \, \ah^2
           + \Bigl( \zeta_2^2 + 3 \, \zeta_4 \Bigr) \ah^3
\nonumber\\ 
&&\hskip0.0cm\null
           - \Bigl( \zeta_2^3 + 6 \, \zeta_2 \, \zeta_4 - \zeta_3^2
                  + {25\over2} \, \zeta_6 \Bigr) \ah^4
\nonumber\\ 
&&\hskip0.0cm\null
           + \Bigl( \zeta_2^4 + 9 \, \zeta_2^2 \, \zeta_4 
                  - 2 \, \zeta_2 \, \zeta_3^2
                  + 25 \, \zeta_2 \, \zeta_6 - 10 \, \zeta_3 \, \zeta_5 
                  + {39\over4} \, \zeta_4^2 
                  + {245\over4} \, \zeta_8 \Bigr) \ah^5 
\ +\ \cdots~~~~~~
\label{gammaKA}\\ 
 &=&         \ah - {\pi^2\over6} \, \ah^2
           + {11\over180} \pi^4 \, \ah^3
           - \Bigl( {73\over2520} \, \pi^6 - \zeta_3^2 \Bigr) \ah^4
\nonumber\\ 
&&\hskip0.0cm\null
           + \Bigl( {887\over56700} \, \pi^8 
                   - {\pi^2\over3} \, \zeta_3^2
                   -10 \, \zeta_3 \, \zeta_5 \Bigr) \ah^5 
           +\ \cdots \,.
\label{gammaKB}
\eea
Here $f_0(\ah)$ can be identified with the first of a series of
three constants (per loop order) appearing in the iterative relation
for the four-gluon amplitude~\cite{Iterate2,Iterate3}.  The first
three terms of \eqn{gammaKB} agree with previously-known
results~\cite{Makeenko,KLV,KLOV,Iterate3,MOS}, so the new predictions
begin with the $\ah^4$ term.  The conjecture~(\ref{gammaKA}) has also been
arrived at recently by Belitsky, using a proposed
generalization of the Baxter equation to all loop orders~\cite{Belitsky}.

In QCD the three-loop cusp (soft) anomalous dimension has been
computed by Moch, Vermaseren and Vogt as part of the impressive
computation of the full leading-twist anomalous dimensions~\cite{MVV}
needed for next-to-next-to-leading order evolution of parton
distribution functions.  (Terms proportional to the number of quark
flavors were obtained first in ref.~\cite{SoftNf}.)  
Kotikov, Lipatov, Onishchenko and Velizhanin (KLOV)~\cite{KLOV}
made the inspired observation, based on evidence at two loops~\cite{KL02},
that the \SYM\ anomalous dimensions 
may be obtained simply from the ``leading-transcendentality''
contributions in QCD.
The cusp anomalous dimensions are
polynomials in the Riemann $\zeta$ values, $\zeta_n \equiv \zeta(n)$,
or their multi-index generalizations, $\zeta_{n_1,n_2,\ldots}$.
(These cannot show up below five loops.)
The degree of transcendentality of $\zeta_n$ is just $n$,
and the transcendentality is additive for products of $\zeta$ values.
At $L$ loops, the leading-transcendentality contributions to the cusp
anomalous dimension have degree (or weight) equal to $2L-2$.
All \SYM\ leading-twist anomalous dimensions computed to date
have had uniform, leading transcendentality, whereas the corresponding
QCD results contain an array of terms of lower transcendentality,
all the way down to rational numbers.  

The KLOV conversion principle applies to the leading-twist anomalous
dimensions for any spin $j$, with an appropriate definition of leading
transcendentality for the harmonic sums $S_{\vec{m}}(j)$ that appear
in the results.  Using assumptions of integrability, Staudacher
confirmed the three-loop KLOV result through
$j=70$~\cite{Staudacher,StaudacherPrivate}, building on earlier work
of Beisert, Kristjansen and Staudacher~\cite{MoreIntegrable} at $j=4$.
Eden and Staudacher extended this analysis to the three-loop cusp
anomalous dimension (the $j\to\infty$ limit), in the course of
arriving at their all-orders proposal~(\ref{gammaKA}) based on
integrability~\cite{EdenStaudacher}.  The three-loop cusp anomalous
dimension in \SYM\ was independently determined from the $1/\e^2$ pole
in the three-loop four-gluon scattering amplitude~\cite{Iterate3},
providing a confirmation of the KLOV result in the limit $j\to\infty$.
An important feature of the ES proposal~(\ref{gammaKA}) is that it is
consistent with KLOV's observation that the \SYM\ anomalous dimensions
are homogeneous in the transcendentality, at least through three
loops.

The ES proposal emerges from mapping single-trace gauge-invariant
operators to spin chains.  The dilatation operator, whose
eigenvalues are anomalous dimensions, 
is mapped to the spin-chain Hamiltonian.  The form of the
$S$ matrix for this spin chain is fixed, up to an overall
phase, called the {\it dressing\/} factor~\cite{AFS04},
by the superconformal [PSU$(2,2|4)$] symmetry of both 
AdS${}_5\times S^5$ and $\NeqFour$ supersymmetric gauge theory.
At one and two loops, superconformal symmetry in combination with explicit
calculations fixes the dressing 
factor to be 1.  At higher loops, it is not so constrained.
A nontrivial dressing factor is required by the strong-coupling 
behavior~\cite{AFS04,OtherDressing,HernandezLopez}, and there have been
several recent investigations of it using properties
such as worldsheet crossing symmetry~\cite{JanikCrossing,BHL}. 
Yet the order in the weak-coupling expansion at which it becomes 
nontrivial is still uncertain.

The presence of a dressing factor at three loops
would lead to a shift in the anomalous dimension at four loops.
For example, ES have proposed~\cite{EdenStaudacher} a modification
of the asymptotic Bethe ansatz~\cite{AsymptoticBA} at this order,
with coefficient $\beta$, which is consistent with the presently 
known integrable structure.  This modification alters the predicted
four-loop anomalous dimension in \eqn{gammaKB} to
\begin{equation}
 - \Bigl( {73\over2520} \, \pi^6 - \zeta_3^2 + 2\beta \zeta_3 \Bigr) \ah^4
\,.
\label{betashifteq}
\end{equation}
Thus a calculation of the four-loop cusp anomalous dimension has 
the potential to probe a nontrivial dressing factor.  In a
new preprint~\cite{BESNew}, Beisert, Eden and Staudacher 
(BES) have shown how to extend the above dressing factor to all orders 
in the coupling, as well as to ensure its consistency with other
constraints.  This leads to the prediction that $\beta = \zeta_3$.

In this paper, we perform this calculation, in order to test whether the
ES all-orders proposal gives the correct result, or to reveal how it needs
to be modified if it doesn't.  We do so by evaluating
the infrared poles of the planar four-loop four-gluon scattering amplitude
through $1/\e^2$, the order at which the four-loop cusp anomalous
dimension appears.  The form of the infrared singularities at all loop
orders are fully understood~\cite{MagneaSterman}, up to a set of constants
that must be computed explicitly.  At $1/\e^2$ this undetermined constant
is precisely the cusp anomalous dimension appearing in the ES formula.

First, though, we need a representation of the planar
four-loop four-gluon amplitude.  Rather than construct this 
representation from Feynman diagrams, we shall employ the unitarity
method~\cite{NeqFourOneLoop,Fusing,UnitarityMachinery,OneLoopReview,
TwoLoopSplitting}.  This method was also used to construct the planar 
two- and three-loop amplitudes~\cite{BRY,Iterate3}.
Because the unitarity method builds amplitudes at
any loop order from on-shell lower-loop amplitudes, structure
uncovered at the tree and one-loop levels can easily
be fed back into the construction of the higher-loop amplitudes.
We will find that the planar four-loop amplitude can be expressed
as a linear combination of just eight four-loop integrals.

A very interesting feature of the integrals appearing in the planar
four-point amplitudes through four loops is that they are all, in
a well-defined sense, conformally invariant.  To analyze the conformal 
properties of potential four-loop integrals, we make use of the recent
description of such integrals by Drummond, Henn, Sokatchev, and one 
of the authors~\cite{DHSS}.  

Once we know what four-loop four-point integrals enter into the amplitude,
we must compute them explicitly through $\Ord(\e^{-2})$.
Here we make use of important recent advances in multi-loop
integration~\cite{SmirnovDoubleBox,%
LoopIntegrationAdvance,SmirnovTripleBox,Tausk,TwoloopOffandMassive,
Buch,AnastasiouDaleo,CzakonMB}.
In particular, we use the program {\tt MB}~\cite{CzakonMB} to 
automatically extract poles in $\e$ from the Mellin-Barnes
representation of loop integrals, and to integrate
the resulting contour integrals.  We carry out the integrations
analytically for coefficients of the first five poles, 
$1/\e^8$ through $1/\e^4$.  These coefficients are expressed 
in terms of well-studied functions, harmonic polylogarithms 
(HPLs)~\cite{HPL,HPL2}, 
making it straightforward to confirm the expected infrared structure
for arbitrary values of the scattering angle.
For the coefficients of the $1/\e^3$ and $1/\e^2$ poles,
our analysis is numerical.  We evaluate the amplitude
at four kinematic points, 
$(s,t)=(-1,-1)$, $(-1,-2)$, $(-1,-3)$, and $(-1,-15)$.
Numerical evaluation suffices because the expected 
behavior is completely specified at order $1/\e^3$, and specified up
to one unknown, but predicted, constant, $f_0^{(4)}$, at order $1/\e^2$.
We find consistent results from all four kinematic points.

We also need to evaluate the infrared singular terms to the same order
$1/\e^2$.  These may be expressed in terms of lower-loop amplitudes.
For this purpose, we must expand the one-, two-, and
three-loop amplitudes to $\Ord(\e^4)$, $\Ord(\e^2)$ and $\Ord(\e^0)$,
respectively.  Fortunately, these are precisely the orders required to
test the full iteration relation at three loops, so the needed
lower-loop expressions can all be found, in analytic form, in
ref.~\cite{Iterate3}.  (These results are based partly on the earlier
evaluation of the double-box~\cite{SmirnovDoubleBox} and
triple-ladder~\cite{SmirnovTripleBox} integrals.)

We find that the ES conjecture is incorrect, although our numerical
results suggest that a simple modification of the four-loop ES
prediction may yield the correct answer.  In particular, if we flip
the sign of the $\zeta_3^2$ term in their prediction, we obtain
a value within the error bars of our result.  
If we choose to interpret the
modification as a dressing factor within the form taken
in ref.~\cite{EdenStaudacher}, it suggests taking
the value $\beta = \zeta_3$ for their parameter.  This value
would violate the apparent ``uniform transcendentality'' observed to date
generally for quantities in the $\NeqFour$ theory,
for example in the anomalous dimension of the higher-twist operator
$\Tr(X^2 Z^3)+\cdots$~\cite{EdenStaudacher}.
The same value, $\beta = \zeta_3$,
was suggested independently by BES~\cite{BESNew}, based on properties
of the spin-chain model.  This coincidence, modulo caveats we shall 
discuss in \sect{LargeCouplingSection}, provides some support
to violation of uniform transcendentality in quantities other 
than the cusp anomalous dimension.
This possibility can be tested via other perturbative computations;
if uniform transcendentality is nonetheless maintained,
our result might instead imply that the form postulated
for the dressing factor in ES and 
elsewhere~\cite{AFS04,OtherDressing,BeisertKlose,BeisertInvariant,%
BeisertDynamic,BeisertPhase,BHL} is not general enough.

We also use our four-loop results to investigate the strong-coupling
limit of the cusp anomalous dimension.  In the AdS/CFT correspondence,
this quantity can be computed from the energy of a long, folded string,
spinning in AdS${}_5$~\cite{StrongCouplingLeadingGKP}. 
The first two coefficients in the strong-coupling, large-$\Nc$ limit 
of the cusp anomalous dimension have been determined using a semi-classical
expansion based on this string
configuration~\cite{StrongCouplingLeadingGKP,Kruczenski,Makeenko,
StrongCouplingSubleading}.  
We shall discuss how our four-loop result can be used to give a
remarkably accurate estimate for these coefficients.
We employ an approximate formula devised by
Kotikov, Lipatov and Velizhanin~\cite{KLV,KLOV}
to interpolate between the weak- and strong-coupling regimes.
Using our four-loop result as input to this formula, the
first two strong-coupling coefficients are estimated to within
2.6\% and 5\%, respectively, of the values computed from string theory.
This agreement provides direct evidence in support of the 
AdS/CFT correspondence as well as a smooth transition between weak and 
strong coupling.

An even better approximation for the cusp anomalous dimension can be found
by incorporating into the interpolating formula the string predictions for
the first two coefficients in the strong-coupling expansion.  Based on our
success in accurately estimating the two leading strong-coupling
coefficients, we expect this improved approximation to be accurate to
within a few percent, for all values of the coupling.  Curiously, our
approximate formula predicts that the third coefficient in the
strong-coupling expansion should be very small, and may even vanish.
The formula also predicts the numerical value of the five-loop
coefficient.  This value turns out to be extremely close to a 
simple modification of the five-loop ES prediction, flipping the 
signs of the terms containing odd $\zeta$ values, as at four loops.
We have confirmed our analysis using Pad\'e approximants, which
also give insight into the complex analytic structure of $f_0(\ah)$.

Our representation of the planar four-loop four-gluon amplitude in 
terms of eight four-loop integrals can be used for more than just the 
extraction of the cusp anomalous dimension.   As mentioned above,
it can also be used to check the proposed iterative structure at four loops.  
In order to do so, one would need to evaluate the integrals all the way
through the finite terms, $\Ord(\ep^0)$, instead of just the level 
carried out in this paper, $\Ord(\ep^{-2})$.  One would also need 
to evaluate all integrals appearing in the lower-loop amplitudes 
to two orders higher in $\ep$.

This paper is organized as follows. In \sect{ReviewSection}, we review
the iterative structure of \SYM\ loop amplitudes, commenting in particular 
on how the cusp anomalous dimension appears in the
infrared singular terms.  In \sect{ConstructionSection}, we present the
construction of the four-loop amplitudes via the unitarity method and
also describe the conformal properties of the resulting integrals.  In
\sect{FourLoopAmpSection}, we give analytical results for the
amplitudes through $\Ord(\ep^{-4})$ and numerical results though
$\Ord(\ep^{-2})$, allowing us to extract a numerical value for the
four-loop cusp anomalous dimension.  The four-loop anomalous dimension
is then used in \sect{LargeCouplingSection} to estimate the
coefficients that appear at strong coupling and also to estimate higher-loop
contributions to the cusp anomalous dimension.  Our conclusions are
given in \sect{AnalysisSection}.  Two appendices are included,
one presenting Mellin-Barnes representations for the integrals 
appearing in the four-loop amplitude, and one reviewing properties 
of harmonic polylogarithms.

\section{Iterative structure of \SYM\ loop amplitudes}
\label{ReviewSection}

In this paper we consider the planar contributions to 
gluonic scattering in \SYM\ with gauge group $SU(N_c)$,
that is, the leading terms as $N_c \to \infty$.
We do not discuss subleading-color contributions; at present
they do not appear to have a simple iterative structure~\cite{Iterate2}.  

The leading-color terms have the same color structure as the 
corresponding tree amplitudes.  The leading-$N_c$
contributions to the $L$-loop $SU(N_c)$ gauge-theory $n$-point
amplitudes may be written as,
\begin{eqnarray}
{\cal A}_n^{(L)} & = & g^{n-2}
 \Biggl[ { 2 e^{- \e \gamma} g^2 N_c \over (4\pi)^{2-\e} } \Biggr]^{L}
 \sum_{\sigma\in S_n/Z_n}
\Tr( T^{a_{\sigma(1)}} 
   \ldots T^{a_{\sigma(n)}} )
               A_n^{(L)}(\sigma(1), \sigma(2), \ldots, \sigma(n))\,,
\label{LeadingColorDecomposition}
\end{eqnarray}
where $\gamma$ is Euler's constant, and
the sum runs over non-cyclic permutations of the external legs.
In this expression we have suppressed the (all-outgoing) momenta $k_i$ 
and helicities $\lambda_i$, leaving only the index $i$ as a label.  This
decomposition holds for all particles in the gauge super-multiplet,
as they are all in the adjoint representation.  The color-ordered 
partial amplitudes $A_n$ are independent of the color factors, 
and depend only on the kinematics.  For \SYM, supersymmetric
Ward identities~\cite{SWI} imply 
that the four-gluon helicity amplitudes 
$({+}{+}{+}{+})$ and $({-}{+}{+}{+})$ (as well as their parity conjugates)
vanish identically.   Furthermore, the nonvanishing four-point
(MHV) amplitudes are all related by simple overall factors.   
Hence we do not need to specify the helicity configuration, 
{\it i.e.} whether the color ordering is $({-}{-}{+}{+})$ 
or $({-}{+}{-}{+})$.

It is convenient to scale out a factor of the tree amplitude,
and work with the quantities $M_n^\Lloop$ defined by
\begin{equation}
M_n^\Lloop(\rho;\e) = A_n^\Lloop(\rho)/A_n^\tree(\rho) \,.
\label{RescaledLoopAmplitude}
\end{equation}
Here $\rho$ indicates the dependence on the external momenta,
$\rho \equiv \{ s_{12}, s_{23}, \ldots \}$, 
where $s_{i(i+1)} = (k_i+k_{i+1})^2$ are invariants built
from color-adjacent momenta. 
The iteration relation proposed in ref.~\cite{Iterate3} 
then takes the form,
\begin{equation}
{\cal M}_n(\rho) \equiv 1 + \sum_{L=1}^\infty a^L M_n^{(L)}(\rho;\e) 
= \exp\Biggl[\sum_{l=1}^\infty a^l 
          \Bigl(f^{(l)}(\e) M_n^{(1)}(\rho;l \e) + C^{(l)} 
               + E_n^{(l)}(\rho;\e)  \Bigr) \Biggr] \,.
\label{ExponentialResum}
\end{equation}
In this expression, the factor,
\be
a \equiv { N_c \alpha_s \over 2\pi } (4\pi e^{-\gamma})^\e\,,
\label{alphaberdef}
\ee
keeps track of the loop order of perturbation theory, and coincides 
with the prefactor in brackets in \eqn{LeadingColorDecomposition}.
(It becomes equal to $\ah$ in \eqn{ahdef} as $\e \to 0$.)
The quantity $M_n^{(1)}(\rho;l\e)$ is the one-loop amplitude,
with the tree amplitude scaled out according 
to~\eqn{RescaledLoopAmplitude}, and with the substitution
$\e \to l\e$ performed.  (That is, it is evaluated in 
$d = 4 - 2l\e$.)  Each $f^{(l)}(\e)$ is given by a three-term series
in $\e$, beginning at $\Ord(\e^0)$,
\be
 f^{(l)}(\e) = f_0^{(l)} + \e f_1^{(l)} + \e^2 f_2^{(l)} \,.
\label{flexp}
\ee
The objects $f_k^{(l)}$, $k=0,1,2$, and $C^{(l)}$ are pure constants,
independent of the external kinematics $\rho$, and also independent of the 
number of legs $n$.  We expect them to be polynomials in the 
Riemann values $\zeta_m$ with rational coefficients, and a uniform degree 
of transcendentality, which is equal to $2l-2+k$ for $f_k^{(l)}$,
and $2l$ for $C^{(l)}$.  (Multiple zeta values $\zeta_{n_1,n_2,\ldots}$
may also appear, but there are no independent ones of weight less than
eight, and so they can only appear starting at five-loop order.)
The $f_k^{(l)}$ and $C^{(l)}$ can be determined by matching
to explicit computations.  The one-loop values are defined to be,
\begin{equation}
f^{(1)}(\e) = 1\,, \hskip 2 cm C^{(1)} = 0\,, 
\hskip 2 cm E_n^{(1)}(\rho;\e) = 0\,.
\label{OneLoopfCE}
\end{equation}
The $E_n^{(l)}(\rho;\e)$ are non-iterating $\Ord(\e)$ contributions to the
$l$-loop amplitudes, which vanish as $\e \rightarrow 0$, $E_n^{(l)}(\rho;0)
= 0$.  These terms contribute to the exponentiated form of the
amplitudes~(\ref{ExponentialResum}) even for $\ep \rightarrow 0$
because they can appear multiplied by the infrared-divergent
parts of the one-loop amplitude $M_n^{(1)}(\rho;l \e)$.
After canceling the infrared divergences between real
emission and virtual contributions, such terms should not contribute
to infrared-safe observables.

The first two values of $f^{(l)}(\e)$ in the three-term 
expansion~(\ref{flexp}), namely $f^{(l)}_0$ and
$f^{(l)}_1$, can be identified with quantities appearing in the
resummed Sudakov form factor~\cite{MagneaSterman},
\begin{eqnarray}
f^{(l)}_0 &=& {1\over4} \, \hat\gamma_K^{(l)} \,,
\label{f0toK}\\
f^{(l)}_1 &=& {l \over 2} \, \hat{\cal G}_0^{(l)} \,.
\label{f1toG}
\end{eqnarray}
The first object, $f^{(l)}_0$, is identified
with the $l$-loop cusp anomalous dimension.
The quantities $f^{(l)}_0$ and $\hat{\cal G}_0^{(l)} = (2/l)f^{(l)}_1$
are known through three loops~\cite{KL02,Makeenko,KLV,KLOV,Iterate3,MOS},
\begin{eqnarray}
f_0^{(1)} &=& 1 \,, \nn\\
f_0^{(2)} &=& - \zeta_2 \,, \label{f0Values}\\
f_0^{(3)} &=& {11\over2} \zeta_4 \nn
\end{eqnarray}
(see also \eqn{gammaKB}), and
\begin{eqnarray}
\hat{\cal G}_0^{(1)} &=& 0 \,, \nn\\
\hat{\cal G}_0^{(2)} &=& - \zeta_3 \,, \label{calGValues}\\
\hat{\cal G}_0^{(3)} &=& 4 \zeta_5 + {10\over 3} \zeta_2 \zeta_3 \,. \nn
\end{eqnarray}
A principal task of this paper is to compute $f_0^{(4)}$
and compare the result with the prediction~(\ref{gammaKB}).

\Eqn{ExponentialResum} is equivalent~\cite{Iterate3} to
\begin{equation}
M_n^{(L)}(\rho;\e) = X_n^{(L)}\bigl[M_n^{(l)}(\rho; \e)\bigr] 
  + f^{(L)}(\e) M_n^{(1)}(\rho; L \e) + C^{(L)} 
               + E_n^{(L)}(\rho; \e) \,,
\label{iterX}
\end{equation}
where the quantities $X_n^{(L)} = X_n^{(L)}[M_n^{(l)}]$ 
only depend on the lower-loop amplitudes $M_n^{(l)}(\rho;\e)$ 
with $l<L$.  
The $X_n^{(L)}$ can be computed simply by performing
the following Taylor expansion,
\begin{equation}
X_n^{(L)}\bigl[ M_n^{(l)} \bigr]
= M_n^{(L)} 
- \ln\Biggl( 1 + \sum_{l=1}^\infty a^l M_n^{(l)} \Biggr) 
\Biggr\vert_{a^L\ {\rm term}} \,.
\label{Xsol}
\end{equation}
\Eqns{iterX}{Xsol} express the $L$-loop amplitude explicitly 
in terms of lower-loop amplitudes, plus constant remainders.
Here we need the values of $X_n^{(L)}$ for $L=2,3,4$,
\begin{eqnarray}
X_n^{(2)}\bigl[M_n^{(l)}\bigr] 
                  &=& {1\over2} \Bigl[ M_n^{(1)} \Bigr]^2  \,, 
\label{X2} \\
X_n^{(3)}\bigl[M_n^{(l)}\bigr]  &=& - {1\over3} \Bigl[ M_n^{(1)} \Bigr]^3 
            + M_n^{(1)} M_n^{(2)}\,, 
\label{X3}\\
X_n^{(4)}\bigl[M_n^{(l)}\bigr]  &=& {1\over4} \Bigl[ M_n^{(1)} \Bigr]^4 
            - \Bigl[ M_n^{(1)} \Bigr]^2 M_n^{(2)}
            + M_n^{(1)} M_n^{(3)} 
            + {1\over2} \Bigl[ M_n^{(2)} \Bigr]^2 \,.
\label{X4}
\end{eqnarray}

We note that the exponentiated result~(\ref{ExponentialResum})
leads to a simple exponentiated form for suitably-defined 
``finite remainders'' $F_n^{(L)}$ associated with the multi-loop
amplitudes~\cite{Iterate3}.  We define
\begin{equation}
F_n^{(L)}(\rho; \e) = M_n^{(L)} 
       -  \sum_{l=0}^{L-1} \hat I_n^{(L-l)} \, M_n^{(l)} \,,
\label{FnLdef}
\end{equation}
where the $\hat I_n^{(L-l)}(\rho;\e)$ are iteratively-defined divergent terms,
and $M_n^{(0)} \equiv 1$.  After some algebra,
one finds that $I_n^{(L)}(\rho;\e)$ and $F_n^{(L)}(\rho;\e)$ obey
iterative relations very similar to \eqn{iterX}.
In the limit as $\e\to0$, the relation for $F_n^{(L)}(\rho;\e)$ becomes
\begin{eqnarray}
F_n^{(L)}(\rho;0) &\equiv& X_n^{(L)}\bigl[ F_n^{(l)}(\rho;0) ]
                  + f^{(L)}(\rho;0) F_n^{(1)}(\rho;0) + C^{(L)} \,. 
\label{Fsolzero}
\end{eqnarray}
Because $\e$ has disappeared from \eqn{Fsolzero}, it can be 
solved neatly for $F_n^{(L)}(\rho;0)$ for any $L$, 
in terms of the one-loop remainder $F_n^{(1)}(\rho;0)$ alone~\cite{Iterate3}.
The solution can be represented as,
\begin{eqnarray}
{\cal F}_n(\rho;0) \equiv 1 + \sum_{L=1}^\infty \ah^L F_n^{(L)}(\rho;0)
&=&  \exp\Biggl[\sum_{l=1}^\infty \ah^l 
          \Bigl(f^{(l)}_0 F_n^{(1)}(\rho;0) + C^{(l)} \Bigr) \Biggr]
\nonumber\\
&\equiv& 
\exp\Biggl[ {1\over 4} \gamma_K(\ah) \ F_n^{(1)}(\rho;0) + C(\ah) \Biggr] \,,
\label{F0Resum}
\end{eqnarray}
where $C(\ah) = \sum_{l=1}^\infty C^{(l)} \ah^l$ and we used the relation
(\ref{f0toK}) of $f^{(l)}_0$ to the cusp anomalous dimension.  The result
for $F_n^{(L)}(\rho;0)$ is given by the $\ah^L$ term in the Taylor
expansion of the exponential. 

Next we present the specific forms of the iterative amplitude
relations~(\ref{iterX}) through four loops, specializing to $n=4$.
The two-loop version is~\cite{Iterate2}
\begin{equation}
M_4^{\twoloop}(\rho;\e) 
= {1 \over 2} \Bigl[ M_4^{\oneloop}(\rho;\e) \Bigr]^2
 + f^\twoloop(\e) \, M_4^{\oneloop}(\rho;2\e) + C^{(2)}
 + \Ord(\e) \,,
\label{TwoLoopOneLoopAgainn}
\end{equation}
where
\begin{equation}
f^\twoloop(\e) = - (\zeta_2 + \zeta_3 \e + \zeta_4 \e^2) \,,
\label{f2def}
\end{equation}
and the constant $C^{(2)}$ is given by
\begin{equation}
 C^{(2)} =  - {1\over 2} \zeta_2^2 \,.
\label{C2def}
\end{equation}
The three-loop version, explicitly verified in ref.~\cite{Iterate3}, is
\begin{eqnarray}
M_4^\threeloop(\rho;\e) &=& - {1\over 3} \Bigl[M_4^\oneloop(\rho;\e)\Bigr]^3
            + M_4^\oneloop(\rho;\e)\, M_4^\twoloop(\rho;\e)
            + f^\threeloop(\e) \, M_4^\oneloop (\rho; 3\,\e) \nn\\
&& \hskip 1cm \null 
   +  C^{(3)} + \Ord(\e) \,,
\label{ThreeLoopFourPtIteration}
\end{eqnarray}
where
\begin{equation}
f^\threeloop(\e) = {11\over 2} \, \zeta_4
                 + \e (6 \zeta_5 + 5 \zeta_2 \zeta_3 ) +
   \e^2 (c_1 \zeta_6 +  c_2\zeta_3^2) \,,
\label{f3def}
\end{equation}
and the constant $C^{(3)}$ is given by
\begin{equation}
 C^{(3)} =  \biggl( {341\over 216} \, + {2\over 9} c_1 \biggl) \zeta_6
            + \biggl( - {17\over 9} + {2\over 9} c_2 \biggr)\zeta_3^2\,.
\label{C3def}
\end{equation}
The constants $c_1$ and $c_2$ are expected to be rational numbers.
They drop out from the right-hand side of 
\eqn{ThreeLoopFourPtIteration} because of a cancellation between
$f_2^{(3)}$ and $C^{(3)}$.  A computation of the three-loop five-point
amplitude, or of the three-loop splitting amplitude, could be used to 
determine them.

The four-loop iteration relation would have the following form,
\begin{eqnarray}
M_4^\fourloop(\rho;\e) &=& {1\over4} \Bigl[ M_4^\oneloop(\rho;\e) \Bigr]^4 
            - \Bigl[ M_4^\oneloop(\rho;\e) \Bigr]^2 M_4^\twoloop(\rho;\e)
            + M_4^\oneloop(\rho;\e) M_4^\threeloop(\rho;\e) 
\nonumber\\
&& \hskip1cm\null 
            + {1\over2} \Bigl[ M_4^\twoloop(\rho;\e) \Bigr]^2
            + f^\fourloop(\e) \, M_4^\oneloop (\rho; 4\,\e) + C^{(4)}
            + \Ord(\e) \,.
\label{FourLoopFourPtIteration}
\end{eqnarray}
As we shall not be computing the $1/\e$ and finite terms in the
present paper, we cannot do more here than verify the (universal)
divergent terms and extract the value of the four-loop cusp anomalous
dimension.  We leave to future work the important task of verifying
\eqn{FourLoopFourPtIteration}, using the integral form of the
four-loop amplitude presented in this paper.

\section{Construction of four-loop planar \SYM\ loop amplitude}
\label{ConstructionSection}

The unitarity
method~\cite{NeqFourOneLoop,Fusing,UnitarityMachinery,OneLoopReview,
TwoLoopSplitting} is an efficient way to determine the representations
of loop amplitudes in terms of basic loop integrals.  The coefficients
of the loop integrals are obtained by sewing sets of on-shell
tree amplitudes.  If we are using a four-dimensional form of
the unitarity method, the tree amplitudes can be significantly
simplified before sewing. At one-loop supersymmetric amplitudes
are fully determined from their four-dimensional cuts, but unfortunately
for higher loops no such theorem has been proven. 
In the present calculation, we will
therefore use $D$-dimensional unitarity, for which we will need the tree
amplitudes to be evaluated
without assuming the four-dimensional helicity states 
for the external legs.  These amplitudes are nonetheless simpler than
the completely off-shell amplitudes that would implicitly arise
in a conventional Feynman-diagram calculation.
For \SYM, a key feature is that the on-shell tree amplitudes have the full
$\NeqFour$ supersymmetry manifest, in the form of simple $S$-matrix
Ward identities~\cite{SWI}.
It is impossible to maintain the full $\NeqFour$ supersymmetry
in any off-shell formalism, because the superspace constraints
imply the equations of motion via the Bianchi identities~\cite{GSWANP}.
The unitarity method derives its efficiency
from the ability to use simplified forms of tree amplitudes to produce
simplified loop integrands.  (To maintain the supersymmetry, 
we apply the four-dimensional helicity (FDH) scheme~\cite{FDH}, 
a variation on dimensional reduction (DR)~\cite{Siegel}, in performing
the sum over intermediate gluon polarization states.) 

\begin{figure}[t]
\centerline{\epsfxsize 5.5 truein \epsfbox{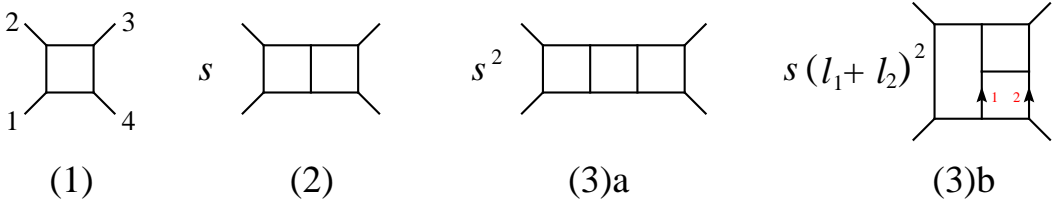}}
\caption{Integrals required for $gg \to gg$ scattering in planar \SYM\
at one loop (1), two loops (2) and three loops ((3)a and (3)b).
The box (1), planar double box (2) and three-loop ladder (3)a 
integrals are scalar integrals, with no loop-momentum dependent
factors in the numerator.  The tennis-court integral (3)b contains a factor
of $(l_1+l_2)^2$, where $l_1$ and $l_2$ are marked with arrows
in the figure.
}
\label{LowerLoopFigure}
\end{figure}

The unitarity method expresses the amplitude in terms of a set of loop
integrals.  In general gauge theories, such as QCD, the number of 
required integrals proliferates rapidly as the number of loops increases, 
and sophisticated algorithms based on integration-by-parts
identities~\cite{IBP} have been devised to relate such integrals to a smaller
class of master integrals, successfully through two 
loops~\cite{IBP2loop}.
Fortunately, the number of required integrals grows much more slowly
for gluon-gluon scattering in planar $\NeqFour$ super Yang-Mills theory.  
At $L=1,2,3$ the respective numbers are $1,1,2$, and the required
integrals are all shown in \fig{LowerLoopFigure}.
We will see that at $L=4$ eight integrals are required.

The result for the one-loop four-point amplitude is~\cite{GSB}
\begin{equation}
M_4^{\oneloop}(\e) =
- {1 \over 2} \, \I^\oneloop(s,t) \,,
\label{OneLoopAmplitude}
\end{equation}
where the Mandelstam variables are $s = (k_1 + k_2)^2$ 
and $t = (k_2 +k_3)^2$.
The factor of $1/2$ in~\eqn{OneLoopAmplitude} follows from our 
normalization convention for $A_n^\Lloop$, which is defined
by \eqn{LeadingColorDecomposition}.
The one-loop scalar box integral $\I_4^\oneloop(s,t)$ (multiplied by a 
convenient normalization factor), depicted in \fig{LowerLoopFigure},
is
\begin{equation}
\I^\oneloop(s,t) \equiv s t \, I_4^\oneloop(s,t) \,,
\end{equation}
where $I_4^\oneloop(s,t)$ is defined in eq.~(B1) of
ref.~\cite{Iterate3}.  We absorb the factor of $st$ into the
definition of the integrals we use, because it cancels a factor of
$1/(st)$ appearing in the explicit expression for $I_4^\oneloop(s,t)$,
and matches the form in which it appears in the \SYM\ amplitudes.
This integral is given in terms of HPLs in eq.~(B1) of that reference,
through the order we require here, $\Ord(\e^4)$.  The higher-order
terms in $\e$ are needed to be able to evaluate terms in the
infrared/iterative representation~(\ref{FourLoopFourPtIteration}) of
the four-loop amplitude through $\Ord(\e^{-2})$.  For example, in the
term $[ M_4^\oneloop(\e) ]^4\ \propto\ [ \I^\oneloop(\e) ]^4$ in
\eqn{FourLoopFourPtIteration}, if one takes the leading $1/\e^2$ term
from three of the four factors, then the coefficient of $\e^4$ in the
fourth factor contributes to the $\Ord(\e^{-2})$ term in the product.

The planar two-loop \SYM\ four-point amplitude is given by~\cite{BRY}
\begin{equation}
M^\twoloop_4(\e) =
{1\over 4} \,
 \Bigl[  \I^\twoloop(s,t) + \I^\twoloop(t,s) \Bigr] \,.
\label{TwoloopPlanarResult}
\end{equation}
The two-loop scalar double-box integral, shown 
in \fig{LowerLoopFigure}, is
\begin{equation}
\I^\twoloop(s,t) \equiv s^2 t \, I_4^\twoloop(s,t) \,, 
\end{equation}
where $I_4^\twoloop(s,t)$ is defined in 
eq.~(B4) of ref.~\cite{Iterate3}.   As at one loop,
we have rescaled the integral to remove the rational prefactor.
This integral was first evaluated through $\Ord(\e^0)$ in terms
of polylogarithms~\cite{SmirnovDoubleBox}.  
Because the infrared/iterative 
expression~(\ref{FourLoopFourPtIteration})
contains, for example,
$[ M_4^\twoloop(\e) ]^2\ \propto\ [ \I^\twoloop(\e) ]^2$,
and because the expansion of $\I^\twoloop(\e)$ begins at
order $1/\e^4$, we need its expansion through $\Ord(\e^2)$.
This expansion is presented in terms of HPLs in eq.~(B5) of
ref.~\cite{Iterate3}.

The three-loop planar amplitude is given by~\cite{BRY,Iterate3}
\begin{equation}
M^\threeloop_4(\e) =
-{1\over8} \,
\Bigl[  \I^{\threeloop{\rm a}}(s,t) +
   2 \, \I^{\threeloop{\rm b}}(t,s) +
        \I^{\threeloop{\rm a}}(t,s) +
   2 \, \I^{\threeloop{\rm b}}(s,t) \Bigr] \,.
\label{ThreeLoopPlanarResult}
\end{equation}
The scalar triple-ladder and non-scalar ``tennis-court'' integrals,
illustrated in \fig{LowerLoopFigure}, are
\begin{eqnarray}
&& \I^{\threeloop{\rm a}} 
\equiv s^3 t \,  I_4^{\threeloop \rm a}(s,t)\,, \nn \\
&& \I^{\threeloop{\rm b}} 
\equiv s t^2 \,  I_4^{\threeloop \rm b}(s,t)\,,
\end{eqnarray}
where $I_4^{\threeloop \rm a}(s,t)$ and  $I_4^{\threeloop \rm b}(s,t)$
are defined in
eqs. (3.1) and (3.2), respectively, of ref.~\cite{Iterate3}.
Because these integrals multiply $\I^\oneloop$ in the term
$M_4^\oneloop(\e) M_4^\threeloop(\e)$ in \eqn{FourLoopFourPtIteration},
we need their expansion through $\Ord(\e^0)$.
These expansions were first carried out in terms of HPLs
for $I_4^{\threeloop \rm a}$ in ref.~\cite{SmirnovTripleBox}, 
and for $I_4^{\threeloop \rm b}$ in ref.~\cite{Iterate3}.
The results are collected in eqs.~(B7) and (B9) of ref.~\cite{Iterate3}.

The coefficients of the integrals in the two- and three-loop
expressions~(\ref{TwoloopPlanarResult}) and
(\ref{ThreeLoopPlanarResult}) were originally determined~\cite{BRY}
using iterated two-particle cuts.  Such cuts can be evaluated to all
orders in $\e$ because $\NeqFour$ supersymmetry relates all nonvanishing
four-point amplitudes; therefore precisely the same algebra enters as
at one loop, for which it leads to the
amplitude~(\ref{OneLoopAmplitude}).  More generally, an ansatz for the
planar contributions to the integrands was proposed in terms of a
``rung insertion rule''~\cite{BRY,BDDPR} to be described below, which
was based largely on the structure of the iterated two-particle cuts.
At three loops, the planar integrals generated by the rung rule can
all be constructed using iterated two-particle cuts.  Also, the
three-loop planar amplitude~(\ref{ThreeLoopPlanarResult}) has the
correct infrared poles and a remarkable iterative
structure~\cite{Iterate3}, so there is little doubt that it is the
complete answer.

However, beyond three loops --- and even at three loops for non-planar 
contributions --- the rung rule generates graphs that cannot 
be obtained using iterated two-particle cuts.  
It is less certain that the rung rule gives the correct results 
for such contributions.  Indeed, we shall see that there are 
additional, non-rung-rule, contributions to the planar 
amplitude beginning at four loops.

Nevertheless, we start constructing the planar 
four-loop \SYM\ amplitude using the diagrams generated by
the rung rule.  According to this rule,
each diagram in the planar $L$-loop amplitude can be used
to generate planar $(L+1)$-loop diagrams as follows:
First, one generates a set of diagrams by inserting a new line 
joining each possible pair of internal lines.  
Next, one removes from this set all diagrams with triangle
or bubble subdiagrams.   Besides the scalar propagator associated
with the new line, one also includes an additional numerator factor for
the diagram, beyond that inherited from the $L$-loop diagram, 
of $i(l_1+l_2)^2$.  Here $l_1$ and $l_2$ are the momenta
flowing through each of the legs to which the new line is joined.  
Each distinct $(L+1)$-loop contribution is counted once, even if it 
can be generated in multiple ways.  (Contributions corresponding 
to identical graphs but with different numerator factors should be 
counted as distinct.)  Rung-rule diagrams have also been referred
to as ``Mondrian diagrams'' because of their visual
similarity~\cite{Iterate3}.

At one loop, the only no-triangle graph for the four-point process
is the box graph depicted in \fig{LowerLoopFigure}.  
In going to two loops, there is only one inequivalent way to 
add a rung to the box graph without creating a triangle.  
Connecting opposing sides of the box, we form the planar double 
box in \fig{LowerLoopFigure}.  
(One might also imagine attaching a propagator between two adjacent external 
legs.  However, this operation yields the same double box.)
To generate the three-loop graphs, we add a rung, either
vertically or horizontally, to one of the boxes of the two-loop
box diagram, thus yielding the triple ladder (3)a and tennis-court
integral (3)b of \fig{LowerLoopFigure}.
(Attaching a propagator between external legs again gives nothing new.)
Of course, there are several permutations of external legs present 
for any given type of graph; all of them are produced by the rung rule.
Here we are identifying only the topologically distinct integrals 
that arise.

%
\begin{figure}[t]
\centerline{\epsfxsize 6.4 truein \epsfbox{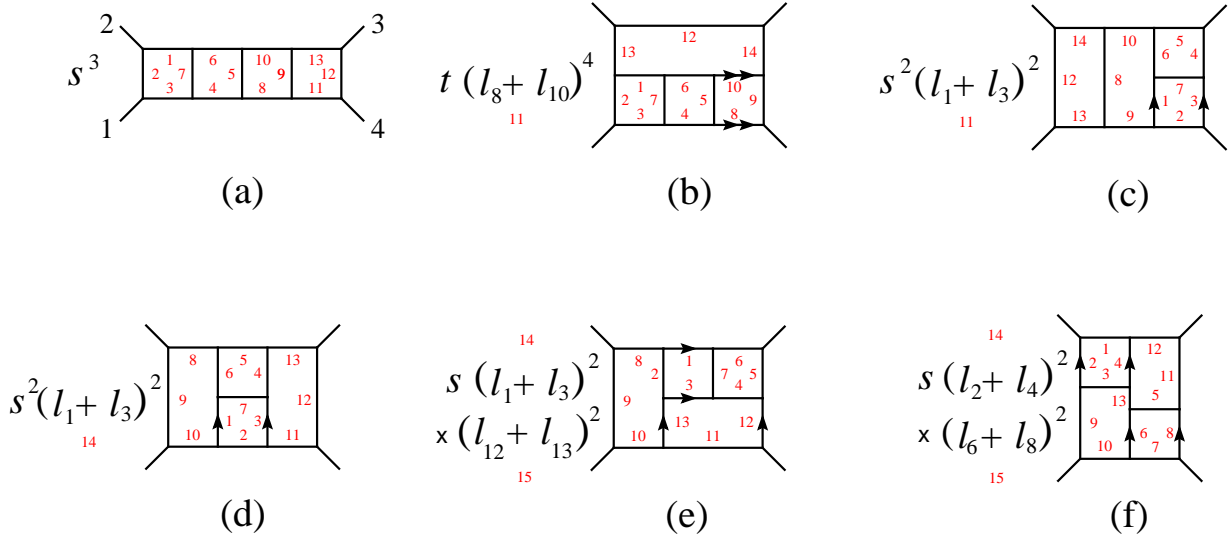}}
\caption{``Rung-rule'' contributions to the leading-color four-loop amplitude,
in terms of integral functions given in 
eqs.~(\ref{MBIntegralA})--(\ref{MBIntegralF}). 
An overall factor of $st$ has been suppressed in each figure,
compared with the definitions in 
eqs.~(\ref{MBIntegralA})--(\ref{MBIntegralF}).}
\label{rrFigure}
\end{figure}

What happens when we try to add rungs to the three-loop integrals?
There are two inequivalent ways to add a rung inside either
the left- or the right-most box of the triple ladder integral
in \fig{LowerLoopFigure}; 
these give the integrals of \fig{rrFigure}(a) and~(c).  
Adding a vertical rung inside the middle
box does not yield a topologically distinct integral.  Adding
a horizontal rung inside the middle box yields the integral
of \fig{rrFigure}(d).  Inserting a vertical rung inside the upper-right
box of the tennis-court integral in \fig{LowerLoopFigure}
gives us \fig{rrFigure}(e), and a horizontal one, 
\fig{rrFigure}(b).  Finally, adding a horizontal rung inside the left-side
box of the tennis-court integral gives us a new kind of integral,
shown in~\fig{rrFigure}(f), which has no two-particle cuts.

\begin{figure}[t]
\centerline{\epsfxsize 5.5 truein \epsfbox{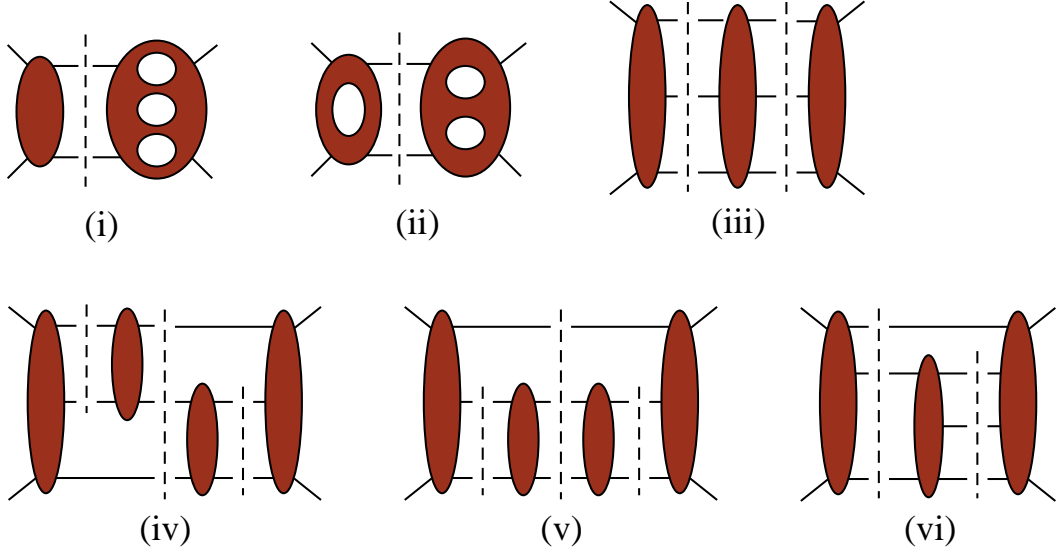}}
\caption{Generalized cuts that provide information about the 
planar four-loop amplitude.  (i) A two-particle cut separating 
a tree amplitude from a three-loop amplitude.  
(ii)  A two-particle cut separating a one-loop amplitude from 
a two-loop amplitude. (iii) A ``3--3'' cut separating the amplitude
into a product of three tree amplitudes.  (iv) An ``upper-2--3--lower-2''
cut separating the amplitude into a product of four tree amplitudes. 
(v) A ``lower-2--3--lower-2'' cut.  (vi) A ``3--lower-3'' cut.
}
\label{CutsFigure}
\end{figure}

The propagators and momentum numerators present in integrals 
\fig{rrFigure}(a)--(e) are determined by the two-particle cuts,
as depicted in \fig{CutsFigure}(i) and (ii).  
That is, the rung rule is guaranteed to be correct for them.  
There is no such guarantee
for integral~\fig{rrFigure}(f), which has no two-particle
cut with real external momenta.  In order to check it, we need
to compute a three-particle cut.  We chose to perform the 
generalized-unitarity cut
of \fig{CutsFigure}(iv), a threefold-cut with a central three-particle
cut and two secondary two-particle cuts.  This cut reveals that the 
rung rule is more robust than might have been expected, 
based on its origin in iterated two-particle cuts:  
The rule does in fact give the correct form for the numerator 
of the integral in~\fig{rrFigure}(f), even though no two-particle cut
can detect it.

Because we have no proof that the $(-2\e)$-dimensional parts of loop
momenta are unimportant to the computation, we perform these
calculations in $D$ dimensions.  For this purpose, we need tree
amplitudes with (some of) the external states kept in general
dimension.  In the three-particle cuts, one has contributions from
three-gluon states, and from states with gluons and fermion (gluino)
pairs crossing the cut, in addition to states with scalar pairs or
fermion pairs and a lone scalar.  In principle, one could
evaluate these cuts by computing all the required amplitudes, and
summing over the particle multiplet.  However, it is easier to use a
trick.  The trick takes advantage of the fact that the \SYM\ multiplet,
consisting of the gluon, four Majorana fermions, and six real scalars,
can also be understood as a single $\NeqOne$ multiplet in ten
dimensions.  Instead of summing over the multiplet seen as the
$\NeqFour$ multiplet in four dimensions, we sum over the $\NeqOne$
multiplet in ten dimensions.  This trick reduces the number of types
of intermediate states one has to consider, though of course the total
number of states is unchanged.  The loop momenta are in any event kept
in $D$ dimensions.  (The trick is compatible with the FDH
regularization scheme~\cite{FDH}, where the momenta are taken to be in $D$
dimensions, but the number of states in the loops is kept at the
four-dimensional value.)

%
\begin{figure}[t]
\centerline{\epsfxsize 3.4 truein \epsfbox{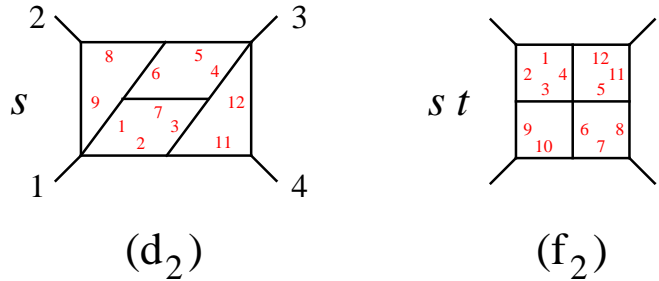}}
\caption{Non-rung-rule contributions to the leading-color four-loop amplitude,
in terms of integral functions defined in
eqs.~(\ref{MBIntegralD2}) and (\ref{MBIntegralF2}).  
Integral (d${}_2$) follows the labeling of integral (d) in \fig{rrFigure},
and integral (f${}_2$) follows the labeling of integral (f).
An overall factor of $st$ has been suppressed in each figure,
compared with eqs.~(\ref{MBIntegralD2}) and (\ref{MBIntegralF2}).}
\label{nonrrFigure}
\end{figure}

The cut of \fig{CutsFigure}(iv) also reveals the presence of a
non-rung-rule integral, shown in \fig{nonrrFigure}(d${}_2$).  
The integral has, of course, no two-particle cuts, but it can be
obtained from the integral topology of \fig{rrFigure}(d) 
by canceling two propagators, labeled 10 and 13.
The integral of \fig{rrFigure}(f) could also have been checked using
another generalized-unitarity cut, the two-fold three-particle cut
of \fig{CutsFigure}(iii).  This cut also reveals the presence of
the ``four-square'' integral, shown in~\fig{nonrrFigure}(f${}_2$).  
This integral can be obtained from the topology of \fig{rrFigure}(f)
by canceling the propagator labeled 13.
The cut of \fig{CutsFigure}(iii)
also detects the integral of \fig{nonrrFigure}(d${}_2$).

Under mild assumptions, which we discuss in \sect{correctnesssection},
it is sufficient to compute two additional multiple cuts,
shown in \fig{CutsFigure}(v) and (vi), in order
to rule out any additional contributions.  We have computed these cuts,
and we indeed find that no additional integrals appear.

\section{Integral representation of the four-loop planar amplitude}
\label{FourLoopAmpSection}

We find that the four-loop planar amplitude is given by
\begin{eqnarray}
M^\fourloop_4(\e) &=&
{1\over16}
\Bigl[ \I^{\rm (a)}(s,t) + \I^{\rm (a)}(t,s)
     + 2 \, \I^{\rm (b)}(s,t) + 2 \, \I^{\rm (b)}(t,s)
     + 2 \, \I^{\rm (c)}(s,t) + 2 \, \I^{\rm (c)}(t,s)
\nonumber\\
&&\hskip0.4cm\null
     + \I_4^{\rm (d)}(s,t) + \I^{\rm (d)}(t,s)
     + 4 \, \I^{\rm (e)}(s,t) + 4 \, \I^{\rm (e)}(t,s)
     + 2 \, \I^{\rm (f)}(s,t) + 2 \, \I^{\rm (f)}(t,s)
\nonumber\\
&&\hskip0.4cm\null
     - 2 \, \I^{\rm (d_2)}(s,t) - 2 \, \I^{\rm (d_2)}(t,s)
     - \I^{\rm (f_2)}(s,t)
 \Bigr] 
\,.
\label{FourLoopPlanarResult}
\end{eqnarray}
The rung-rule integrals $\I^{\rm (a)}$ through $\I^{\rm (f)}$ are depicted
in \fig{rrFigure}.  The two additional integrals, $\I^{\rm (d_2)}$ and
$\I^{\rm (f_2)}$, depicted in \fig{nonrrFigure}, do not follow from
the rung rule, and were detected using generalized cuts with at 
least one three-particle channel, as discussed in 
\sect{ConstructionSection}.

The integrals appearing in the four-point amplitude are defined
generically, for diagram ``$(x)$'' by
\begin{equation}
\I^{(x)}(s,t) \equiv
(-i e^{\e \gamma} \pi^{-d/2})^4
\int \dd^d p\,\dd^d q\,\dd^d u\, \dd^d v\
  { s t \, {\cal N}^{(x)} \over \prod_j p_j^2 } \,,
\label{genericintdef}
\end{equation}
where $p,q,u,v$ are the four independent loop integration variables,
and $d=4-2\e$.  The product in the denominator of \eqn{genericintdef}
runs over the labels of internal lines in the graph $(x)$.  (For
graphs (b) and (c), line 11 corresponds to a numerator factor, so it
should be excluded from this product.  Similarly, lines 10 and 13 are
to be omitted from the denominator product for graph (d${}_2$), and
line 13 from the product for graph (f${}_2$).)  Each line carries
momentum $p_j$, which is some linear function of $p,q,u,v$ and the
external momenta.  The line label $j$ is shown next to each internal
line.  The numerator factor ${\cal N}^{(x)}$ is also shown explicitly,
to the left of the graph for $(x)$.  We have omitted an overall factor
of $st$ from ${\cal N}^{(x)}$ associated with each integral from the
figure, in order to avoid cluttering it.

For the quadruple-ladder graph (a), and for the non-rung-rule graphs
(d${}_2$) and (f${}_2$), ${\cal N}^{(x)}$ is completely independent of the 
loop momentum, and so it may be pulled outside of the integral.
For the other graphs, at least one factor in ${\cal N}^{(x)}$ 
depends on the loop momenta.  In generating a Mellin-Barnes 
representation for these integrals, we think of these factors
as additional ``propagators'' appearing in the numerator instead
of the denominator.  Accordingly, we
attach a line label $j$ to each such 
factor.   The presence of such a factor is also indicated graphically
by a pair of parallel arrows, marking the lines whose momenta are
summed, then squared, to generate the numerator factor.

For example, the quadruple ladder integral (a) is defined by
\begin{eqnarray}
\I^{\rm (a)} 
&=& (-i e^{\e \gamma} \pi^{-d/2})^4 \, s^4 t
 \int
 {\dd^d p\,\dd^d r\,\dd^d u\,\dd^d v
 \over p^2 \, (p - k_1)^2 \, (p - k_1 - k_2)^2 
    \, q^2 \, (p - q)^2 \, (q-k_1-k_2)^2 }
\nonumber \\
&& \hskip3cm \times
{1\over u^2 \, (q - u)^2 \, (u-k_1-k_2)^2
     \, v^2 \, (u - v)^2 \, (v-k_1-k_2)^2 \, (v+k_4)^2 }
\nonumber \\
&=& (-i e^{\e \gamma} \pi^{-d/2})^4
 \int
 {\dd^d p\,\dd^d r\,\dd^d u\,\dd^d v \ s^4 t
 \over \prod_{j=1}^{13} p_j^2}\,.
\label{QuadrupleLadder}
\end{eqnarray}
Similarly, integral (b) is defined by
\begin{eqnarray}
\I^{\rm (b)} 
&=& (-i e^{\e \gamma} \pi^{-d/2})^4 \, s t^2
 \int
 {\dd^d p\,\dd^d r\,\dd^d u\,\dd^d v  \ [(v+k_1)^2]^2
 \over p^2 \, (p - k_1)^2 \, (p - v - k_1)^2 \, 
    \, q^2 \, (p - q)^2 \, (q-v -k_1)^2 }
\nonumber \\
&& \hskip3cm \times
{1\over u^2 \, (q - u)^2 \, (u-v-k_1)^2 \, (u+k_4)^2
     \, v^2 \, (v - k_2)^2 \, (v-k_2-k_3)^2 }
\nonumber \\
&=& (-i e^{\e \gamma} \pi^{-d/2})^4
 \int
 {\dd^d p\,\dd^d r\,\dd^d u\,\dd^d v \ s t^2 \, (p_{11}^2)^2
 \over p_{12}^2 \, p_{13}^2 \, p_{14}^2 \prod_{j=1}^{10} p_j^2}\,.
\label{Integralb}
\end{eqnarray}
The double arrows indicate that in this case the numerator factor appears
squared,  $(p_{11}^2)^2 \equiv (l_8 + l_{10})^4 = [(v+k_1)^2]^2$.

\section{Establishing the correctness of the integrand}
\label{correctnesssection}

In this section we justify the result~(\ref{FourLoopPlanarResult}) for
the four-loop planar amplitude, based upon our evaluation of the
unitarity cuts.  Because we have not evaluated all possible unitarity
cuts, we have to impose some mild assumptions about the types of
integrals that should be present.  We will see that another, stronger
assumption of conformal invariance also holds for the individual 
integrals that appear, although we do not require it.  In addition, as
we shall see in the next section, the agreement of the infrared
singularities through $\Ord(\ep^{-2})$ with their known
form~\cite{MagneaSterman} --- up to the one unknown constant at 
$\Ord(\ep^{-2})$ --- provides a non-trivial consistency check on our
construction.

The analysis determining the integrand of the four-loop four-point 
amplitude proceeds in several steps:
\begin{itemize}
\item We assume that there are no integrals with triangle or 
bubble subgraphs.
\item We classify the four-loop planar integrals
of this type topologically.  We begin with the subset of 
graphs having only cubic vertices, from which we can obtain 
the remaining graphs.
\item We construct a set of generalized cuts capable of detecting
all such integrals.  That is, each such integral, when restricted 
to the generalized cut kinematics, should be
nonvanishing for at least one cut in the set.  For it to
be nonvanishing, it must have a propagator present for each
line being cut.
We used this set of cuts to deduce the 
terms in the expression~(\ref{FourLoopPlanarResult}).  Indeed, we
find that each such cut of the expression is completely consistent
with our evaluation of the cut.  
This step confirms the result, under the ``no triangle'' assumption.
\item Alternatively, we consider the result of assuming that only
{\it conformally-invariant integrals} contribute, when the external 
legs are taken off-shell so that the integrals become well-defined 
(finite) in four dimensions.
Such integrals were considered recently in ref.~\cite{DHSS}.
We find that the conformal-invariance assumption is a powerful one;
it allows all eight integrals contributing to \eqn{FourLoopPlanarResult} 
to be present, while forbidding all but two of the additional
no-triangle integrals.  The potential contributions of these remaining 
two integrals are easily ruled out by examining the two-particle cuts.
\end{itemize}
Next we elaborate on each step in the analysis.

\subsection{Unitarity construction}
\label{unitarity subsection}

Our assumption, that there are no integrals with triangle or 
bubble subgraphs in multi-loop $\NeqFour$ super-Yang-Mills theory,
is sometimes referred to as the ``no triangle hypothesis''.
(Such an assumption also appears to be applicable to $\NeqEight$ 
supergravity, at least at one loop~\cite{GravNoTriangle}.)
We now discuss evidence in favor of this hypothesis.
First, notice that a bubble subgraph would lead to an 
ultraviolet subdivergence.  In the absence of cancellations 
between different integral topologies, such subdivergences 
are forbidden by the finiteness of \SYM~\cite{Finiteness}. 
Keep in mind that all cancellations between different particles
in the supermultiplet have already been taken into account, so
the coefficients of all bubble integrals should indeed be zero.

Next, suppose there were a triangle subgraph in some multi-loop
integral. Excise a region around the triangle, cutting through open
propagators attached to the integral.   This excised region represents 
a one-particle-irreducible triangle-type contribution to a
one-loop $n$-particle scattering amplitude.
If all $n$ cut legs are gluons, then we know that 
such a contribution is forbidden in \SYM, for arbitrary 
$n$, by applying loop-momentum power-counting to a computation
of the one-loop amplitude using background-field 
gauge~\cite{SuperspaceBook,NeqFourOneLoop}.  
However, in the present case some of the $n$ legs might be 
associated with other fields of the $\NeqFour$ supermultiplet.  
Supersymmetry Ward identities~\cite{SWI} typically relate 
many such amplitudes to the gluonic case, but we do not 
know of a general proof.  Thus we do not claim to 
have a full proof that all topologies with triangle subgraphs 
are absent, although we strongly suspect that it is the case.

We now wish to classify the four-loop planar integrals
containing no triangle or bubble subgraphs.  
We can perform this classification first
for the subset of such graphs that have only cubic (three-point)
vertices, for the following reason:
If a graph contains a quartic or higher-point vertex, we can ``resolve''
the vertex into multiple three-point vertices by moving some of the
lines attached to the vertex.  Such a procedure never decreases the
number of propagators associated with a given loop.  Therefore a
no-triangle graph will remain a no-triangle graph under this procedure.
So, if we know all the no-triangle graphs with only cubic 
vertices, we can get every remaining no-triangle graph by sliding
cubic vertices together until they coincide, a procedure known
as ``canceling propagators''.

%
\begin{figure}[t]
\centerline{\epsfxsize 1.3 truein \epsfbox{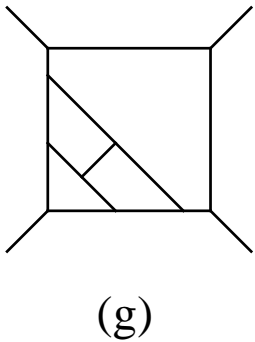}}
\caption{The only one-particle-irreducible purely-cubic
four-loop four-point graph with no triangle or bubble subgraphs,
besides the rung-rule graphs in \fig{rrFigure}. }
\label{gFigure}
\end{figure}

The cubic subset can be classified iteratively in the number 
of loops using the Dyson-Schwinger equation.  In order to
use the Dyson-Schwinger approach for the case at hand, 
the planar four-loop four-point amplitude, we also need to 
classify the planar cubic no-triangle graphs of the following types:
one loop for the number of legs $n$ up to 7, two loops 
for $n$ up to 6, and three loops for $n$ up to 5.  
The result is that there are a total of 13 planar cubic four-loop
four-point no-triangle graphs.
Of the 13 graphs, seven are one-particle-irreducible.
Six of these seven have the topology of the rung-rule graphs 
shown in \fig{rrFigure}.  The seventh is shown in \fig{gFigure}.
Note that here we are only classifying
graphs, and not yet specifying the loop-momentum polynomials
associated with each graph.

The six one-particle-reducible graphs can be obtained
by sewing external trees onto four-loop graphs with
either two or three external legs.  (There are no planar cubic
no-triangle graphs at four loops with fewer than two 
external legs.)  From the point of view of the cuts,
these one-particle-reducible graphs are equivalent to 
certain of the non-cubic graphs obtained by canceling external
propagators, so we shall defer their description briefly.

%
\begin{figure}[t]
\centerline{\epsfxsize 6.5 truein \epsfbox{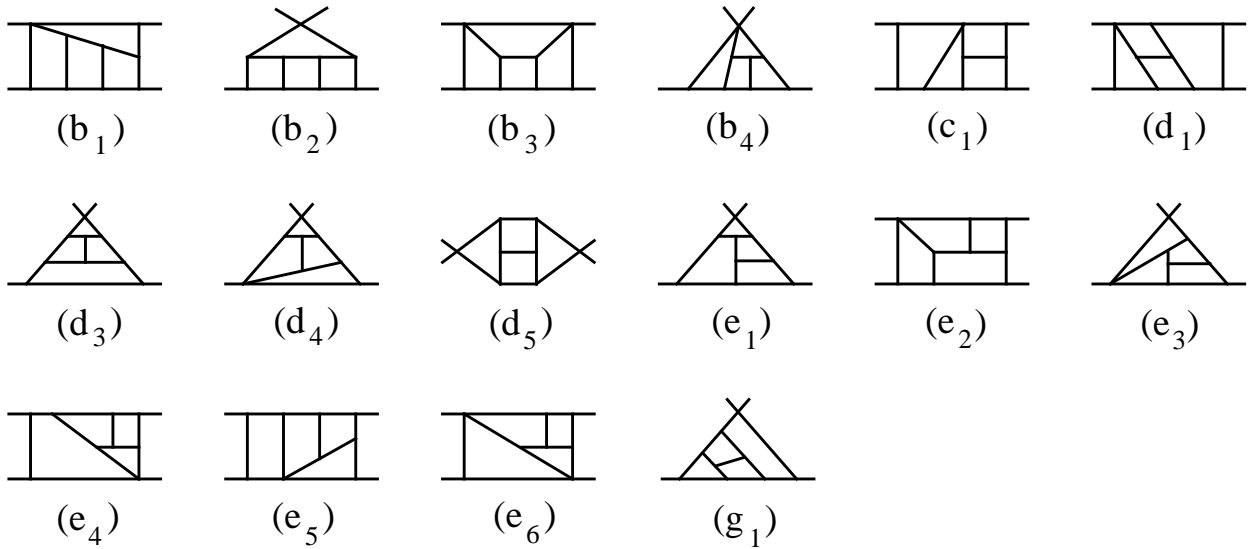}}
\caption{No-triangle planar four-loop graphs, in addition
to those given in \figs{rrFigure}{nonrrFigure}.  The notation
indicates the rung-rule graph in \fig{rrFigure}, or graph (g) in
\fig{gFigure}, that a given graph here is derived from
by canceling propagators. }
\label{NoTriangleFigure}
\end{figure}

The next step is to cancel propagators between vertices 
in the cubic graphs in \figs{rrFigure}{gFigure}.  That is, we merge adjacent
three-point vertices into four-point (or higher-point) vertices 
by eliminating the line(s) between them, while insisting on
at least four propagators around every sub-loop.
This procedure is also straightforward to carry out.
It generates the two non-rung-rule graphs in the 
expression~(\ref{FourLoopPlanarResult}), shown in \fig{nonrrFigure}, 
as well as the 16 additional graphs shown in \fig{NoTriangleFigure}.
We have given each graph a notation which indicates the rung-rule
graph (or graph (g)) from which it can be generated by canceling one or more
propagators.  For example, graph (b${}_3$) is found by canceling
propagators 13 and 14 in rung-rule graph (b), and graph (e${}_6$)
is found by canceling propagators 8 and 11 in rung-rule graph (e).

Some graphs in \fig{NoTriangleFigure} can be generated from more 
than one cubic graph.  In fact, five of the six
one-particle-reducible cubic graphs mentioned earlier are equivalent, 
upon canceling external propagators, to diagrams in \fig{NoTriangleFigure},
namely graphs (b${}_2$), (d${}_3$), (d${}_5$), (e${}_1$), and (g${}_1$). 
Hence we do not provide a separate figure for the 
one-particle-reducible cubic graphs.
The sixth graph has the form of a massless version of graph (d${}_5$),
sewn as an external bubble to a four-point tree graph.  Such
integrals vanish in dimensional regularization, so we need
not consider them.

To complete the direct justification of the four-loop
result~(\ref{FourLoopPlanarResult}), given the no-triangle assumption,
we just need to show that each graph appearing in
figs.~\ref{rrFigure}, \ref{nonrrFigure}, \ref{gFigure} 
and \ref{NoTriangleFigure} can be detected by at least one of the 
generalized cuts we have computed, 
out of the six cuts shown in \fig{CutsFigure}. 
Table~\ref{CutDetectTable} summarizes
some of the cuts that detect the no-triangle graphs.
For some of the graphs, other cuts also detect them,
but for brevity they were not listed in the table. 
Almost all graphs appear in multiple cuts.  
Only two of the graphs, (b${}_4$) and (e${}_6$), appear 
in a unique cut, the ``3--lower-3'' cut labeled (vi) in \fig{CutsFigure}.

\begin{table}
\caption{\label{CutDetectTable} 
The no-triangle graphs, and some of the cuts from \fig{CutsFigure}
that detect them.
(In some cases, the diagram must be rotated or flipped first.)}

\vskip .4 cm

\begin{tabular}{||c|c||c|c||c|c||}
\hline
\hline
Graph & Cuts & Graph & Cuts & Graph & Cuts \\
\hline
\hline
(a) &  (i), (ii), (iii) & 
(b${}_1$) & (v), (vi) & 
(d${}_5$) & (i), (iii), (iv) \\
\hline
(b) &  (i), (v) & 
(b${}_2$) & (i), (vi) & 
(e${}_1$) & (i), (iii), (iv) \\
\hline
(c) &  (i), (ii), (iii), (iv) & 
(b${}_3$) & (v), (vi) & 
(e${}_2$) & (iii), (iv) \\
\hline
(d) &  (i), (iii), (v) & 
(b${}_4$) & (vi) & 
(e${}_3$) & (i), (vi) \\
\hline
(e) &  (i), (iii), (iv), (vi) & 
(c${}_1$) & (i), (iii), (iv)& 
(e${}_4$) & (i), (vi) \\
\hline
(f) &  (iii), (iv), (vi) & 
(d${}_1$) & (i), (iii), (iv) & 
(e${}_5$) & (i), (iii), (iv) \\
\hline
(d${}_2$) &  (iii), (iv) & 
(d${}_3$) & (i), (iii), (iv) & 
(e${}_6$) & (vi) \\
\hline
(f${}_2$) & (iii), (vi) & 
(d${}_4$) & (i), (iii), (iv) & 
(g) & (i), (vi) \\
\hline
          &             & 
          &             &
(g${}_1$) & (i), (vi) \\
\hline
\hline
\end{tabular}
\end{table}

\subsection{Conformal properties}
\label{conformalsubsection}

We have finished our justification of the
representation~(\ref{FourLoopPlanarResult}) for the four-loop planar
amplitude. In the rest of this section we would like to examine the
consequences of making a stronger assumption than the no-triangle
hypothesis.  This assumption is that each of the integral functions
that appears is conformally invariant.  Here we are inspired by the
discussion of conformally-invariant integrals by Drummond, Henn,
Sokatchev, and one of the authors~\cite{DHSS}.  Although the
requirement of conformal invariance is natural because of the
conformal invariance of the theory in four dimensions, we do not 
have a proof that these integrals are the only ones that can 
appear in the amplitudes. Nevertheless, as we shall see, the 
conformal properties offer a rather useful guide.
(It is possible that extensions of the conformal-invariance
analysis in ref.~\cite{GKS} could be used to prove that only
such conformally invariant integrals can be present.)

We actually wish to study the conformal properties of the integrals in
four dimensions, yet they are ill-defined there because of the infrared
divergences associated with on-shell, massless external legs.  We
therefore adopt a different infrared regularization of the integrals by taking
the external legs off shell, letting $k_i^2\neq0$, $i=1,2,3,4$, instead of
using a dimensional regulator as in the rest of the paper.  We demand
that each integral that appears be conformally invariant.  Actually,
the integrals need only transform covariantly,
carrying conformal weights associated with each of the external legs,
in such a way that they can be made invariant by multiplying by appropriate
overall factors of $s$ or $t$.  In canceling the conformal weights
using the external invariants, we should not use any factors that vanish
as the external legs return on shell, in the limit $k_i^2 \to 0$.  
Such factors would lead to power-law divergences or to vanishings 
of the integrals that are too severe, compared to the typical 
logarithmic dependences on $k_i^2$ from the known form of 
infrared singularities.  Thus integrals that require powers of $k_i^2$
to be conformally invariant should not appear in any on-shell amplitude.
This on-shell restriction turns out to be a powerful one.

The net result of the conformal-invariance requirement
will be that, besides the
eight integrals already present in \eqn{FourLoopPlanarResult} and
\figs{rrFigure}{nonrrFigure}, remarkably only two other potential
conformal integrals survive.  The first of them is the (d${}_5$) graph
from \fig{NoTriangleFigure}.  This graph has the structure of a
potential propagator correction at four loops.  However, in the
context of the $gg\to gg$ scattering amplitude of the $\NeqFour$
theory, it can be excluded very simply, without computing any
generalized cuts.  One only has to use the structure of the three-loop
amplitude, and the simplest two-particle cut, cut (i) in
\fig{CutsFigure}, to see that it cannot be present.  The second of the
additional potential integrals has the topology of integral (d) 
in \fig{rrFigure}, but it has a different numerator factor,
to be described below.
This new integral, (d${}'$), is also easy to exclude using 
the same two-particle cut (i) in \fig{CutsFigure}.

It is rather striking that every integral identified via the unitarity 
cuts is conformally invariant, and that there are only two other conformal 
integrals, which can be eliminated easily via two-particle cuts.
Furthermore, the two integrals that are not present differ from the
eight that are present in how the conformal invariance is achieved,
as we shall discuss below.

To analyze the conformal invariance properties, we shall use changes
of variables as suggested in ref.~\cite{DHSS}. (The same conformal
integrals appear in coordinate-space correlators of gauge invariant
operators~\cite{Schubert}; the coincidence is presumably an accident
of there being a limited number of conformal integrals.)  As an
example, consider the two-loop double box depicted in
\fig{LowerLoopFigure},
\begin{eqnarray}
\I^{(2)}(s,t) &=&  (-i e^{\epsilon \gamma} \pi^{-2})^2 s^2 t 
\nonumber\\
 &&\hskip 3mm\times 
  \int { \d^4 p\ \d^4 q
   \over  p^2 (p-k_1)^2 (p-k_1-k_2)^2  q^2 (q-k_4)^2 (q-k_3-k_4)^2 
             (p+q)^2}\,.\hskip 8mm
\label{DoubleBox}
\end{eqnarray}
We have taken $d=4$, with the $k_i$ off shell to serve as 
an infrared regulator.  Next, use the change of variables,
\begin{equation}
k_1 = x_{41} \,, \hskip .8cm  k_2 = x_{12} \,, 
\hskip .8cm k_3 = x_{23} \,, \hskip .8cm  k_4 = x_{34} \,, \hskip .8cm
p = x_{45}, \hskip .8cm q = x_{64} \,, 
\label{twoloopktox}
\end{equation}
where $x_{ij} \equiv x_i - x_j$.  This choice of variables
automatically ensures that momentum is conserved, 
$k_1 + k_2 + k_3 + k_4 = 0$.  Note that the external invariants
become
\begin{equation}
s = (k_1 + k_2)^2 = x_{24}^2 \,, 
\hskip 1cm t = (k_2 + k_3)^2 = x_{13}^2 \,.
\label{sttox}
\end{equation}
Performing the change of variables~(\ref{twoloopktox}) 
in the double box, we obtain,
\begin{equation}
\I^{(2)}(x_1,x_2,x_3,x_4) = (-i e^{\epsilon \gamma} \pi^{-2})^2
 x_{24}^4 x_{13}^2 \int \d^4 x_5\ \d^4 x_6 \,
{1\over x_{45}^2 x_{15}^2 x_{25}^2  x_{46}^2 x_{36}^2 x_{62}^2 x_{56}^2}
\,.
\label{TwoLoopDualForm}
\end{equation}
The principal conformal-invariance constraints on integrals 
constructed from the invariants $x_{ij}^2$ are exposed by 
performing an inversion on all points, 
$x_i^\mu \rightarrow {x_i^\mu / x_i^2}$.
(We cannot impose such an inversion on the $k_i$ directly,
because it would violate the constraint of momentum conservation.)
Under the inversion, we have
\begin{equation}
x_{ij}^2 \rightarrow {x_{ij}^2 \over x_i^2 x_j^2} \,, \hskip 1 cm
 {\d^4 x_5} \rightarrow {\d^4 x_5 \over x_5^{8}} \,, \hskip 1 cm
{\d^4 x_6} \rightarrow {\d^4 x_6 \over x_6^{8}} \,.
\end{equation}
It is easy to see that the planar double-box
integral is invariant under inversion, {\it i.e.}
$
\I^{(2)}(x_1,x_2,x_3,x_4)\ \rightarrow\ \I^{(2)}(x_1,x_2,x_3,x_4).
$
For this result to hold, it is important that the unintegrated points
$x_1,x_2,x_3,x_4$ appear in the numerator just enough times to 
cancel their appearance in the denominator.  The integrated points
$x_5,x_6$ each appear four times in the denominator.
The dimensionally-regulated version of the
conformally-invariant integral $\I^{(2)}(x_1,x_2,x_3,x_4)$
is precisely the form in which the two-loop double box appears 
in the two-loop planar amplitude (\ref{TwoloopPlanarResult}).

In order to analyze the conformal properties of integrals beyond two
loops, it is helpful to follow the discussion of ref.~\cite{DHSS}, and
introduce a set of dual diagrams~\cite{Nakanishi}.  We construct the
dual to a diagrammatic representation of a planar loop-momentum
integral by placing vertices corresponding to the $x_i$ at the centers
of the loops and in between pairs of external lines.  Denominator
factors of $x_{ij}^2$ are denoted by drawing dark solid (blue) lines
between the corresponding vertices.  Numerator factors are denoted by
drawing dotted lines between the corresponding vertices. One solid
line crosses each propagator in a loop.  The conformal weight in each
$x_i$ variable is then given by the number of solid lines entering the
corresponding vertex, less the number of dotted lines.  A
conformally-invariant integral will have weight four at each internal
vertex (to balance the weight of the integration measure), and weight
zero at each external vertex.  In the diagrams, we shall omit one
dotted line connecting external vertices $x_2$ and $x_4$, and another
one connecting $x_1$ and $x_3$, in order to simplify the presentation.
These two omitted lines correspond to the overall factor of $st =
x_{24}^2 x_{13}^2$ omitted from the momentum-space diagrams in
\figs{rrFigure}{nonrrFigure}.  Expressions for integrals in terms of
the $x_i$ variables can be read off quickly from the dual diagrams
(and vice versa).

%
\begin{figure}[t]
\centerline{\epsfxsize 2.4 truein \epsfbox{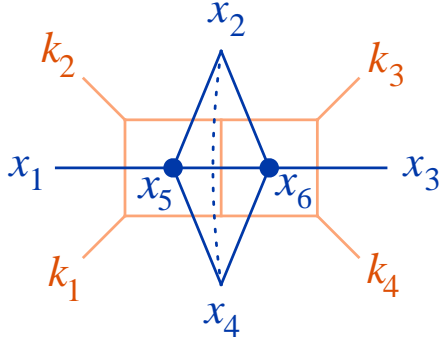}}
\caption{The two-loop planar double box and its dual diagram. 
The double box is represented by light colored lines and the 
dual diagram by dark (blue) lines.
A dark line connecting $x_i$ with $x_j$ represents the factor $1/x_{ij}^2$.
A dotted line signifies a numerator factor of $x_{ij}^2$.  
The momentum corresponding to any $x_{ij}$ is given by the sum of 
momenta of the light lines crossing the dark line joining $x_i$ and $x_j$.
An overall factor of $st$ has been removed for clarity.
}
\label{twoloopdualFigure}
\end{figure}

For example, \fig{twoloopdualFigure} contains the diagram dual to the
double box.  Each of the solid lines starting at an $x_i$ and ending at
an $x_j$ corresponds to a factor of $1/x_{ij}^2$ appearing in
\eqn{TwoLoopDualForm}, while the dotted line corresponds to a factor
of $x_{24}^2 = s$.  With this identification the dual figure is in
direct correspondence with \eqn{TwoLoopDualForm}, after removing one
overall factor of $st = x_{24}^2 x_{13}^2$ (in order to reduce the
visual clutter in the diagram).  Because the number of solid lines
minus the number of dotted lines at each of the two internal vertices
$x_5$ and $x_6$ is four, the integral is conformally invariant with
respect to these points.  Similarly, since each of the external points
$x_1,x_2, x_3, x_4$ has one more solid line than dotted line
emanating from it, the conformal weight is unity.  
If we multiply back by the $x_{24}^2 x_{13}^2$ factor removed previously, 
then we obtain an integral which is conformally
invariant with respect to the external as well as internal points.

The assumption of conformal invariance for the integrals immediately
implies the ``no-triangle'' rule for the momentum-space diagrams. 
A loop with only three propagators would necessarily result in a 
negative weight for the $x$ point corresponding to the loop momentum, 
because only three lines enter the dual diagram vertex.  
That negative weight can only be eliminated by additional denominator 
powers of $x$ --- that is, by additional propagators which would 
turn the triangle subgraph into at least a box subgraph.

By placing the dual diagrams on top of the original momentum-space 
diagrams, we can read off directly the change of variables between 
the $x_{ij}$ and the momenta:
$x_{ij}$ is just the sum of the momenta of the each momentum-space
line crossed by the dual line running from $x_i$ to $x_j$.
In the rung-insertion rule, when a rung is inserted between two 
parallel lines with momenta $l_1$ and $l_2$, to go from
an $L$-loop contribution to an $(L+1)$-loop one, the loops on 
either side of the parallel lines have acquired a new propagator.
Hence each of their dual $x$ vertices has a new solid line emanating
from it.  The rung-rule momentum-insertion factor of $i(l_1+l_2)^2$
is represented by a dotted line stretching between the two vertices,
so it restores the conformal invariance for those two loops.
This property may help to explain the form of the rung rule.

%
\begin{figure}[t]
\centerline{\epsfxsize 6.2 truein \epsfbox{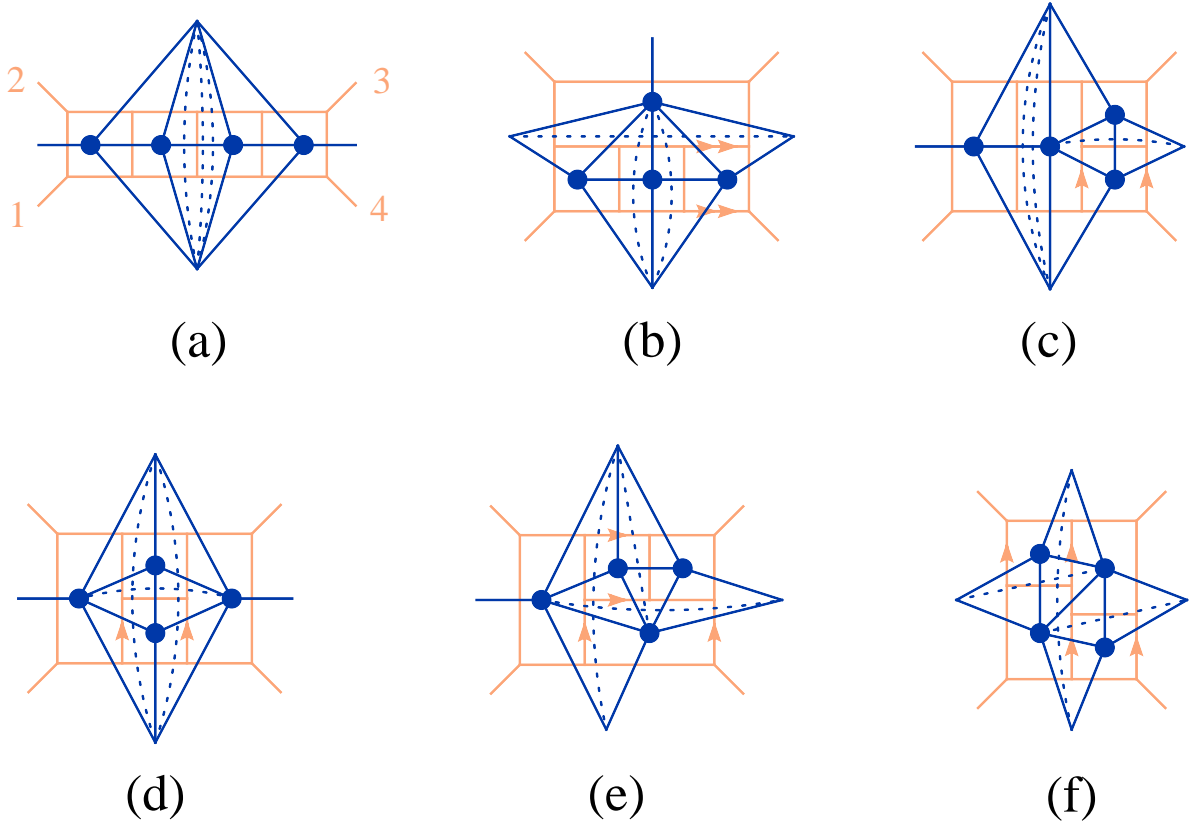}}
\caption{The rung-rule dual diagrams. A factor of $st$ has been 
removed.}
\label{rrdualFigure}
\end{figure}

%
\begin{figure}[t]
\centerline{\epsfxsize 3.8 truein \epsfbox{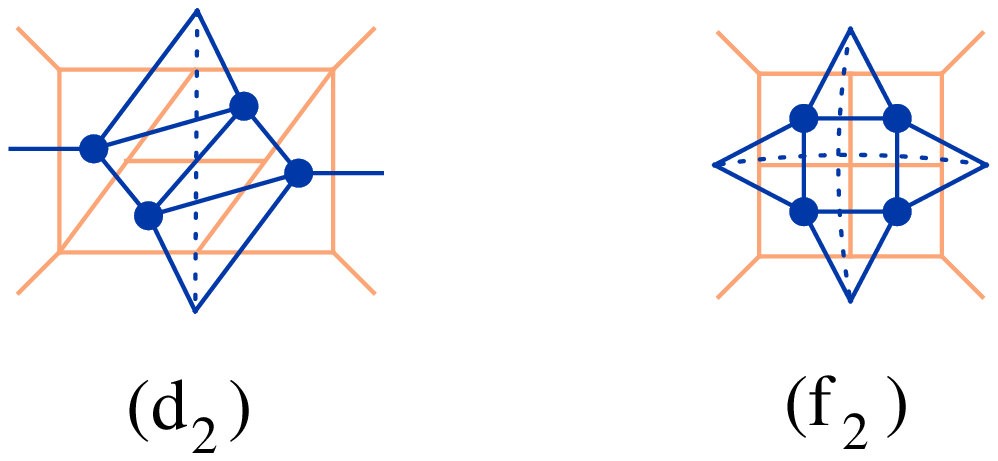}}
\caption{The non-rung rule dual diagrams. A factor of $st$ has been 
removed.}
\label{nonrrdualFigure}
\end{figure}

%
\begin{figure}[t]
\centerline{\epsfxsize 3.8 truein \epsfbox{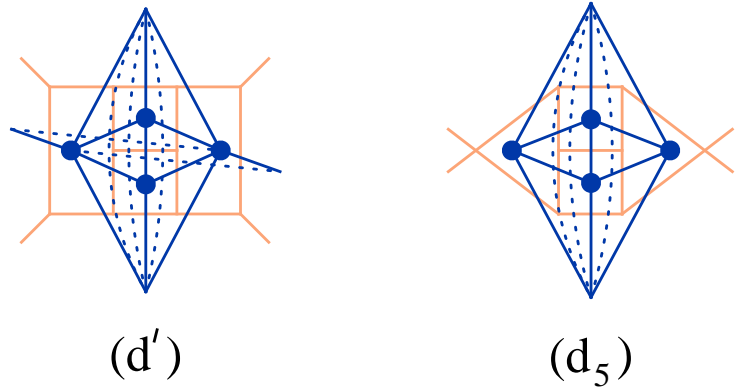}}
\caption{The two four-loop dual integrals, in addition
to those given in \figs{rrFigure}{nonrrFigure}, that survive
the requirement of conformal invariance.  Both integrals are 
ruled out by two-particle cuts.  In this case, no factor of 
$st$ has been removed.
}
\label{confdualFigure}
\end{figure}

Let us now focus on  four-loop four-point integrals.  We may find all
conformally-invariant integrals by drawing the set of all dual
diagrams that have conformal weight zero with respect to all $x_i$.
Again to prevent cluttering the diagrams in
\figs{rrdualFigure}{nonrrdualFigure} with dotted lines we have removed
a factor of $st$ from the figures. The complete list of
conformal four-loop integrals, as it turns out, contains the rung-rule
diagrams of \fig{rrdualFigure}, the non-rung-rule diagrams of
\fig{nonrrdualFigure}, and the two extra integrals shown in 
\fig{confdualFigure}.\footnote{It is possible to dress 
diagram (f) in \fig{rrdualFigure} with dotted lines in a second way, 
but that dressing is related simply to the one shown by the 
exchange of legs $k_1$ and $k_3$, or in other words $s \lr t$.}

%
\begin{figure}[t]
\centerline{\epsfxsize 5.5 truein \epsfbox{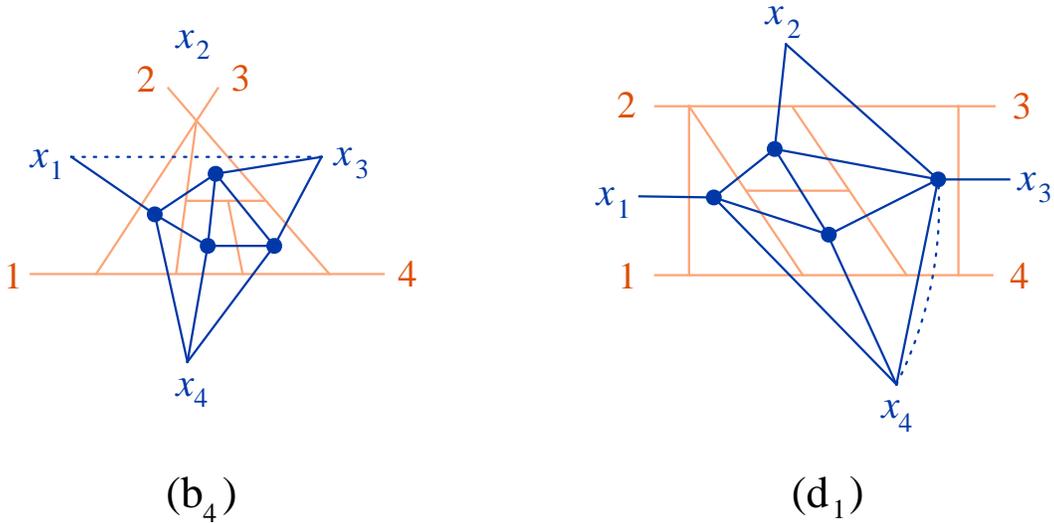}}
\caption{Two examples of dual diagrams which do not lead to
new conformal integrals in the on-shell limit, for reasons 
discussed in the text. }
\label{nonconfexamplesFigure}
\end{figure}

Because conformal invariance in the sense discussed above implies the
``no-triangle'' rule, the complete list of candidate graphs for
conformal integrals are given by the same no-triangle list described
above, namely the diagrams in figs.~\ref{rrFigure}, \ref{nonrrFigure},
\ref{gFigure} and \ref{NoTriangleFigure}.  Upon attempting to draw the dual
diagrams to \figs{gFigure}{NoTriangleFigure}, we find that in all cases except
(d${}_5$), either it is impossible to add dotted or even solid lines
so as to obtain a conformally-invariant integral, 
or the obtained conformally-invariant cases are equivalent to 
previously-included cases, or else
conformal invariance can only be achieved by adding solid or dotted 
lines connecting neighboring external vertices.  The latter lines 
are not admissible, however, because differences of neighboring 
external $x$ points correspond to individual external
momenta $k_i$.  The corresponding factors are then $k_i^2$.
Their presence would lead to an unwanted power-law vanishing or
divergence of the integral in the on-shell limit.

\Fig{nonconfexamplesFigure} illustrates two examples of dual diagrams
that cannot be made conformal, or that reduce to previous cases.
First consider the graph labeled (b${}_4$).  The internal dual points
all have weight four, so no dotted lines can attach to them.
The dotted line shown between external points $x_1$ and $x_3$ 
reduces the $x_1$ weight to zero and the $x_3$ weight to one.
It is nonvanishing in the massless limit.  However, there is 
no other numerator factor available that is nonvanishing in this limit, 
to further reduce the conformal weights of external legs $x_3$
and $x_4$.  In particular, $x_{34}^2 = k_4^2 \to 0$ in the on-shell
limit.  

As a second example, consider the graph labeled (d${}_1$)
in \fig{nonconfexamplesFigure}.  Here there is one pentagon subgraph,
and hence one internal $x$ point to which a dotted line can attach.
The dotted line shown can be used to reduce the conformal weight 
of $x_4$ from three to two, which would then balance its weight
with that of the opposite external point $x_2$.  (Balanced opposing
weights can always be reduced to zero using powers of $s=x_{24}^2$ or
$t=x_{13}^2$.)
However, this choice of dotted line merely cancels the propagator 
that it crosses, and thereby reduces the (d${}_1$) graph to 
the (d${}_2$) graph in \fig{nonrrdualFigure}, which we already
know is conformally invariant and present in the four-loop planar
amplitude.  Without using the dotted line shown, it is impossible to
balance the $x_4$ conformal weight, in the massless limit, so the only
surviving possibility reduces to an existing integral.

It is curious that the two conformally-invariant integrals 
represented in \fig{confdualFigure}, which are
not present in the planar four-loop four-point amplitude,
can be distinguished from the eight 
in \figs{rrdualFigure}{nonrrdualFigure} that are present,
by the fact that they do not have explicit overall factors
of both $s$ {\it and} $t$.  As drawn in \fig{confdualFigure},
they have three powers of $s$, but no powers of $t$.
The integral (d${}'$) has the same basic topology as (d)
in \fig{nonrrdualFigure}, but the dotted lines emanating
from the two pentagon loops are connected to external legs
$x_1$ and $x_3$ instead of to each other.  So it has two
``rung-rule-type'' numerator factors involving the squares
of the sums of three loop momenta, but no power of $t$.
At the moment, however, we have no good argument why explicit
factors of both $s$ and $t$ have to be present in order that an 
integral be present in the four-point amplitude.

\section{Analytic and numerical results}
\label{IntegralResultsSection}

Our next task is to evaluate the integrals entering the four-loop 
planar amplitude~(\ref{FourLoopPlanarResult}) in a Laurent
expansion around $\e=0$.
For the case at hand, massless gluon-gluon scattering in $\NeqFour$
super-Yang-Mills theory, all of the integrals encountered
can be evaluated through three loops in terms of a class of functions 
known as harmonic polylogarithms (HPLs)~\cite{HPL}.  
We expect this class of functions
to continue to suffice at four loops.  We know it suffices through
$\Ord(\e^{-4})$, for which we have analytic results.

\subsection{Analytic expressions through $\Ord(\ep^{-4})$}
\label{AnalyticSubsection}

The analytic results for the four-loop integrals were obtained
with the help of the {\tt MB} program~\cite{CzakonMB}.
We let $x=-t/s$ and $L=\ln(-x)$.  Through $\Ord(\ep^{-4})$ the
results, expressed in terms of the HPLs defined in 
\app{HarmonicPolyLogAppendix}, are,
\bea
\I^{\rm (a)}(s,t) &=& (-t)^{-4\e} \Biggl\{
  { 4 \over 9 \, \e^8 } 
 + { 35\over 72 \, \e^7 } L 
 - { 187 \, \pi^2 \over 432 \, \e^6 }
\nonumber\\
&&\hskip0.1cm
 + \, { 10\over 9 \, \e^5} \Biggl[
        H_{0,0,1}(x) - L \, H_{0,1}(x)
      + {1\over2} ( L^2 + \pi^2) H_{1}(x) 
      + {23\over48} \pi^2 L - {1169\over240} \zeta_3 \Biggr]
\nonumber\\
&&\hskip0.1cm
 + \, {10\over 9 \, \e^4} \Biggl[ 
      - 22 \, H_{0,0,0,1}(x) - H_{0,0,1,1}(x)
      - H_{0,1,0,1}(x) - H_{1,0,0,1}(x)
\nonumber\\
&&\hskip1.3cm
      + {L} {} ( 16 \, H_{0,0,1}(x) + H_{0,1,1}(x) + H_{1,0,1}(x) )
      - {L^2 \over2} ( 10 \, H_{0,1}(x) + H_{1,1}(x) )
\nonumber\\
&&\hskip1.3cm
      - {\pi^2\over2} ( 7 \, H_{0,1}(x) + H_{1,1}(x) - L \, H_{1}(x) )
\nonumber\\
&&\hskip1.3cm
      + {2\over3} L^3 \, H_{1}(x) + \zeta_3 \, H_{1}(x)
      - {97\over12} \zeta_3 \, L 
      - {2339\over3600} \pi^4 \Biggr] 
\nonumber\\
&&\hskip0.1cm
 +\ \Ord(\e^{-3}) \Biggr\} \,,
\label{IaAnalytic}
\eea
\bea
\I^{\rm (b)}(s,t) &=& (-t)^{-4\e} \Biggl\{
  { 4 \over 9 \, \e^8 } 
 + { 17\over 18 \, \e^7 } L 
 + {1\over \e^6} \Biggl[ { 53 \over 72 } L^2 
                      - { 211 \over 432 } \pi^2 \Biggr]
\nonumber\\
&&\hskip0.1cm
 + \, { 8 \over 9 \, \e^5} \Biggl[
        H_{0,0,1}(x) - L \, H_{0,1}(x)
      + {1\over2} ( L^2 + \pi^2) H_{1}(x) 
      + {25\over96} L^3
      - {45\over64} \pi^2 L - {601\over96} \zeta_3 \Biggr]
\nonumber\\
&&\hskip0.1cm
 + \, { 8 \over 9 \, \e^4} \Biggl[ 
      - 10 \, H_{0,0,0,1}(x) - H_{0,0,1,1}(x)
      - H_{0,1,0,1}(x) - H_{1,0,0,1}(x)
\nonumber\\
&&\hskip1.3cm
      + L {}\Bigl( {71\over8} \, H_{0,0,1}(x) 
               + H_{0,1,1}(x) + H_{1,0,1}(x) \Bigr)
      - {L^2 \over2} \Bigl( {31\over4} \, H_{0,1}(x) + H_{1,1}(x) \Bigr)
\nonumber\\
&&\hskip1.3cm
      - {\pi^2\over2} \Bigl( 3 \, H_{0,1}(x) + H_{1,1}(x)
                           - {15\over 8} L \, H_{1}(x)
                           + {73\over96} L^2 \Bigr)
\nonumber\\
&&\hskip1.3cm
      + {53\over48} L^3 \, H_{1}(x) + \zeta_3 \, H_{1}(x)
      - {L^4\over48}
      - {1579\over96} \zeta_3 \, L 
      - {743\over4608} \pi^4 \Biggr] 
\nonumber\\
&&\hskip0.1cm
 +\ \Ord(\e^{-3}) \Biggr\} \,,
\label{IbAnalytic}
\eea
\bea
\I^{\rm (c)}(s,t) &=& (-t)^{-4\e} \Biggl\{
  { 4 \over 9 \, \e^8 } 
 + { 13\over 24 \, \e^7 } L 
 + {1\over 18 \, \e^6} \Biggl[ L^2 
                      - { 29 \over 3 } \pi^2 \Biggr]
\nonumber\\
&&\hskip0.1cm
 + \, { 8 \over 9 \, \e^5} \Biggl[
        H_{0,0,1}(x) - L \, H_{0,1}(x)
      + {1\over2} ( L^2 + \pi^2) H_{1}(x) 
      - {L^3\over24}
      + {\pi^2\over12} L - {1175\over192} \zeta_3 \Biggr]
\nonumber\\
&&\hskip0.1cm
 + \, { 8 \over 9 \, \e^4} \Biggl[ 
      - {73\over4} \, H_{0,0,0,1}(x) - H_{0,0,1,1}(x)
      - H_{0,1,0,1}(x) - H_{1,0,0,1}(x)
\nonumber\\
&&\hskip1.3cm
      + L {}\Bigl( {107\over8} \, H_{0,0,1}(x) 
               + H_{0,1,1}(x) + H_{1,0,1}(x) \Bigr)
      - {L^2 \over2} \Bigl( {17\over2} \, H_{0,1}(x) + H_{1,1}(x) \Bigr)
\nonumber\\
&&\hskip1.3cm
      - {\pi^2\over2} \Bigl( {23\over4} \, H_{0,1}(x) + H_{1,1}(x)
                           - {7\over 8} L \, H_{1}(x)
                           - {5\over12} L^2 \Bigr)
\nonumber\\
&&\hskip1.3cm
      + {29\over48} L^3 \, H_{1}(x) + \zeta_3 \, H_{1}(x)
      + {L^4\over48}
      - {253\over32} \zeta_3 \, L 
      - {5663\over23040} \pi^4 \Biggr] 
\nonumber\\
&&\hskip0.1cm
 +\ \Ord(\e^{-3}) \Biggr\} \,,
\label{IcAnalytic}
\eea
\bea
\I^{\rm (d)}(s,t) &=& (-t)^{-4\e} \Biggl\{
  { 4 \over 9 \, \e^8 } 
 + { 11\over 18 \, \e^7 } L 
 + {1\over 8 \, \e^6} \Biggl[ L^2 
                      - { 169 \over 54 } \pi^2 \Biggr]
\nonumber\\
&&\hskip0.1cm
 + \, { 10 \over 9 \, \e^5} \Biggl[
        H_{0,0,1}(x) - L \, H_{0,1}(x)
      + {1\over2} ( L^2 + \pi^2) H_{1}(x) 
      - {3\over40} L^3
      - {11\over80} \pi^2 \, L - {521\over240} \zeta_3 \Biggr]
\nonumber\\
&&\hskip0.1cm
 + \, { 10 \over 9 \, \e^4} \Biggl[ 
      - {11\over5} \, H_{0,0,0,1}(x) - H_{0,0,1,1}(x)
      - H_{0,1,0,1}(x) - H_{1,0,0,1}(x)
\nonumber\\
&&\hskip1.3cm
      + L {}\Bigl( {5\over2} \, H_{0,0,1}(x) 
               + H_{0,1,1}(x) + H_{1,0,1}(x) \Bigr)
      - {L^2 \over2} \Bigl( {14\over5} \, H_{0,1}(x) + H_{1,1}(x) \Bigr)
\nonumber\\
&&\hskip1.3cm
      - {\pi^2\over2} \Bigl( {2\over5} \, H_{0,1}(x) + H_{1,1}(x)
                           - {7\over 10} L \, H_{1}(x)
                           - {27\over40} L^2 \Bigr)
\nonumber\\
&&\hskip1.3cm
      + {31\over60} L^3 \, H_{1}(x) + \zeta_3 \, H_{1}(x)
      + {3\over80} L^4
      - {271\over120} \zeta_3 \, L 
      - {101\over600} \pi^4 \Biggr] 
\nonumber\\
&&\hskip0.1cm
 +\ \Ord(\e^{-3}) \Biggr\} \,,
\label{IdAnalytic}
\eea
\bea
\I^{\rm (e)}(s,t) &=& (-t)^{-4\e} \Biggl\{
  { 4 \over 9 \, \e^8 } 
 + { 3\over 4 \, \e^7 } L 
 + {1\over \e^6} \Biggl[ {25 \over 72} L^2 
                      - { 49 \over 108 } \pi^2 \Biggr]
\nonumber\\
&&\hskip0.1cm
 + \, { 8 \over 9 \, \e^5} \Biggl[
        H_{0,0,1}(x) - L \, H_{0,1}(x)
      + {1\over2} ( L^2 + \pi^2) H_{1}(x) 
      - {L^3\over24}
      - {41\over96} \pi^2 \, L - {1657\over384} \zeta_3 \Biggr]
\nonumber\\
&&\hskip0.1cm
 + \, { 8 \over 9 \, \e^4} \Biggl[ 
      - {25\over4} \, H_{0,0,0,1}(x) - H_{0,0,1,1}(x)
      - H_{0,1,0,1}(x) - H_{1,0,0,1}(x)
\nonumber\\
&&\hskip1.3cm
      + L {}\Bigl( {47\over8} \, H_{0,0,1}(x) 
               + H_{0,1,1}(x) + H_{1,0,1}(x) \Bigr)
      - {L^2 \over2} \Bigl( {11\over2} \, H_{0,1}(x) + H_{1,1}(x) \Bigr)
\nonumber\\
&&\hskip1.3cm
      - {\pi^2\over2} \Bigl( {7\over4} \, H_{0,1}(x) + H_{1,1}(x)
                           - {11\over 8} L \, H_{1}(x)
                           - {5\over12} L^2 \Bigr)
\nonumber\\
&&\hskip1.3cm
      + {41\over48} L^3 \, H_{1}(x) + \zeta_3 \, H_{1}(x)
      - {13\over384} L^4
      - {107\over16} \zeta_3 \, L 
      - {7153\over46080} \pi^4 \Biggr] 
\nonumber\\
&&\hskip0.1cm
 +\ \Ord(\e^{-3}) \Biggr\} \,,
\label{IeAnalytic}
\eea
\bea
\I^{\rm (f)}(s,t) &=& (-t)^{-4\e} \Biggl\{
  { 8 \over 9 \, \e^8 } 
 + { 107\over 72 \, \e^7 } L 
 + {1\over \e^6} \Biggl[ {49 \over 72} L^2 
                      - { 235 \over 216 } \pi^2 \Biggr]
\nonumber\\
&&\hskip0.1cm
 + \, { 4 \over 3 \, \e^5} \Biggl[
        H_{0,0,1}(x) - L \, H_{0,1}(x)
      + {1\over2} ( L^2 + \pi^2) H_{1}(x) 
      - {7\over144} L^3
      - {11\over12} \pi^2 \, L - {1001\over144} \zeta_3 \Biggr]
\nonumber\\
&&\hskip0.1cm
 + \, { 4 \over 3 \, \e^4} \Biggl[ 
      - {13\over2} \, H_{0,0,0,1}(x) - H_{0,0,1,1}(x)
      - H_{0,1,0,1}(x) - H_{1,0,0,1}(x)
\nonumber\\
&&\hskip1.3cm
      + L {}\Bigl( {23\over4} \, H_{0,0,1}(x) 
               + H_{0,1,1}(x) + H_{1,0,1}(x) \Bigr)
      - {L^2 \over2} ( 5 \, H_{0,1}(x) + H_{1,1}(x) )
\nonumber\\
&&\hskip1.3cm
      - {\pi^2\over2} \Bigl( {11\over6} \, H_{0,1}(x) + H_{1,1}(x)
                           - {13\over12} L \, H_{1}(x)
                           - {19\over48} L^2 \Bigr)
\nonumber\\
&&\hskip1.3cm
      + {17\over24} L^3 \, H_{1}(x) + \zeta_3 \, H_{1}(x)
      - {5\over288} L^4
      - {1405\over144} \zeta_3 \, L 
      + {4253\over17280} \pi^4 \Biggr] 
\nonumber\\
&&\hskip0.1cm
 +\ \Ord(\e^{-3}) \Biggr\} \,,
\label{IfAnalytic}
\eea
\bea
\I^{\rm (d_2)}(s,t) &=& (-t)^{-4\e} \Biggl\{
 - { 2 \over 3 \, \e^5 } \, \zeta_3 
 + {1\over \e^4} \Biggl[ - {4\over3} \, \zeta_3 \, L
                      + {11\over432} \, \pi^4 \Biggr]
\ +\ \Ord(\e^{-3}) \Biggr\} \,, \hskip 3.5 cm 
\label{Id2Analytic}
\eea
\bea
\I^{\rm (f_2)}(s,t) &=& (-t)^{-4\e} \Biggl\{
  { 16 \over 9 \, \e^8 } 
 + { 32 \over 9 \, \e^7 } L 
 + {1\over \e^6} \Biggl[  {91 \over 36} L^2 
                      - { 235 \over 108 } \pi^2 \Biggr]
\nonumber\\
&&\hskip-0.1cm
 + \, { 8 \over 3 \, \e^5} \Biggl[
        H_{0,0,1}(x) - L \, H_{0,1}(x)
      + {1\over2} ( L^2 + \pi^2) H_{1}(x) 
      + {29\over144} L^3
      - {199\over144} \pi^2 \, L - {1073\over144} \zeta_3 \Biggr]
\nonumber\\
&&\hskip-0.1cm
 + \, { 8 \over 3 \, \e^4} \Biggl[ 
      - H_{0,0,0,1}(x) - H_{0,0,1,1}(x)
      - H_{0,1,0,1}(x) - H_{1,0,0,1}(x)
\nonumber\\
&&\hskip1.0cm
      + L {}\Bigl( {5\over2} \, H_{0,0,1}(x) 
               + H_{0,1,1}(x) + H_{1,0,1}(x) \Bigr)
      - {L^2 \over2} ( 4 \, H_{0,1}(x) + H_{1,1}(x) )
\nonumber\\
&&\hskip1.0cm
      - {\pi^2\over2} \Bigl( H_{1,1}(x)
                           - {3\over2} L \, H_{1}(x)
                           + {115\over144} L^2 \Bigr)
\nonumber\\
&&\hskip1.0cm
      + {11\over12} L^3 \, H_{1}(x) + \zeta_3 \, H_{1}(x)
      - {L^4\over36}
      - {1037\over72} \zeta_3 \, L 
      + {6467\over17280} \pi^4 \Biggr] 
\nonumber\\
&&\hskip-0.1cm
 +\ \Ord(\e^{-3}) \Biggr\} \,.
\label{If2Analytic}
\eea

Using \eqn{FourLoopPlanarResult}, together with the above results for 
the integrals, the total four-loop planar amplitude, 
$M_4^\fourloop$, has the expansion,
\bea
M_4^\fourloop(s,t) &=& (-t)^{-4\e} \Biggl\{
   { 2 \over 3 \, \e^8 } 
 + { 4 \over 3 \, \e^7 } L 
 + {1\over \e^6} \Biggl[ L^2 
                      - { 13 \over 18 } \pi^2 \Biggr]
\nonumber\\
&&\hskip0.1cm
 + \, { 4 \over 3 \, \e^5} \Biggl[
        H_{0,0,1}(x) - L \, H_{0,1}(x)
      + {1\over2} ( L^2 + \pi^2) H_{1}(x) 
      + {L^3\over4}
      - {5\over6} \pi^2 \, L - {59\over12} \zeta_3 \Biggr]
\nonumber\\
&&\hskip0.1cm
 + \, { 4 \over 3 \, \e^4} \Biggl[ 
      - H_{0,0,0,1}(x) - H_{0,0,1,1}(x)
      - H_{0,1,0,1}(x) - H_{1,0,0,1}(x)
\nonumber\\
&&\hskip1.3cm
      + L {}\Bigl( {5\over2} \, H_{0,0,1}(x) 
               + H_{0,1,1}(x) + H_{1,0,1}(x) \Bigr)
      - {L^2 \over2} ( 4 \, H_{0,1}(x) + H_{1,1}(x) )
\nonumber\\
&&\hskip1.3cm
      - {\pi^2\over2} \Bigl( H_{1,1}(x)
                           - {3\over2} L \, H_{1}(x)
                           + {15\over16} L^2 \Bigr)
\nonumber\\
&&\hskip1.3cm
      + {11\over12} L^3 \, H_{1}(x) + \zeta_3 \, H_{1}(x)
      + {L^4\over32}
      - {28\over3} \zeta_3 \, L 
      + {637\over17280} \pi^4 \Biggr] 
\nonumber\\
&&\hskip0.1cm
 +\ \Ord(\e^{-3}) \Biggr\} \,.
\label{M4Analytic}
\eea
%

\begin{table}
\caption{\label{FourLoopTable} 
Numerical values of individual four-loop integrals, and $M_4^\fourloop$, 
at $(s,t)=(-1,-1)$.  The uncertainties at orders $\e^{-3}$ and $\e^{-2}$
are indicated in parentheses.  (The presence of two digits in parentheses 
signifies the uncertainty in the last two digits of the central value.)}
\vskip .4 cm
\def\hs{\hskip .001 cm $\!\!$}
\begin{tabular}{||c||c|c|r|r|r|r|r||}
\hline
\hline
Integral \hs & $\e^{-8}$ \hs & $\e^{-7}$ 
\hs & $\e^{-6}$\hbox{~~~~~~~} \hs & $\e^{-5}$\hbox{~~~~~~~}
 \hs & $\e^{-4}$\hbox{~~~~~~~} \hs & $\e^{-3}$\hbox{~~~~~~~~}
 \hs & $\e^{-2}$\hbox{~~~~~~~} \\
\hline
\hline
(a)   & $ 
    4/9 
   $  & $
       0     
   $  & $
      -4.27225931 
   $ \hs & $
      -11.30789527
   $ \hs & $
      -18.44325855
   $ \hs & $
      -58.84504
       (10)
   $ \hs &  $
      -180.852
       (3) \hphantom{1}    
     \, $ \\
\hline
(b)  &  $
      4/9          
   $ & $
       0      
   $\hs & $
      -4.82057067 
   $ \hs & $
      -10.53107909
    $ \hs & $
      3.00827162 
    $ \hs & $
      67.62584
       (13) 
    $ \hs & $
      190.235
       (3) \hphantom{1} 
     \, $
     \\
\hline
(c) & $
    4/9          
    $ & $      
     0      
    $ & $ 
     -5.30034310 
    $ \hs & $
     -10.38082198
    $ \hs & $
      12.55376125 
    $ \hs & $
       81.91311
      (64)  
    $ \hs & $
      99.292
      (5) \hphantom{1}  
      \, $ \\
\hline
(d)  & $ 
       4/9   
     $ & $ 
    0      
    $ & $
       -3.86102580 
     $ \hs & $
       -7.70172456
     $ \hs & $ 
       -16.94003184 
    $  \hs& $
       -80.03212
        (51) 
     $ \hs & $
       -52.555
        (21) 
       \, $  \\
\hline
(e) & $
       4/9         
   $ & $
     0     
   $ & $
       -4.47787607 
   $  \hs & $ 
      -8.45252237
   $  \hs& $ 
       -4.13769237 
   $ \hs & $ 
     -11.60392
       (20) 
   $ \hs & $
       28.823
        (9) \hphantom{1}  
      \,  $   \\
\hline
(f) & $
     8/9   
   $ & $
      0     
  $ & $
     -10.73776405 
   $ \hs  & $
      -16.90406921
   $ \hs & $
      46.68731190 
   $ \hs & $
      219.08111 
       (12)
   $ \hs & $ 
      364.167
       (7) \hphantom{1}
     \, $\\
\hline
(d${}_2$) 
   & $ 
       0    
   $ \hs & $
       0     
    $ \hs & $
      0\hbox{~~~~~~~~} 
    $ \hs & $
      -0.80137127
    $ \hs & $ 
       2.48032408 
    $ \hs & $
       36.23672
        (11) 
    $ \hs & $
       132.811
         (13)
    \, $  \\
\hline
(f${}_2$) 
       & $  
      16/9    
    $ \hs & $
        0      
    $ \hs & $
      -21.47552810 
   $ \hs & $
      -35.41088096
   $ \hs & $
      92.92365579 
   $ \hs & $
      521.48787
       (31) 
   $ \hs & $
      1314.856
        (12)
        \, $  \\
\hline
\hline
$M_4^\fourloop$ 
     & $
        2/3
     $ \hs  & $
           0      
     $ \hs  & $
        -7.12804762 
      $ \hs & $
       -13.64293336
      $ \hs & $
         2.64276920 
     $ \hs  & $
         27.34123
       (13) 
      $ \hs & $
         33.278 
       (7) \hphantom{1}
      \, $  \\
\hline
\hline
\end{tabular}
\end{table}

In table~\ref{FourLoopTable} we present the numerical values of the
eight integrals appearing in the four-loop planar amplitude, through
$\Ord(\e^{-2})$. 
The values through $\Ord(\e^{-4})$ can be
found easily from the analytic expression given above. 
The numerical values for $\ep^{-3}$ and $\ep^{-2}$
were obtained using the {\tt CUBA} numerical integration
package~\cite{CUBA}, which is incorporated into the {\tt MB}
program~\cite{CzakonMB}.   In the table, we also give
the total value of the amplitude $M_4^\fourloop$, according to
\eqn{FourLoopPlanarResult}.


\subsection{Values of lower-loop amplitudes at $(s,t)=(-1,-1)$}
\label{LowerLoopSymmetricSubsection}

Next we need to compare our results for \eqn{FourLoopPlanarResult}
with the prediction~(\ref{FourLoopFourPtIteration}) based
on the known structure of the infrared poles.
To do this, we evaluate the lower-loop amplitudes, using formulas from 
ref.~\cite{Iterate3}, at 
the symmetric kinematical point $(s,t)=(-1,-1)$, through the accuracy
needed to evaluate \eqn{FourLoopFourPtIteration} to $\Ord(\e^{-2})$.
At this point, a limited number of analytic expressions appear,
built out of $\ln2$, $\pi$, $\zeta_3$, $\zeta_5$, $\Li_4(\hf)$, 
$\Li_5(\hf)$, $\Li_6(\hf)$, and the harmonic sum~\cite{HPLMaitre},
\be
s_6 \equiv S(\{-5,-1\},\infty) 
= \sum_{i_1=1}^\infty { (-1)^{i_1} \over i_1^5 }
  \sum_{i_2=1}^{i_1} { (-1)^{i_2} \over i_2 }
= 0.98744142640329971377\ldots.
\label{s6def}
\ee

The one-loop amplitude $M_4^\oneloop(s,t;\e)$ in \eqn{OneLoopAmplitude}, 
evaluated at $(s,t)=(-1,-1)$, has the $\e$-expansion,
\bea
M_4^\oneloop(-1,-1;\e) &=&
 - {2 \over \e^2} + {2\over3} \pi^2
 + \e {}\biggl( {\pi^2\over2} \ln2 + {77\over12}  \zeta_3 \biggr)
\nonumber\\ &&\hskip0.1cm
 +\ \e^2  \biggl( - 2 \Li_4(\hf)
         - {1\over12}  \ln^{4}2
         + {\pi^2\over3} \ln^{2}2 + {49\over720} \pi^4 \biggr)
\nonumber\\ &&\hskip0.1cm
 +\ \e^3  \biggl( 2 \Li_5(\hf)
         - {1\over60} \ln^{5}2 + {\pi^2\over9} \ln^{3}2
         + {\pi^4\over360} \ln2
         - {245\over144} \pi^2 \zeta_3 + {62\over5} \zeta_5 \biggr)
\nonumber\\ &&\hskip0.1cm
 +\ \e^4  \biggl( - 2 \Li_6(\hf)
           + {\pi^2\over6} \Li_4(\hf)
           - {1\over360} \ln^{6}2 + {5\over144} \pi^2 \ln^{4}2
           - {\pi^4\over180} \ln^{2}2
\nonumber\\ &&\hskip1.0cm
 -\ {7\over6} \pi^2 \zeta_3 \ln2
           - {343\over36} \zeta_3^2 - {\pi^6\over10080} \biggr) 
 +\ \Ord(\e^5)
\label{M1_11_A}\\
&=&
 - {2\over\e^2} + 6.5797362673929057461 
\nonumber\\ &&\hskip0.1cm
 +\ 11.133742693869288271\ \e + 7.1556624851455749140\ \e^2 
\nonumber\\ &&\hskip0.1cm
 -\ 5.760188577405266135\ \e^3 - 23.794568007684383734\ \e^4 
+ \Ord(\e^5)\,. \nonumber\\
\label{M1_11_Num}
\eea
The two-loop amplitude $M_4^\twoloop(s,t;\e)$ 
in~\eqn{TwoloopPlanarResult}, is given by
\bea
M_4^\twoloop(-1,-1;\e) &=&
  {2 \over \e^4} - {5 \, \pi^2 \over4 \, \e^2 } 
 +\ {1\over\e}  \biggl( - \pi^2 \ln2 - {37\over3} \zeta_3 \biggr)
\nonumber\\ &&\hskip0.1cm
+\ 4 \Li_4(\hf) + {1\over6} \ln^{4}2
- {2\over3} \pi^2 \ln^{2}2 - {\pi^4\over30} 
\nonumber\\ &&\hskip0.1cm
+\ \e {}\biggl( - 4 \Li_5(\hf) + {1\over30} \ln^{5}2 
        - {2\over9} \pi^2 \ln^{3}2 + {43\over360} \pi^4 \ln2 
        + {77\over12} \pi^2 \zeta_3 - {3919\over80} \zeta_5 \biggr)
\nonumber\\ &&\hskip0.1cm
+\ \e^2 \biggl( - 7 s_6 + 4 \Li_6(\hf) + {22\over3} \pi^2 \Li_4(\hf)
        + {1\over180} \ln^{6}2 + {\pi^2\over4} \ln^{4}2
        - {59\over240} \pi^4 \ln^{2}2 
\nonumber\\ &&\hskip1.0cm
        +\ {307\over24} \pi^2 \zeta_3 \ln2
        + {4319\over288} \zeta_3^2 - {541\over6480} \pi^6 \biggr) 
+\ \Ord(\e^3)
\label{M2_11_A}\\
&=&
   {2\over\e^4} - {12.337005501361698274 \over \e^2} 
\nonumber\\ &&\hskip0.1cm
 -\ {21.666456936158779398 \over \e} - 4.2998350584631215560
\nonumber\\ &&\hskip0.1cm
 +\ 30.635795346547106621\ \e + 68.218654436238118625\ \e^2
+ \Ord(\e^3)\,. \nonumber\\
\label{M2_11_Num}
\eea
The three-loop amplitude $M_4^\threeloop(s,t;\e)$ 
in~\eqn{ThreeLoopPlanarResult} is given by
\bea
M_4^\threeloop(-1,-1;\e) &=&
  - {4\over 3\,\e^6} + {7\,\pi^2 \over 6 \, \e^4} 
  + {1\over\e^3} \biggl( \pi^2 \ln2 + {71\over6} \zeta_3 \biggr)
\nonumber\\ &&\hskip0.1cm
  +\ {1\over\e^2} \biggl( - 4 \Li_4(\hf) - {1\over6} \ln^{4}2 
      + {2\over3} \pi^2 \ln^{2}2 - {89\over3240} \pi^4 \biggr)
\nonumber\\ &&\hskip0.1cm
  +\ {1\over\e} \biggl( 4 \Li_5(\hf) - {1\over30} \ln^{5}2
          + {2\over9} \pi^2 \ln^{3}2 - {73\over360} \pi^4 \ln2
          - {3779\over432} \pi^2 \zeta_3 + {8621\over120} \zeta_5 \biggr)
\nonumber\\ &&\hskip0.1cm
  +\ 14 s_6 - 4 \Li_6(\hf) - {91\over6} \pi^2 \Li_4(\hf)
  - {1\over180} \ln^{6}2 - {83\over144} \pi^2 \ln^{4}2
  + {191\over360} \pi^4 \ln^{2}2 
\nonumber\\ &&\hskip1.0cm
  - 23\ \pi^2 \zeta_3 \ln2
  - {1385\over144} \zeta_3^2 + {43159\over233280} \pi^6 
+\ \Ord(\e)
\label{M3_11_A}\\
&=&
 - {4\over 3\,\e^6} + {11.514538467937585056\over \e^4}
\nonumber\\ &&\hskip0.1cm
 +\ {21.065428484578982255\over\e^3} - {1.6228781926783846589\over\e^2}
\nonumber\\ &&\hskip0.1cm
 -\ {40.219043687209842734\over\e} - 67.305777557207060997
+\ \Ord(\e)\,.   
\label{M3_11_Num}
\eea
The iterative formula for the four-loop amplitude in terms
of the lower-loop amplitudes is given in \eqn{FourLoopFourPtIteration}.
Inserting the values at $(s,t)=(-1,-1)$ of $M_4^\oneloop$,
$M_4^\twoloop$ and $M_4^\threeloop$, we obtain
\bea
M_4^\fourloop(-1,-1;\e)\Bigr|_{\rm iter.} &=&
    {2\over3\,\e^8} - {13\,\pi^2 \over 18\,\e^6}
  + {1\over\e^5} \biggl( - {2\over3} \pi^2 \ln2 - {68\over9} \zeta_3 \biggr)
\nonumber\\ &&\hskip0.1cm
  +\ {1\over\e^4} \biggl( {8\over3} \Li_4(\hf) + {1\over9} \ln^{4}2
            - {4\over9} \pi^2 \ln^{2}2 + {89\over2592} \pi^4 \biggr)
\nonumber\\ &&\hskip0.1cm
  +\ {1\over\e^3} \biggl( - {8\over3}  \Li_5(\hf) + {1\over45}  \ln^{5}2
              - {4\over27} \pi^2 \ln^{3}2 + {22\over135} \pi^4  \ln2
\nonumber\\ &&\hskip1.0cm
              + {251\over36} \pi^2 \zeta_3 - {7469\over120} \zeta_5 \biggr)
\nonumber\\ &&\hskip0.1cm
  +\ {1\over\e^2} \biggl( - 14 s_6 + {8\over3} \Li_6(\hf) 
              + {139\over9} \pi^2 \Li_4(\hf)
              + {1\over270} \ln^{6}2 + {131\over216} \pi^2 \ln^{4}2
\nonumber\\ &&\hskip1.0cm
              -\ {607\over1080} \pi^4 \ln^{2}2 
              + {791\over36} \pi^2 \zeta_3 \ln2
              + {895\over432}  \zeta_3^2 - {20759\over102060}  \pi^6 
              - {1\over8} f_0^\fourloop \biggr)
\nonumber\\ &&\hskip0.1cm
+\ \Ord(\e^{-1})
\label{M4_iter_11_A}\\
&=&
  {0.66666666666666666667\over\e^8} - {7.1280476230089812249\over\e^6}
\nonumber\\ &&\hskip0.1cm
-\ {13.642933355332790075\over\e^5} + {2.6427691992903098962\over\e^4}
\nonumber\\ &&\hskip0.1cm
+\ {27.341074205440151100\over\e^3} 
\nonumber\\ &&\hskip0.1cm
+\ {1\over\e^2} \Bigl( 29.611139840724282137 - {1\over8} f_0^\fourloop \Bigr)
+\ \Ord(\e^{-1})\,. 
\label{M4_iter_11_Num}
\eea

Comparing \eqn{M4_iter_11_Num} with the last row of
table~\ref{FourLoopTable}, we verify the pole behavior of the
four-loop amplitude $M_4^\fourloop(-1,-1;\e)$ precisely through
$\Ord(\e^{-4})$.  The agreement at $\Ord(\e^{-3})$ is good to 5 digits.
At $\Ord(\e^{-2})$, we can extract the value of $f_0^\fourloop$.
We obtain,
\be
f_0^\fourloop = 
-29.335 \pm 0.052
\,.
\label{f04ans}
\ee
This value should be compared with that predicted by Eden and Staudacher,
from \eqn{gammaKB},
\be
f_0^\fourloop\Bigr|_{\rm ES}\ =\
- {73\over2520} \, \pi^6 + \zeta_3^2 
\ =\ 
- 26.404825523390660965
\ldots\,.
\label{f04ESans}
\ee
The results do not agree.  The difference can be expressed as, 
\bea
f_0^\fourloop &=& f_0^\fourloop\Bigr|_{\rm ES}\ +\ \Delta f_0^\fourloop\,,
\label{DeltafDef}\\
\Delta f_0^\fourloop &=& 
- 2.930 \pm 0.052
\ldots\,.
\label{DeltafNum}
\eea
We can also parametrize the difference $\Delta f_0^\fourloop$
as a multiple of the weight-6 expression $\zeta_3^2$.
Making this parametrization, we find that 
\bea
\Delta f_0^\fourloop &=& r \, \zeta_3^2 \,,
\label{rDef}\\
r &=& 
- 2.028 \pm 0.036
\ldots\,.
\label{rNum}
\eea
This result is quite suggestive.  To about 1.5\% precision 
on the value of the correction term $\Delta f_0^\fourloop$,
it is equal to $-2\zeta_3^2$, a result which would have
the net effect of {\it flipping the sign of the $\zeta_3^2$
term in the Eden--Staudacher prediction~(\ref{gammaKA}), 
while leaving the $\pi^6$ term unaltered}.  Of course,
the ES prediction follows directly from an integral equation,
and so flipping the sign  of the $\zeta_3^2$ term
is not possible without other modifications.
In \sect{AnalysisSection} we discuss possible reasons
for the discrepancy.


\subsection{Cross checks at asymmetrical kinematical points}

In order to cross check our numerical evaluation of the integrals
at the symmetric kinematical point $(s,t)=(-1,-1)$,
as well as check the behavior of the $\Ord(\e^{-3})$ and 
$\Ord(\e^{-2})$ terms in the planar four-loop amplitude as a function of
the scattering angle, we have performed the numerical analysis
of the last subsection at three additional kinematical points, 
$(s,t)=(-1,-2)$, $(-1,-3)$ and $(-1,-15)$.

We have used the expressions for the lower-loop amplitudes
in ref.~\cite{Iterate3} to numerically evaluate the 
infrared-based iterative formula~(\ref{FourLoopFourPtIteration})
at the asymmetric kinematical points $(s,t)=(-1,-2)$, 
$(-1,-3)$ and $(-1,-15)$.  
The results are,
\bea
M_4^\fourloop(-1,-2;\e)\Bigr|_{\rm iter.} &=&
  {2\over3\,\e^8} - {0.92419624075\over\e^7} - {6.64759460909\over\e^6}
- {4.23222757233\over\e^5}
\nonumber\\ &&\hskip0cm\null
+ {15.89245103368\over\e^4} + {16.11914613046\over\e^3} 
\nonumber\\ &&\hskip0cm\null
+ {1\over\e^2} \Bigl( 1.31283053842 - {1\over8} f_0^\fourloop \Bigr)
+\ \Ord(\e^{-1})\,, 
\label{M4_iter_12_Num}
\eea
\bea
M_4^\fourloop(-1,-3;\e)\Bigr|_{\rm iter.} &=&
  {2\over3\,\e^8} - {1.46481638489\over\e^7} - {5.92109866220\over\e^6}
+ {0.72092946726\over\e^5}
\nonumber\\ &&\hskip0cm\null
+ {19.05722201166\over\e^4} + {4.86152575608\over\e^3} 
\nonumber\\ &&\hskip0cm\null
+\ {1\over\e^2} \Bigl( - 5.61581265989 - {1\over8} f_0^\fourloop \Bigr)
+\ \Ord(\e^{-1})\,, 
\label{M4_iter_13_Num}
\eea
\bea
M_4^\fourloop(-1,-15;\e)\Bigr|_{\rm iter.} &=&
  {2\over3\,\e^8} - {3.61073360147\over\e^7} + {0.20548826868\over\e^6}
+ {14.13192416428\over\e^5}
\nonumber\\ &&\hskip0cm\null
+ {7.41700629511\over\e^4} - {48.55010675803\over\e^3} 
\nonumber\\ &&\hskip0cm\null
+\ {1\over\e^2} \Bigl( 43.61197714 - {1\over8} f_0^\fourloop \Bigr)
+\ \Ord(\e^{-1})\,. 
\label{M4_iter_115_Num}
\eea

Numerical evaluation of the eight four-loop integrals
from \figs{rrFigure}{nonrrFigure} gives the following
results for the amplitude $M_4^\fourloop$ (we omit the
$1/\e^8$ through $1/\e^4$ poles, as they agree analytically),
\begin{eqnarray}
M_4^\fourloop(-1,-2;\e) &=&
\Ord(\e^{-8}\!\cdots \e^{-4})+
 {16.11929 \pm 0.00008 \over\e^3} 
 +  {4.985   \pm 0.006 \over\e^2}
+ \Ord(\e^{-1})\,,\hskip 10mm
\label{M4_amp_12_Num}\\
M_4^\fourloop(-1,-3;\e) &=&
\Ord(\e^{-8}\!\cdots \e^{-4})+
  {4.8617 \pm 0.0003 \over\e^3} 
 -  {1.943  \pm 0.008 \over\e^2}
+ \Ord(\e^{-1})\,,\hskip7mm
\label{M4_amp_13_Num}\\
M_4^\fourloop(-1,-15;\e) &=&
\Ord(\e^{-8}\!\cdots \e^{-4})
- {48.5499 \pm 0.0002 \over\e^3} 
 +  {47.29   \pm 0.02 \over\e^2}
+ \Ord(\e^{-1})\,.\hskip 7mm
\label{M4_amp_115_Num}
\end{eqnarray}
Comparing these sets of numbers at $\Ord(\e^{-3})$, 
we observe good agreement at all points; the 
$(s,t)=(-1,-2)$ point is slightly off, at 1.9\,$\sigma$,
but the other two points are within 1\,$\sigma$.

At $\Ord(\e^{-2})$, we can express the agreement
in terms of the parameter $r$ introduced in \eqn{rDef}.
At the asymmetric kinematical points, we extract the values,
\bea
r &=&
 - 2.059 \pm  0.036
, \qquad (s,t)=(-1,-2),  
\label{rNum12}\\
r &=&
 - 2.062 \pm  0.045
, \qquad (s,t)=(-1,-3),  
\label{rNum13}\\
r &=&
 - 2.074 \pm  0.104
, \qquad (s,t)=(-1,-15). 
\label{rNum115}
\eea
These values are all consistent, within errors, with the 
value~(\ref{rNum}) extracted at $(s,t)=(-1,-1)$.  
(The values at different points, however, have an unknown
correlation between them, because the integrals contain pieces
that are independent of the kinematics, and the numerical 
integration for each value of $(s,t)$ was performed with 
the same sequence of quasi-random integration points.
This means the results for $r$ from the various kinematic points
cannot be combined to reduce the error.)

In summary, our numerical integration of the four-loop planar 
integrand results in a value of the cusp anomalous dimension,
\bea
 f_0(\ah)
  &=&    \ah - {\pi^2\over6} \, \ah^2
           + {11\over180} \pi^4 \, \ah^3
           - \Bigl( {73\over2520} \, \pi^6 - (1+r) \zeta_3^2 \Bigr) \ah^4
           +\ \cdots \,.
\label{gammaKNum}
\eea
where $r$ is given in eqs.~(\ref{rNum}), (\ref{rNum12}),
(\ref{rNum13}) and (\ref{rNum115}).  All values are consistent with
the appealing value of $r=-2$, which corresponds to the value $\beta =
\zeta_3$ for the dressing-factor parameter $\beta$ in
\eqn{betashifteq}.  As noted above, this value would merely flip the
sign of the $\zeta_3^2$ term in the ES prediction~(\ref{gammaKA}).
However, we obviously cannot exclude, on numerical grounds alone,
nearby rational or transcendental numbers.  For example, it is
conceivable that $r$ takes on the value $r = - 5/(2\zeta_3) = -
2.0797\ldots$.  This value would correspond to a rational
dressing-factor parameter, $\beta = 5/4$, and would violate the KLOV
maximum-transcendentality principle.  On the other hand, additional
evidence points toward $r=-2$ as the correct analytical value, as we
shall discuss in the next section.

\section{Estimating strong-coupling behavior}
\label{LargeCouplingSection}

Kotikov, Lipatov and Velizhanin (KLV)~\cite{KLV} made an intriguing proposal
for approximating the cusp anomalous dimension (or equivalently,
$f_0(\ah)$), for all values of the coupling $\ah$.  
They suggested combining perturbative information with the 
knowledge from string theory~\cite{StrongCouplingLeadingGKP} that 
at large values of the coupling $f_0$ has square-root behavior,
$f_0 \sim \sqrt{\ah}$.  They proposed the following approximate relation
as a means for incorporating the known analytic behavior,
\begin{equation}
\ah^n = \sum_{r=n}^{2n} C_r \,  [\tilde{f}_0(\ah)]^r \,,
\label{KLVapproxn}
\end{equation}
where the constants $C_r$ can be fixed using perturbative information.
As we shall discuss below, they can also be fixed using
strong-coupling information.  As more information
becomes available the integer $n$ can be increased.
The strong-coupling square-root behavior of $f_0$
is automatically imposed by the fact that 
$\ah^n \sim C_{2n} [f_0(\ah)]^{2n}$ at large $\ah$.
Similarly, the weak-coupling linear behavior follows
from $\ah^n \sim C_{n} [f_0(\ah)]^{n}$ at small $\ah$.

KLV used the approximation~(\ref{KLVapproxn}) for $n=1$, 
together with the one- and two-loop expressions for the 
cusp anomalous dimension, to write (for the case of
a supersymmetric regulator)
\begin{equation}
\ah = 
\tilde{f}_0 + {\pi^2\over6}\ (\tilde{f}_0)^2 
\,.  \label{fapprox1}
\end{equation}
This formula makes the weak-coupling predictions (beyond
two loops),
\begin{eqnarray}
\tilde{f}_0 &=& 
\ah - {\pi^2\over6}\ \ah^2 
+ {\pi^4\over18}\ \ah^3
- {5\over216} \pi^6\ \ah^4 + {7\over648} \pi^8\ \ah^5
- {7\over1296} \pi^{10}\ \ah^6
 + \ldots \,, \nonumber\\
&& \hskip1cm \hbox{as $\ah\to 0$,}
\label{fapprox1weak}
\end{eqnarray}
and it predicts coefficients in the strong-coupling
expansion, as
\begin{eqnarray}
\tilde{f}_0 &=& 
{ 2 \sqrt{3} \over \pi}\ \sqrt{{\ah\over2}}\
-\ { 3 \over \pi^2 }
\ +\ \Ord( \ah^{-1/2} )
\label{fapprox1strong} \\
&\approx& 
1.1027\ \sqrt{{\ah\over2}}\ -\ 0.30396 
\ +\ \Ord( \ah^{-1/2} ) \,.
\label{fapprox1strongnum}
\end{eqnarray}

The coefficients of the
leading~\cite{StrongCouplingLeadingGKP,Kruczenski,Makeenko} and
subleading~\cite{StrongCouplingSubleading} terms in this expansion are
predicted from string theory to be,
\begin{eqnarray}
f_0 &=& 
\sqrt{{\ah\over2}}\ -\ { 3\ln2 \over 4\pi }
\ +\ \Ord( \ah^{-1/2} )
\label{fstringstrong} \\
&\approx& 
\sqrt{{\ah\over2}}\ -\ 0.16547670011448
\ +\ \Ord( \ah^{-1/2} ) \,.
\label{fstringstrongnum}
\end{eqnarray}
As noted by KLV, the leading coefficient is estimated correctly
to 10\% by the formula~(\ref{fapprox1}).  The subleading coefficient
is off by almost a factor of two, however. 

What happens as we incorporate more perturbative information?
Using the three-loop value for the cusp anomalous dimension,
and setting $n=2$ in \eqn{KLVapproxn}), gives the approximation
\begin{equation}
\ah^2 = 
(\tilde{f}_0)^2
+ {\pi^2\over3}\ (\tilde{f}_0)^3
+ {\pi^4\over60}\ (\tilde{f}_0)^4 
\,.  \label{fapprox2}
\end{equation}
This approximation makes the weak-coupling predictions (beyond three
loops),
\begin{eqnarray}
\tilde{f}_0 &=& 
\ah - {\pi^2\over6}\ \ah^2 
+ {11\over180} \pi^4\ \ah^3
- {31\over1080} \pi^6\ \ah^4 
+ {329\over21600} \pi^8\ \ah^5
- {169\over19440} \pi^{10}\ \ah^6
 + \ldots \,, \nonumber\\
&&  \hskip1cm \hbox{as $\ah\to 0$,}
\label{fapprox2weak}
\end{eqnarray}
and has the strong-coupling expansion,
\begin{eqnarray}
\tilde{f}_0 &=& 
{2\over \pi}\, 15^{1/4}\ \sqrt{{\ah\over2}}\
-\ { 5 \over \pi^2 }
\ +\ \Ord( \ah^{-1/2} )
\label{fapprox2strong} \\
&\approx& 
1.2529\ \sqrt{{\ah\over2}}\ -\ 0.50661 
\ +\ \Ord( \ah^{-1/2} ) \,.
\label{fapprox2strongnum}
\end{eqnarray}
For both the leading and next-to-leading coefficients in the
strong-coupling expansion, the three-loop approximation~(\ref{fapprox2})
leads to a larger disagreement with the string
prediction~(\ref{fstringstrongnum}) than does the two-loop 
version~(\ref{fapprox1}). 
We also note that the numerical value of the four-loop
coefficient predicted by \eqn{fapprox2} is $-27.595$,
which is a bit closer to our result than is the ES prediction, 
but still about 6\% off.

Despite the somewhat discouraging results from including the three-loop
values, we proceed to incorporate our four-loop
cusp anomalous dimension into the $n=3$ version of the
approximation, obtaining
\begin{equation}
\ah^3 = 
(\tilde{f}_0)^3
+ {\pi^2\over2}\ (\tilde{f}_0)^4
+ {\pi^4\over15}\ (\tilde{f}_0)^5 
+ \left( {\pi^6\over378} - 3(1+r) \zeta_3^2 \right) (\tilde{f}_0)^6 
\,.  \label{fapprox3}
\end{equation}
Here we have introduced the same coefficient $r$ defined in \eqn{rDef},
which is constrained to be quite close to $-2$ by our 
numerical result~(\ref{rNum}).
The weak-coupling expansion of this formula predicts (beyond four loops),
\begin{eqnarray}
\tilde{f}_0 &=& 
\ah - {\pi^2\over6}\ \ah^2 
+ {11\over180} \pi^4\ \ah^3
- \left( {73\over2520} \pi^6 - (1+r) \zeta_3^2 \right) \ah^4 
\nonumber\\
&& \hskip0cm \null
+ \left( {1769\over113400} \pi^8 
       - {4\over3} (1+r) \pi^2 \zeta_3^2 \right) \ah^5
- \left( {4111\over453600} \pi^{10}
       - {13\over10} (1+r) \pi^4 \zeta_3^2 \right) \ah^6
 + \ldots \,, \nonumber\\
&&  \hskip1cm \hbox{as $\ah\to 0$.}
\label{fapprox3weak}
\end{eqnarray}
The strong-coupling expansion is given by,
\begin{eqnarray}
\tilde{f}_0 &=& 
\alpha^{-1/6}\ \sqrt{{\ah\over2}}
\ -\ { \pi^4 \over 720 \alpha }\ 
+\ {\pi^2\over256} \left( {\pi^6\over2835} + (1+r)\zeta_3^2 \right) 
 \alpha^{-11/6}\ \sqrt{{2\over\ah}}
\ +\ \Ord( \ah^{-1} ) \,,~~~
\label{fapprox3strong}
\end{eqnarray}
where
\begin{equation}
\alpha\ =\ {\pi^6\over3024} - {3\over8} (1+r) \zeta_3^2 \,, 
\label{alphaDef}
\end{equation}
and we have given one more term in the expansion than before.
Curiously, with the ES value $r=0$, the approximate relation~(\ref{fapprox3})
breaks down, because $\alpha$ is negative, and hence $\alpha^{-1/6}$ is 
not real.

For $r=-2$, formula~(\ref{fapprox3strong}) becomes,
\begin{eqnarray}
\tilde{f}_0 &\approx& 
1.02550\ \sqrt{{\ah\over2}}\ -\ 0.157356\
-\ 0.0562398\ \sqrt{{2\over\ah}}
\ +\ \Ord( \ah^{-1} ) \,.
\label{fapprox3strongnum}
\end{eqnarray}
The numerical agreement between \eqn{fapprox3strongnum} and
the string-theory result in \eqn{fstringstrongnum} is
quite impressive: The leading coefficient agrees within 2.6\%, 
and the subleading coefficient within 5\%.
The coefficient of the term proportional to $\sqrt{{2/\ah}}$ 
in \eqn{fapprox3strongnum} is fairly small.  As we discuss below,
an improved estimate suggests that it may be considerably smaller,
or perhaps even vanish.

In the KLV type of approximation, the predicted value
of the strong-coupling coefficients depends quite sensitively
on the value of the four-loop contribution to the anomalous dimension.
For example, if we scale the numerical value of the four-loop contribution as
follows, 
\begin{equation}
f_0^{(4)}\ \rightarrow\  - (1+ \delta) 
  \Bigl( {73\over2520} \, \pi^6 + \zeta_3^2 \Bigr)
\end{equation}
instead of \eqn{fapprox3strongnum}, we find at strong coupling,
\begin{eqnarray}
\tilde{f}_0 &\approx& 
(0.8597725 +  10.9855 \, \delta)^{-1/6}
\sqrt{{\ah\over2}}\ - {1 \over 6.33501 + 81.1995 \, \delta}
\ +\ \Ord( \ah^{-1/2} ) \,,
\end{eqnarray}
which exhibits a strong sensitivity under just a few percent change
in the four-loop contribution. 
More generally, the sensitivity of the
strong-coupling prediction to the higher-loop orders used in 
a KLV approximation, allows us to test whether a given ansatz appears
compatible with strong coupling, as we discuss below.

%
\begin{figure}[t]
\centerline{\epsfxsize 6.5 truein \epsfbox{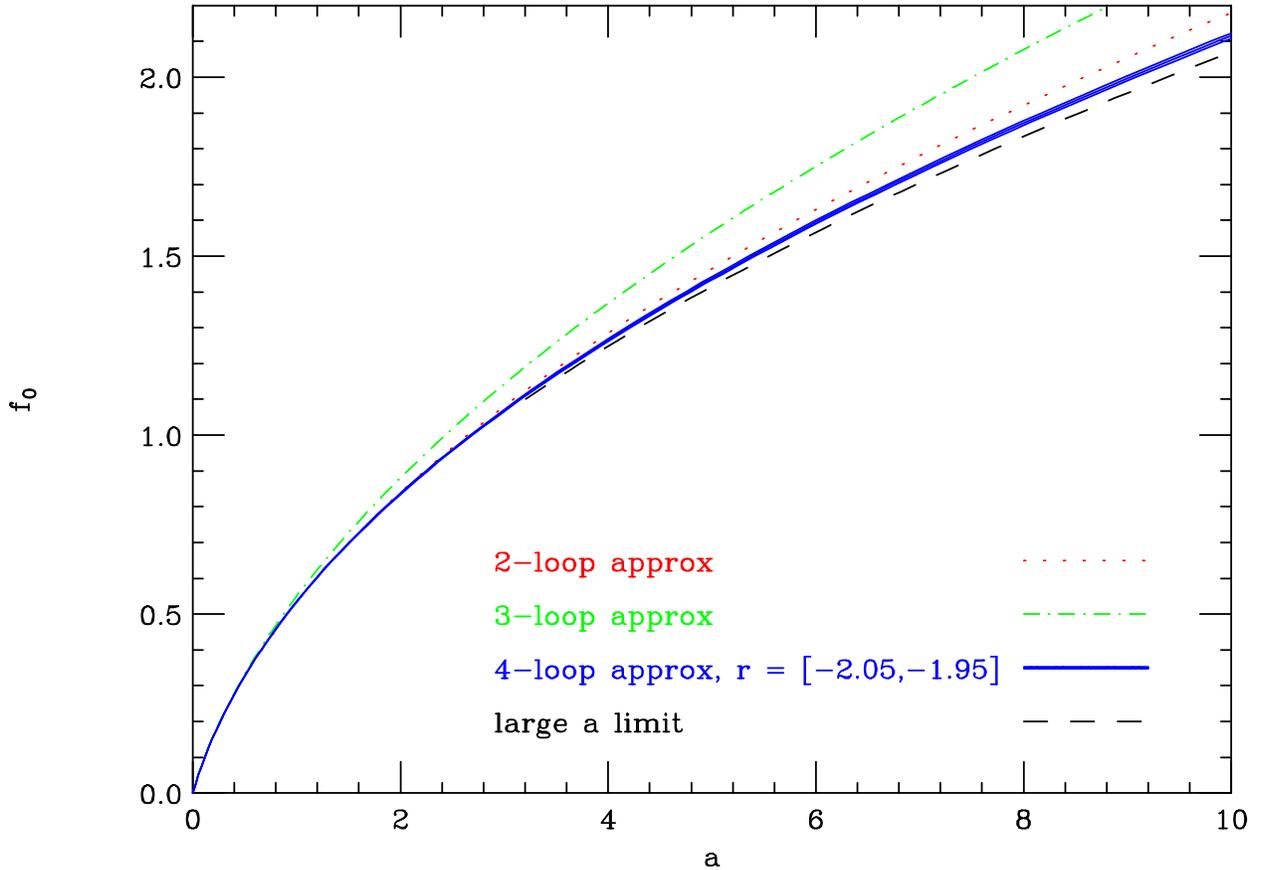}}
\caption{Approximations to the cusp anomalous dimension in planar \SYM\
based on the formula~(\ref{KLVapproxn}) and perturbative information
through two loops (dotted line), three loops (dot-dash line) 
and four loops (solid line).  Also shown is the strong-coupling
prediction from string theory (dashed line).}
\label{cuspapprox123Figure}
\end{figure}

In \fig{cuspapprox123Figure} we plot these estimates as a
function of the coupling, and we also display the strong-coupling 
limit~(\ref{fstringstrongnum}) predicted by string theory.
As noted above, the approximation~(\ref{fapprox1}) using only two-loop
information works quite well, in fact better than the three-loop
approximation~(\ref{fapprox2}).  However, the behavior of the four-loop 
formula~(\ref{fapprox3}) is clearly extremely close to the string theory
prediction.  For this curve, or rather band, we have varied the parameter 
$r$ between $-2.05$ and $-1.95$, consistent with \eqn{rNum}.

We have constructed two further approximations by incorporating the knowledge
of the precise strong-coupling coefficients.  Matching the four-loop 
perturbative information and the leading strong-coupling coefficient
gives the approximation
\begin{eqnarray}
\ah^4 &=& 
(\tilde{f}_0)^4
+ {2\over3} \pi^2\ (\tilde{f}_0)^5
+ {13\over90} \pi^4\ (\tilde{f}_0)^6 
+ \left( {23\over1890} \pi^6 - 4(1+r) \zeta_3^2 \right) (\tilde{f}_0)^7
+ 16\ (\tilde{f}_0)^8
\,. \nonumber\\
&&\hskip0cm \null
 \label{fapprox4}
\end{eqnarray}
It has the weak-coupling expansion,
\begin{eqnarray}
\tilde{f}_0 &=& 
\ah - {\pi^2\over6}\ \ah^2 
+ {11\over180} \pi^4\ \ah^3
- \left( {73\over2520} \pi^6 - (1+r) \zeta_3^2 \right) \ah^4 
\nonumber\\
&& \hskip0cm \null
+ \left( {4747\over302400} \pi^8 
       - {3\over2} (1+r) \pi^2 \zeta_3^2 - 4 \right) \ah^5
\nonumber\\
&& \hskip0cm \null
- \left( {5023\over544320} \pi^{10}
       - {19\over12} (1+r) \pi^4 \zeta_3^2 - {20\over3} \pi^2 \right) \ah^6
 + \ldots \,, \nonumber\\
&&  \hskip1cm \hbox{as $\ah\to 0$.}
\label{fapprox4weak}
\end{eqnarray}
If we add the next-to-leading strong-coupling coefficient 
as another constraint, we obtain the approximation,
\begin{eqnarray}
\ah^5 &=& 
(\tilde{f}_0)^5
+ {5\over6} \pi^2\ (\tilde{f}_0)^6
+ {\pi^4\over4}\ (\tilde{f}_0)^7 
+ \left( {17\over504} \pi^6 - 5(1+r) \zeta_3^2 \right) (\tilde{f}_0)^8
\nonumber\\
&&\hskip0cm \null
+ {240\, \ln2 \over \pi}\ (\tilde{f}_0)^9
+ 32\ (\tilde{f}_0)^{10}
\,, \label{fapprox5}
\end{eqnarray}
with the weak-coupling expansion,
\begin{eqnarray}
\tilde{f}_0 &=& 
\ah - {\pi^2\over6}\ \ah^2 
+ {11\over180} \pi^4\ \ah^3
- \left( {73\over2520} \pi^6 - (1+r) \zeta_3^2 \right) \ah^4 
\nonumber\\
&& \hskip0cm \null
+ \left( {727\over45360} \pi^8 
       - {5\over3} (1+r) \pi^2 \zeta_3^2 - {48\,\ln2\over\pi} \right) \ah^5
\nonumber\\
&& \hskip0cm \null
- \left( {13387\over1360800} \pi^{10}
       - {341\over180} (1+r) \pi^4 \zeta_3^2 
       - 88 \, \pi \, \ln2 + {32\over5}  \right) \ah^6
 + \ldots \,, \nonumber\\
&&  \hskip1cm \hbox{as $\ah\to 0$.}
\label{fapprox5weak}
\end{eqnarray}

We may also use the improved approximation (\ref{fapprox5}) to determine
subleading terms in the strong-coupling expansion.  We have
\begin{eqnarray}
\tilde f_0 &=& 
\sqrt{{\ah\over2}}\ -\ { 3\ln2 \over 4\pi }\ -\ 
\Bigl(  {17\over 161280} \pi^6 - {81\over32 \pi^2} \ln^2 2
          - {1\over 64} (1+r) \zeta_3^2 \Bigr) \sqrt{{2\over \ah}}
       \ +\ \Ord(\ah^{-1}) \,.~~~~
\label{fapprox5strong}
\end{eqnarray}
For $r = -2$, this expression evaluates to 
\begin{eqnarray}
\tilde f_0 &\approx& 
\sqrt{{\ah\over2}}\ -\ 0.16548 \
   -\ 0.000693\ \sqrt{{2\over \ah}}\ 
 +\ \Ord(\ah^{-1}) \,.
\end{eqnarray}
The first two numerical coefficients automatically reproduce 
the input~\cite{StrongCouplingLeadingGKP,Kruczenski,Makeenko,%
StrongCouplingSubleading} 
from string theory, \eqn{fstringstrongnum}, so the content
of this equation is in the coefficient of $\sqrt{2/\ah}$.
Quite interestingly, this coefficient is very small, 
suggesting that it may even vanish in the exact expression.  
It is an intriguing question whether such a tiny or vanishing result 
could be obtained from string theory.

\begin{figure}[t]
\centerline{\epsfxsize 6.5 truein \epsfbox{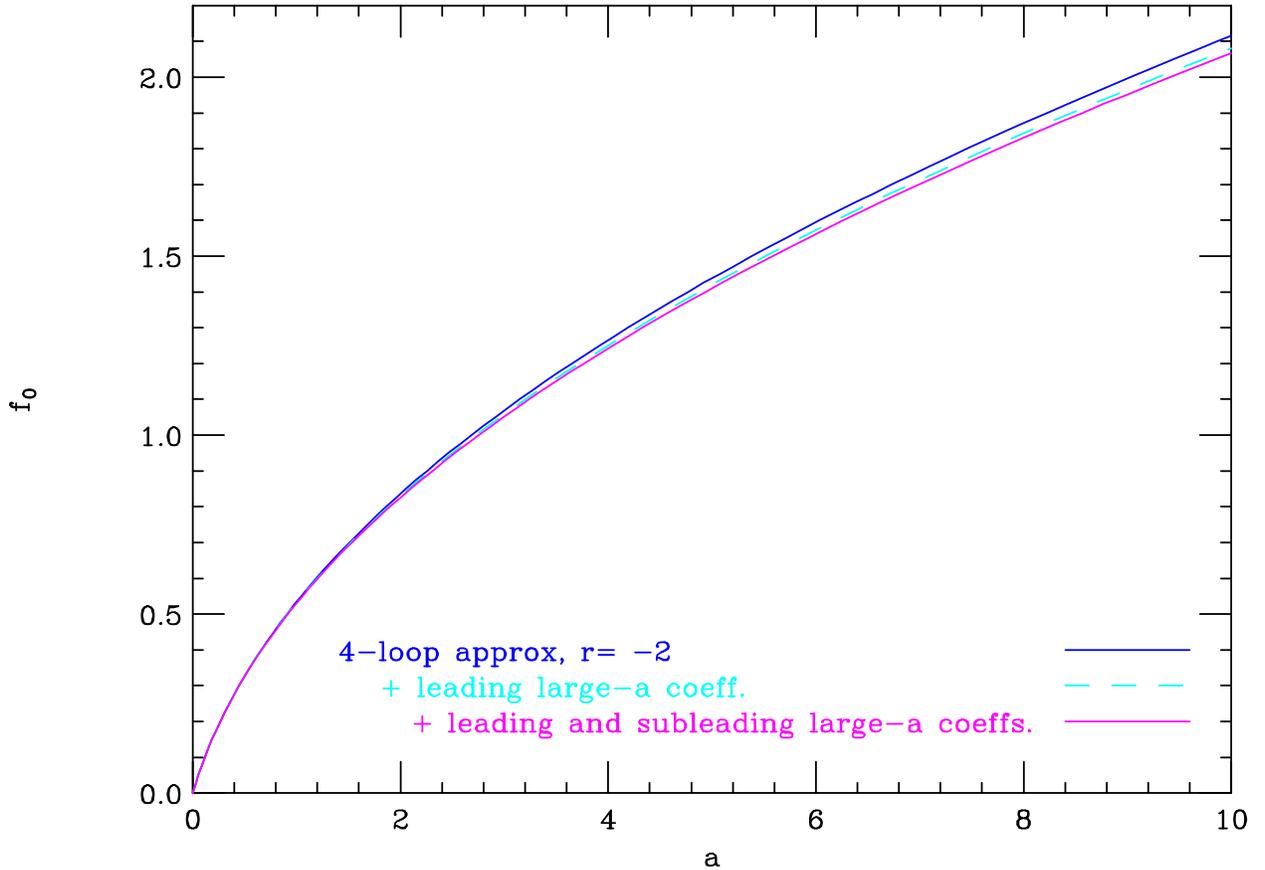}}
\caption{Approximations to the cusp anomalous dimension in planar \SYM\
based on the formula~(\ref{KLVapproxn}), four-loop
perturbative information, and zero, one or two constants from the 
strong-coupling expansion.}
\label{cuspapprox345Figure}
\end{figure}

We plot these two approximations, for $r=-2$, along with the four-loop
approximation~(\ref{fapprox3}), in \fig{cuspapprox345Figure}.
Obviously, they are extremely close to one another.
As a measure of that, the values they predict for the
five-loop planar cusp anomalous dimension (for $r=-2$)
are 167.03, 166.34 and 165.25, corresponding to the three approximations
in eqs.~(\ref{fapprox3weak}), (\ref{fapprox4weak}) and (\ref{fapprox5weak}). 
We expect the last of these numbers to be the most accurate,
given that \eqn{fapprox5weak} uses all available information
at both strong and weak coupling.

%

Can we use these approximations to guide corrections to the ES prediction
at higher loops?  As a first step, consider the five-loop ES prediction in
\eqn{gammaKB}. The ES numerical value is $131.22$, which disagrees
significantly with the prediction of our approximate
formula (\ref{fapprox5weak}).  However, we can modify the 
five-loop ES prediction (\ref{gammaKB}) in a manner analogous
to the modification required at four loops to fit the value $r=-2$.
We flip the signs of the terms containing $\zeta_n$ values for
odd $n$, and leave untouched the terms containing only even $\zeta$ 
values (those terms containing only $\pi$'s and rational numbers).
Then the terms containing $\zeta_3$ and/or $\zeta_5$ in 
the five-loop coefficient in \eqn{gammaKB} acquire the
same sign as the $\pi^8$ term, instead of the opposite sign, giving
an analytic expression through five loops:
\begin{eqnarray}
f_0^{\rm modified}(\ah)
 &=&         \ah - {\pi^2\over6} \, \ah^2
           + {11\over180} \pi^4 \, \ah^3
           - \Bigl( {73\over2520} \, \pi^6 + \zeta_3^2 \Bigr) \ah^4
\nonumber\\ 
&&\hskip2.0cm\null
           + \Bigl( {887\over56700} \, \pi^8 
                   + {\pi^2\over3} \, \zeta_3^2
                   +10 \, \zeta_3 \, \zeta_5 \Bigr) \ah^5 
           +\ \cdots \,.
\label{f0modified}
\end{eqnarray}
The numerical value of the five-loop coefficient in \eqn{f0modified}, 
$165.65$, 
agrees with the five-loop coefficient in our approximate 
formula~(\ref{fapprox5weak}) to a remarkable 0.25\%.
This excellent agreement in turn reinforces the notion
that $r=-2$ gives the correct analytical value of the
four-loop coefficient.  (We remark that a similar procedure
at three loops, using the approximation~(\ref{KLVapproxn})
for $n=4$ and incorporating the two leading strong-coupling
coefficients, works very well to estimate the next
perturbative term:  It predicts a four-loop 
coefficient of 30.22, which compares nicely with our
result~(\ref{f04ans}).)

We now continue this procedure to higher loops, comparing 
various predictions to our approximate formula.  To obtain 
the ES prediction to higher-loop order, we use the integral 
equation from which eq.~(\ref{gammaKB}) is 
derived~\cite{EdenStaudacher}, 
\be
f(\ah)\ =\ \ah - 4 \, \ah^2 \int_0^\infty dt \, \hat \sigma(t) \,
     {J_1(\sqrt{2 \ah}\, t) \over \sqrt{2 \ah}\, t} \,,
\label{ffromsigma}
\ee
where the fluctuation density $\hat \sigma(t)$ is obtained by 
solving the integral equation,
\be
\hat \sigma(t) = {t \over e^t -1} \Biggl[
    {J_1(\sqrt{2 \ah} \, t)\over  \sqrt{2 \ah}\, t} 
- 2 \, \ah \int_0^\infty dt' \hat K(\sqrt{2 \ah}\, t, \sqrt{2 \ah}\, t') \,
     \hat \sigma(t') \Biggr] \,,
\label{IntegralEquation}
\ee
with the kernel,
\be
\hat K(t, t') = {J_1(t) J_0(t') - J_0(t) J_1(t') \over t - t'} \,,
\ee
where $J_0$ and $J_1$ are standard Bessel functions.

Solving this equation perturbatively through 12 loops, 
we find the numerical values of the ES prediction to be, 
\begin{eqnarray}
f_0^{\rm ES} & = & 
\ah - 1.6449 \, \ah^2 + 5.9528 \,\ah^3 - 26.405 \, \ah^4 
         + 131.22 \, \ah^5 - 705.54 \, \ah^6  \nn \\
 && \null
 +  4021.9\, \ah^7 - 23974. \, \ah^8 
   + 1.4800 \, 10^5  \, \ah^9 - 9.3958 \, 10^5  \, \ah^{10}  \nn \\
 && \null
      + 6.1024 \, 10^6 \, \ah^{11} - 4.0387 \, 10^7 \, \ah^{12} 
+ \cdots \,, \nonumber\\
&&  \hskip1cm \hbox{as $\ah\to 0$.}
\label{fESweaknum}
\end{eqnarray}
The values in \eqn{fESweaknum} may be contrasted to the ones obtained 
from the weak coupling expansion of 
our approximate formula (\ref{fapprox5}) (with $r=-2$), 
\begin{eqnarray}
f_0^{\rm approx} & = & 
\ah - 1.6449 \, \ah^2 + 5.9528 \, \ah^3 - 
 29.295 \, \ah^4 + 165.25 \, \ah^5 - 1002.7 \, \ah^6 \nn \\
 && \null
 + 6379.3 \, \ah^7 - 41997. \, \ah^8 
 + 2.8371\, 10^5 \, \ah^9 - 
 1.9555\, 10^6 \, \ah^{10} \nn \\
 && \null 
+ 1.3699\, 10^7 \, \ah^{11} - 
  9.7237\, 10^7 \, \ah^{12}
+ \cdots \,, \nonumber\\
&&  \hskip1cm \hbox{as $\ah\to 0$.}
\label{ApproximateNumerical}
\end{eqnarray}
The agreement between \eqns{fESweaknum}{ApproximateNumerical} at five
loops and beyond is rather poor.  This disagreement is not surprising, 
given that the ES four-loop value, used as input into our 
approximate formula~(\ref{ApproximateNumerical}), differs significantly 
from our calculation.

We have evaluated further terms in the weak-coupling 
expansion of our approximate formula~(\ref{fapprox5}),
up to 75 loops.  The ratio of the $n^{\rm th}$ 
term to the $(n-1)^{\rm st}$ term in the series slowly settles down to a 
value near $(-8)$ --- at 75 loops it is
$(-7.95)$ --- suggesting a radius of convergence of $1/8$,
and a nearest singularity on the negative real axis at $\ah_c = -1/8$.
This value does appear to agree with the location of the nearest
singularity in the original ES equation~\cite{LipatovPotsdam,BESNew}.

What happens if we generalize the four- and five-loop sign
flips in the ES prediction, and require that all contributions
at a given loop order come in with the same sign? 
Doing so, we obtain,
\begin{eqnarray}
f_0^{\rm modified\ ES}
&= &
\ah - 1.6449 \, \ah^2 + 5.9528 \, \ah^3 - 
 29.295 \, \ah^4 + 165.65 \, \ah^5 - 1007.2 \, \ah^6 \nn \\
 && \null
+  6404.7 \, \ah^7 - 42020. \, \ah^8 
+ 2.8223 \, 10^5 \, \ah^9 - 
  1.9307 \,10^6 \, \ah^{10} 
\nn \\
 && \null
+ 1.3406 \,10^7 \, \ah^{11} - 
  9.4226 \, 10^7 \, \ah^{12} 
+ \cdots \,, \nonumber\\
&&  \hskip1cm \hbox{as $\ah\to 0$.}
\label{ESmodifiedNumerical}
\end{eqnarray}
This modified formula is much closer to our approximate expression
(\ref{ApproximateNumerical}), differing even at 12 loops by only 3
percent.  Of course, we have no reason to expect this naive
modification of signs in the ES formula to be correct to all loop
orders, though it appears to get the numerically largest contributions
correct.  Indeed, if we use this series to systematically construct
KLV approximations with larger values of $n$ in \eqn{KLVapproxn}
as more terms are kept, the large $\ah$
coefficients do not settle to the string values~(\ref{fstringstrong}).
For example, using the weak-coupling series~(\ref{ESmodifiedNumerical})
up to eight loops as input to the KLV approximation~(\ref{KLVapproxn})
with $n=7$ leads to a prediction at strong coupling,
\begin{eqnarray}
\tilde f_0 &\approx& 
0.95968 \, \sqrt{{\ah\over2}}\ -\ 0.081182
\ +\ \Ord(\ah^{-1/2}) \,.
\label{fapprox8strongnum}
\end{eqnarray}
which compares poorly with either the string result
(\ref{fstringstrongnum}) or with the four-loop prediction
(\ref{fapprox3strongnum}).  It is noteworthy that the series
(\ref{ESmodifiedNumerical}) corresponds to the contemporaneous
proposal in ref.~\cite{BESNew}. (We have checked 
agreement through 30 loops.)
In order to improve the agreement of the higher-loop 
KLV approximations with the string theory strong-coupling coefficients,
further modifications to the proposal are needed.
Similar conclusions follow from the investigation of a
sequence of Pad\'e approximants, as discussed below.

The surprisingly good agreement of
\eqns{ApproximateNumerical}{ESmodifiedNumerical} does, however, suggest
that a simple repair of the integral equation is possible.  As a
trivial example, by modifying the kernel in \eqn{IntegralEquation} to
$\hat K(\sqrt{2 \ah}\, t, -\sqrt{2 \ah}\, t')$, we obtain
\begin{eqnarray}
f_0^{\rm modified\ K} 
&= & 
\ah - 1.6449 \, \ah^2 + 5.9528 \,\ah^3
 -  29.295 \, \ah^4 + 165.65 \, \ah^5 - 1011.9 \, \ah^6 \nn \\
 && \null
 +  6490.0 \, \ah^7 - 43050. \, \ah^8 + 2.9271 \, 10^5 \, \ah^9 - 
    2.0282 \, 10^6 \, \ah^{10} \nn \\
 && \null 
 + 1.4265 \, 10^7 \, \ah^{11} - 1.0156 \, 10^8 \, \ah^{12} 
 + \cdots \,, 
\nonumber\\
&&  \hskip1cm \hbox{as $\ah\to 0$.}
\label{KmodifiedNumerical}
\end{eqnarray}
which is in fairly good agreement with our approximate result.
Although this ad hoc modification should only be taken as an
illustration, it does show how one can use our approximate formula
(\ref{ApproximateNumerical}) to guide corrections to the ES integral
equation. (It also suffers from the problem that strong-coupling
extrapolations do not match the string values.)  
A proposed modification of the integral equation, valid
through $\Ord(\ah^2)$ and leading to a modification in the anomalous
dimension at $\Ord(\ah^4)$, is given in eqs.~(89) and (91) of
ref.~\cite{EdenStaudacher}. The choice $\beta = \zeta_3$ corresponds
to $r = -2$ in \eqn{rDef}, as can be seen from eq.~(92) of the same
reference.  This is also the choice preferred by
crossing symmetry in the strong-coupling limit of the dressing factor,
within the family of modifications newly studied in ref.~\cite{BESNew}.

It is useful to investigate other approximation schemes, to make
sure that our results are not significantly biased by
the form of the KLV approximate formula~(\ref{KLVapproxn}).
A well-known method for incorporating information from different
expansion regions --- here from both weak and strong coupling ---
is that of Pad\'e approximants, which fit a function to a ratio 
of polynomials in a suitable variable.  In the complex
$\ah$ plane, 
we expect that $f_0(\ah)$ has poles or branch cuts.  As
discussed earlier, there is evidence from the behavior of the KLV
approximation and the ES kernel that the singularity nearest the origin
is at $\ah_c = -1/8$.  For the purposes of constructing the Pad\'e
approximant, however, we will model the singularities by a branch cut
terminating on the negative real axis at the point $-\xi^2/8$.
This assumption leads us to introduce a variable 
$u \equiv \sqrt{1+8\ah/\xi^2}$, and define the $[(m+1)/m]$ Pad\'e
approximant by
\be
f_0^{[(m+1)/m]}(u) = (u-1) { N_0 + N_1 \, u + \ldots + N_m \, u^m
                   \over 1 + D_1 \, u + \ldots + D_m \, u^m } \,.
\label{Padem}
\ee
The relative degrees of the numerator and denominator polynomials
are fixed by the strong-coupling requirement $f_0 \sim \sqrt{\ah} \sim u$.
The factor of $(u-1)$ comes from the vanishing of $f_0$ at $\ah=0$.
The remaining $2m+1$ constants can be fixed by perturbative data,
or by a mix of weak- and strong-coupling data.
Note that when $m$ increases by 1, two more input numbers are required.

%
\begin{figure}[t]
\centerline{\epsfxsize 6.5 truein \epsfbox{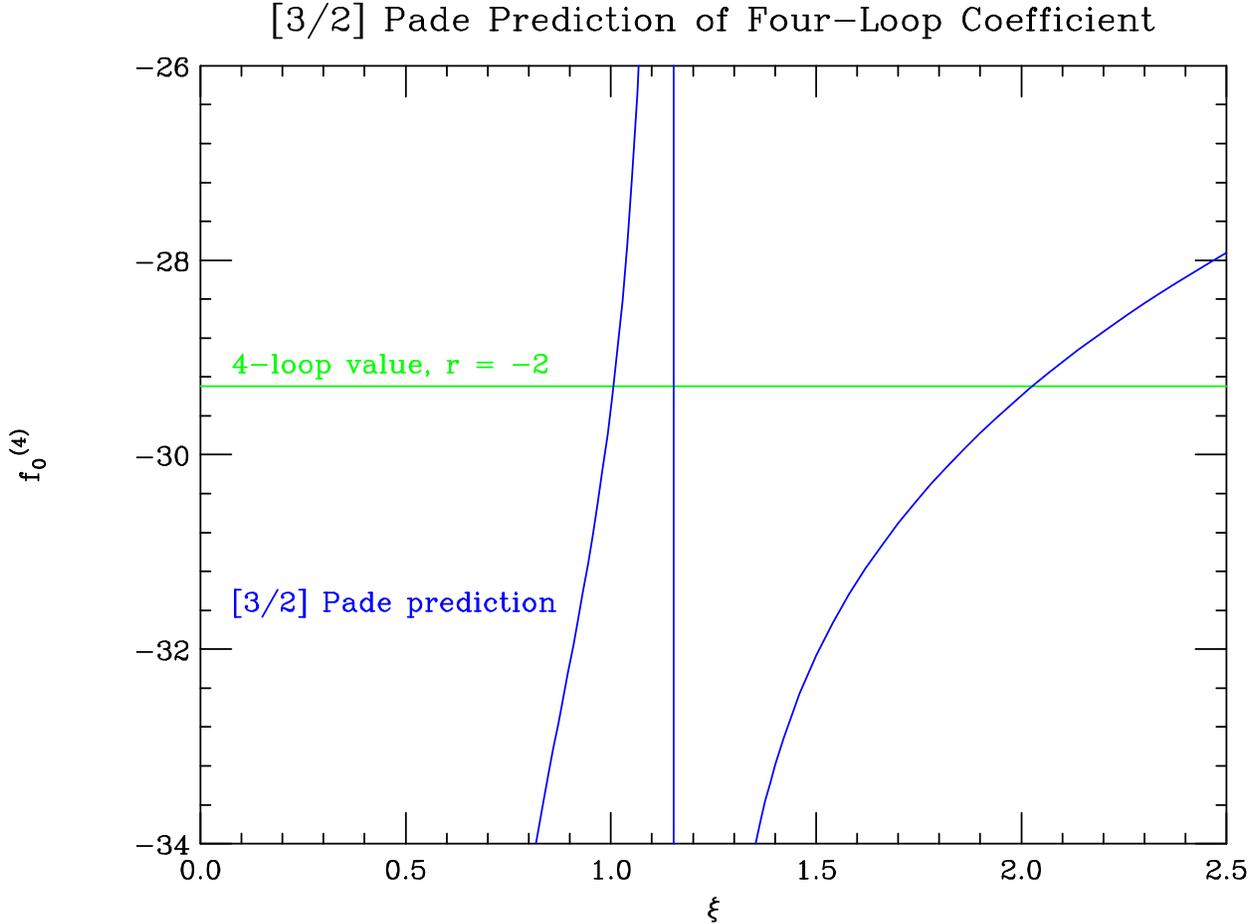}}
\caption{Value estimated for the four-loop cusp anomalous dimension
$f_0^{(4)}$ by the $[3/2]$ Pad\'e approximant~(\ref{Pade32}),
as a function of the parameter $\xi$ controlling the point
at which the cut is assumed to terminate, $\ah = -\xi^2/8$.}
\label{Pade4loopcoeffFigure}
\end{figure}

Consider first the $[3/2]$ Pad\'e approximant, 
\be
f_0^{[3/2]}(u) = (u-1) { N_0 + N_1 \, u + N_2 \, u^2
                   \over 1 + D_1 \, u + D_2 \, u^2 } \,,
\label{Pade32}
\ee 
which is implicitly a function of $\xi$, which parametrizes the 
location of the cut termination on the negative real axis.
We solve for $N_0$, $N_1$, $N_2$, $D_1$ and $D_2$ as a function of $\xi$,
using the two strong-coupling coefficients in \eqn{fstringstrong},
plus the one-, two- and three-loop coefficients.  We then expand
the result in $\ah$ in order to estimate the four-loop coefficient.
This procedure will give us information about $\xi$.  The four-loop 
coefficient is plotted in \fig{Pade4loopcoeffFigure}.  The value $\xi=1$
is picked out rather clearly by this plot, as the (approximate)
first location where the curve crosses the value 
$f_0^{(4)} = -(73\,\pi^6/2520 + \zeta_3^2) = -29.2947$
(for $r=-2$).  At $\xi=1$, formula~(\ref{Pade32}) predicts $-29.521$.
The second crossing is a spurious feature of the Pad\'e,
and the third crossing, near $\xi=2$, is highly implausible, 
based on the strong growth in the perturbative coefficients already at
four loops, not to mention at higher orders in the KLV approximate 
formulas.

\begin{figure}[t]
\centerline{\epsfxsize 6.5 truein \epsfbox{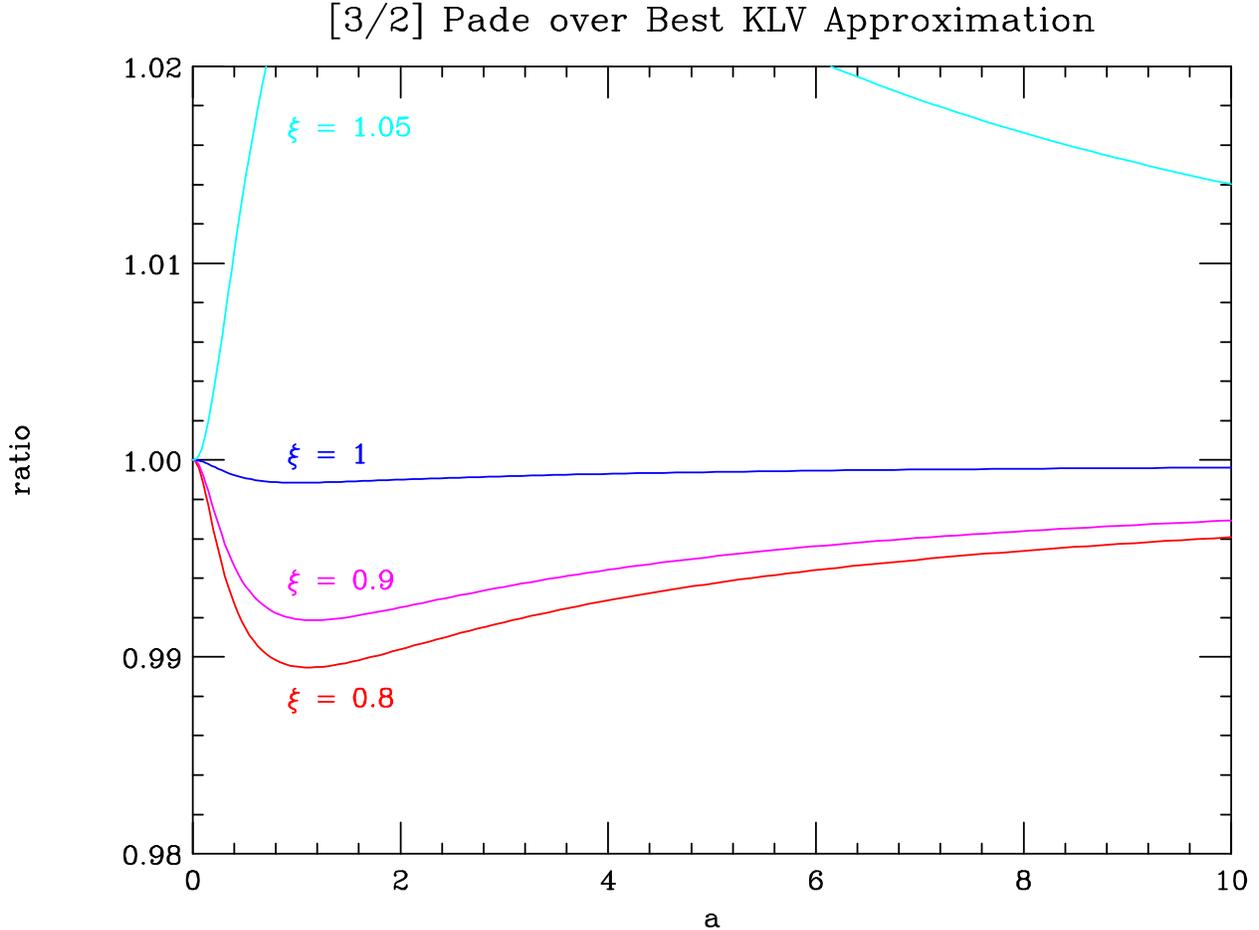}}
\caption{Ratio of the $[3/2]$ Pad\'e approximant~(\ref{Pade32}),
as a function of the parameter $\xi$, to the KLV approximate
formula~(\ref{fapprox5}), as a function of the coupling $\ah$.}
\label{PadeKLVratioFigure}
\end{figure}

In \fig{PadeKLVratioFigure} we plot the ratio of 
the $[3/2]$ Pad\'e to the KLV approximate formula~(\ref{fapprox5}),
for various values of $\xi$ near 1 that produce four-loop
coefficients that are not too far from $-29.2947$.
Except for the pathological case of $\xi=1.05$, all the ratios
are within about 1\% of unity.  For the preferred value of
$\xi=1$, the $[3/2]$ Pad\'e agrees with the KLV approximate formula
everywhere to within about 0.1\%.  We also investigated
the $[4/3]$ Pad\'e, for $\xi=1$.  This approximant requires two 
more input terms, namely the four- and five-loop coefficients.
We used the ``sign-flipped'' values in \eqn{f0modified}.  
The resulting expression predicts the six-loop coefficient to be 
$- 1005.5$, in quite good agreement with the value of $-1002.7$
in \eqn{ApproximateNumerical}.  On other hand, a plot of this
function reveals a singularity on the positive real axis, at $u=4.791$,
which is an artifact of a positive root to the cubic denominator polynomial.
(The quadratic denominator polynomial in the $[3/2]$ Pad\'e
for $\xi=1$ has complex roots, relatively far from the real axis.)
We also constructed a sequence of Pad\'e approximants~(\ref{Padem})
to see how the strong-coupling coefficients are predicted by the 
proposed sign-flipped sequence~(\ref{f0modified}).  As with
the KLV approximation, we do not find any convergence to the 
string values~(\ref{fstringstrong}).  For example, the $[4/3]$ Pad\'e,
based on the first seven loops in \eqn{f0modified}, estimates
\begin{eqnarray}
\tilde f_0 &\approx& 
0.95696 \, \sqrt{{\ah\over2}}\ -\ 0.061164
 \ +\ \Ord(\ah^{-1/2}) \,,
\end{eqnarray}
which is close to the KLV estimation~(\ref{fapprox8strongnum}),
and significantly different from the string values.
 
In summary, the different nature, yet very similar numerical
results from the Pad\'e approximation method gives us confidence
that the KLV approximation~(\ref{fapprox5}) is good to better than 
a percent for all values of the coupling.  It also gives confirming
evidence that the singularity nearest the origin is located
at $\ah_c=-1/8$.

%

\section{Analysis and Conclusions}
\label{AnalysisSection}

Using unitarity, we constructed the planar four-loop four-gluon
amplitude in \SYM, and verified the structure of its infrared
divergences through $1/\e^2$.  At order $1/\e^2$ we were able to
numerically test the prediction of Eden and Staudacher for the
four-loop cusp anomalous dimension.  Our result disagrees with their
predictions.  Using an approximate interpolating formula due to
Kotikov, Lipatov and Velizhanin~\cite{KLV}, with the first four loop
orders as input, we are able to estimate the first two coefficients of the
strong-coupling expansion, coming to within 2.6 and 5 percent of the string
theory prediction.  Improving the approximate formula by using the
string predictions as additional input, we obtain a formula for the
cusp anomalous dimension which we expect to be valid for all
couplings, to better than one percent. We have confirmed this
using Pad\'e approximations.
Not surprisingly, given the
disagreement at four loops, at higher orders the weak-coupling
expansion of our approximate formula~(\ref{fapprox5}) has
large disagreements with the ES prediction~(\ref{fESweaknum}).

There are a number of conceivable reasons for the discrepancy at
four loops, which we shall examine briefly in turn:
\begin{itemize}
\item Different definition of the coupling $g$.
\item Difference between the cusp anomalous dimension defined
for space-like {\it vs.} time-like kinematics.
\item Breakdown of integrability.
\item The wrapping problem.
\item The dressing factor.
\end{itemize}
We expect the latter reasons to be more likely than the former.

In principle our result could differ from the ES result simply because
we are expanding in coupling constants that differ beginning at three
loops, say $\ah^\prime = \ah + x_0 \, \zeta_3^2 \, \ah^4$ for the
appropriate value of $x_0$.  We use the classical coupling as our
expansion parameter, so if this explanation is correct, the asymptotic
Bethe ansatz would have to be using a different coupling.  For
example, in the magnon dispersion relation, discussed in eq.~(2.18) of
ref.~\cite{BeisertDynamic}, a parameter is adjusted to match the
perturbative coupling of \SYM, via the relation 
$\alpha\beta = \ah/2$.  (Here $\alpha$ and $\beta$ 
describe the action of two central charges adjoined to an 
SU$(2|2)$ algebra acting on the spin chain.)
If this simple substitution were to have corrections starting at order $\ah^4$,
it would lead to a four-loop discrepancy between the ES prediction from
integrability and our direct calculation in \SYM.  

Suppose our result and the ES result corresponded to cusp 
anomalous dimensions of different signature, {\it i.e.} one 
space-like and one time-like.  
Could this difference account
for the different values?  According to the results of
Dokhshitzer, Marchesini and Salam~\cite{DMS}, it cannot,
because the leading large-$x$ limits of the space-like and time-like
DGLAP kernels should be identical, to all loop orders.

Integrability of the dilatation operator, interpreted as a spin model
Hamiltonian, has not been proven to hold to all orders.  However,
the fact that there are very similar integrable structures at
very strong gauge coupling --- in the classical sigma 
model~\cite{BPR} and even in its quantum corrections~\cite{Berkovits} ---
 suggests that integrability should persist.

Generically, a wrapping problem can occur if one tries to apply
a Bethe ansatz solution for a spin-interaction that has longer
range than the number of spin sites, which is equal to the twist $J$.
Because the range of the interaction increases with the loop order,
at fixed $J$ this problem becomes more severe with increasing
loop order.   It may well be that the wrapping problem prevents the
asymptotic Bethe ansatz from being applied to short (twist-two) 
operators, even if it can be used for longer ones.  The wrapping
problem is generically supposed to arise at order $\ah^{J-2}$ 
for operators of twist $J$.  It is only other symmetries 
that protect the one-, two- and three-loop cusp anomalous
dimensions from being affected by the wrapping problem.

It is quite possible that the discrepancy is resolved by the so-called
``dressing factor'', an overall phase for the AdS/CFT $S$-matrix,
which is consistent with integrability, the PSU$(2,2|4)$ symmetry,
crossing symmetry, {\it etc.}  The strong-coupling expansion of the
dressing factor is known to be nontrivial.  Indeed, the first two
terms in the semi-classical expansion of the dressing factor on the
string side have been worked
out~\cite{AFS04,OtherDressing,HernandezLopez}.  There
have also been very interesting recent analyses of the properties of
the dressing factor under the worldsheet crossing
symmetry~\cite{JanikCrossing,BHL}.  However, it remains unclear at
which order in the weak-gauge-coupling expansion the dressing factor
becomes non-trivial.  If the reason for the discrepancy is indeed the
dressing factor, then we now know it begins at four loops.

Eden and Staudacher made a specific proposal for modifying
the asymptotic Bethe
ansatz.  They proposed a dressing factor containing a parameter $\beta$, 
thereby modifying the kernel of their integral
equation.  This leads to a shift in the four-loop cusp anomalous 
dimension proportional to $\beta$.   If this proposal 
is correct, we determine the value of this parameter to be close to,
if not exactly, $\beta = \zeta_3$.

Anomalous dimensions of many other classes of operators are linked
through the PSU$(2,2|4)$ symmetry, and are therefore affected by a
nontrivial dressing factor.  As just one example, the anomalous
dimension of the operator ${\cal O} \equiv \Tr(X^2 Z^3)+\ldots$, where
$X$ and $Z$ are two of the three complex scalar fields in \SYM, is
altered by an amount proportional to $\beta$, to have the
form~\cite{EdenStaudacher},
\be
4\ah - 6 \, \ah^2 + 17 \, \ah^3 
- \biggl( {115\over2} + 8 \, \beta \biggr) \ah^4 + \cdots \,.
\label{X2Z3anomdim}
\ee
Thus the Eden--Staudacher proposal for the resolution of the discrepancy 
could be tested by a direct computation of the four-loop anomalous 
dimension of ${\cal O}$.  For $\beta = \zeta_3$,
it would appear to result in a non-uniform transcendentality
for this quantity~\cite{EdenStaudacher}.  (A caveat is that 
one needs to account also for the transcendentality assignment of
harmonic sums~\cite{KLOV}, which are sums of rational numbers.)
As pointed out by Eden and Staudacher, a nonzero value of $\beta$, 
through its effect on the anomalous dimension of ${\cal O}$, 
apparently rules out the Hubbard model Hamiltonian as a candidate 
for the $SU(2)$ dilatation operator beyond three loops~\cite{RSS}.

The Eden--Staudacher modification is the lowest-order term of a
general form for the dressing factor $S_{12}=\exp(2i\theta(x_1,x_2))$,
with
\begin{equation}
\theta(x_k,x_j) = \sum_{r=2}^\infty \sum_{v=r+1}^\infty
c_{r,v} \bigl( q_r(x_k) q_{v}(x_j)-q_{v}(x_k) q_{r}(x_j)\bigr)\,,
\label{GeneralDressingPhase}
\end{equation}
suggested in the literature~\cite{AFS04,OtherDressing,BeisertKlose,
BeisertInvariant,BeisertDynamic,BeisertPhase,BHL}.
In this equation, the $q_r$ are the spin-chain charges, 
\begin{equation}
q_r(x_k) = {i\over r-1}
  \biggl( {1\over (x^+_k)^{r-1}}- {1\over (x^-_k)^{r-1}}\biggr)\,,
\end{equation}
and the $x^\pm_i$ are rapidities entering
into the Bethe ansatz by which the spin-chain has been solved at
lower orders.

All known quantities, anomalous dimensions or amplitudes, in
the $\NeqFour$ supersymmetric theory, have a uniform transcendentality.
That is, all polylogarithms or zeta constants at a given order in the
perturbative expansion have the same polylog weight or transcendentality.
The ES proposal deviates from this observed property
in some quantities, such as \eqn{X2Z3anomdim}.

Is the dressing phase in \eqn{GeneralDressingPhase} general
enough to accommodate our expectations for the $\NeqFour$ theory?
As noted by Eden and Staudacher, maintaining uniform transcendentality
with purely rational coefficients
requires every power of $t$ or $t'$ in their integral equation to
come along with a power of the coupling $g$. Now, the coefficient
of the $\zeta_3^2$ terms, which we wish to modify, arises from terms
in the kernel proportional to $t t'$.  This term is {\it odd\/} in 
$t'$.  Such a term cannot arise from Fourier transforming a lone charge
$q$, because we are interested in symmetric densities,
and hence will ultimately take the symmetric part of this transform.
The symmetric part is either zero (for $q_r$ with $r$ odd) or a symmetric
function of $t'$.  In \eqn{GeneralDressingPhase}, every term is
linear in a charge $q(x_j)$, and so it cannot lead to the modifications
we need.  We must therefore seek a more general dressing
factor, or else introduce non-rational coefficients.

Very interestingly, in a contemporaneous
paper, Beisert, Eden and Staudacher~\cite{BESNew} have 
done just the latter, in such a way that, at
least for the anomalous dimension under consideration, 
uniform transcendentality is maintained.  
This leads to a modified integral equation of the
ES type for the cusp anomalous dimension. 
At four loops the resulting anomalous dimension is in
complete agreement with our direct computation of this quantity.
Their proposal is compatible with integrability and with the KLOV
transcendentality principle for the cusp anomalous dimension, 
and violates perturbative Berenstein-Maldacena-Nastase (BMN)
scaling~\cite{BMN} starting at four loops.  
At five loops their proposal also matches our
prediction of this coefficient, given in \eqn{f0modified}, based on
using both KLV~\cite{KLV} and Pad\'e approximations.  At higher-loop
orders, their proposal corresponds to \eqn{ESmodifiedNumerical}
and appears to properly incorporate the numerically largest contributions.
That is, it matches reasonably well our approximate expression
(\ref{ApproximateNumerical}).  
However, we find that successive KLV and Pade approximations, based on
truncations at increasingly higher orders through 13 and 11 loops
respectively, do not match the strong-coupling string results.  This
indicates a tension between the weak- and strong-coupling results, and
suggests to us that further modifications may be necessary.  The question
merits further study.

We remark that, assuming the KLOV conversion principle, our
result for the leading-color four-loop cusp anomalous dimension
also predicts a piece of the corresponding result in QCD.
Which piece?  At three loops and below, the QCD result is
a polynomial in the $SU(N_c)$ Casimir operators $C_A$ and $C_F$,
while the \SYM\ result is composed solely of $C_A = \Nc$,
so it has no subleading-color terms.
The \SYM\ result provides one constraint on the leading-transcendentality
parts of the coefficients of the color factors in QCD, such that, 
after the group theory Casimirs
have been set to the values $C_F = C_A = \Nc$, the leading-color terms 
are equal to the \SYM\ result.  Starting at four loops, however, 
there are color factors that cannot be reduced to polynomials 
in $C_A$ and $C_F$. 
The relevant color factors are those of $L$-loop propagator diagrams.

In \SYM, any triangle subdiagram leads to a group-theory factor of $C_A$,
times a lower-loop group-theory factor.  So the question is, when
do no-triangle propagator diagrams first appear?  At three loops, 
there is one nonplanar no-triangle propagator diagram, but its color factor
vanishes identically using the Jacobi identity 
(see {\it e.g.} ref.~\cite{TwoLoopSplitting}).
At four loops, \fig{NoTriangleFigure}(d${}_5$) illustrates 
the unique such planar graph (when the two pairs of external lines 
on the left and right sides of the graph are each replaced with single lines).
There are a number of nonplanar graphs as well.
Hence the cusp anomalous dimension in \SYM\ can now have subleading-color 
terms.  Presumably the KLOV principle will apply with respect
to the leading-color, planar terms, once the fermionic color factors in
the QCD result are shifted from the fundamental to adjoint representation.
But will it also apply to subleading-color terms?  If conformal invariance
is the main issue~\cite{KLOV}, then it should.  But if planarity is
important, perhaps it will not.  Of course, the question is a moot one
until the cusp anomalous dimension in QCD, and the subleading-color part
of that in \SYM, are both known at four loops.  Similar remarks apply
to the four-loop form-factor quantity ${\cal G}_0^{(4)}$.

In order to confirm that the four-loop cusp anomalous dimension is
given exactly by the $\Ord(\ah^4)$ term in \eqn{f0modified}, it would
be important to evaluate it analytically.  To do so, the four-loop
integrand~(\ref{FourLoopPlanarResult}) would have to be evaluated
analytically through $\Ord(\ep^{-2})$ instead of only through
$\Ord(\ep^{-4})$ as done here.  It would also be extremely
interesting to evaluate the four-loop integrand through
$\Ord(\ep^{0})$, to explicitly check the four-loop iteration relation
(\ref{FourLoopFourPtIteration}) for scattering
amplitudes~\cite{Iterate2,Iterate3}.

The program {\tt MB}~\cite{CzakonMB} can be used to express the coefficients
at each order in the Laurent expansion in $\e$ as a sum of finite
contour integrals.  However, the number of such integrals increases 
rapidly with decreasing inverse power of $\e$.
This property, along with an increase in dimensionality of the 
integrals, makes it harder to compute the integrals purely numerically, 
requiring a substantial increase in computational resources for 
the same relative level of accuracy.  For example, the order $\e^{-2}$ 
term in $M_4^\fourloop$ in table~\ref{FourLoopTable} has a relative 
precision of $2\times 10^{-4}$, 40 times larger than that of the
order $\e^{-3}$ term, despite having had significantly more computing
resources applied to it.
Although a brute-force computation of the $1/\e$ and $\e^0$ terms
could probably be completed with sufficient resources, it would be
considerably lengthier than the present computation.  Accordingly, 
we believe that additional analytic work would be rather desirable
before proceeding further.

The information provided in this paper offers a guide to further
progress in determining the integrable structure of planar \SYM, as
well as in studying the transition from weak to strong coupling.  For
the cusp anomalous dimension, the striking match between the first two
coefficients of the strong-coupling expansion, as estimated by us,
using our four-loop result as input to the KLV
approximation~\cite{KLV}, and as obtained from string
theory~\cite{StrongCouplingLeadingGKP,Kruczenski,Makeenko,
StrongCouplingSubleading}, provides good evidence that we have an
excellent  numerical
understanding of this anomalous dimension at any coupling.
The same approximation strongly suggests that the
correct analytic forms of the four- and five-loop perturbative
coefficients are the ones given in \eqn{f0modified}.  These
types of approximations can also be useful for checking whether a given
ansatz for higher-order terms in the weak-coupling expansion are consistent
with the known string theory strong-coupling results.
This should help in finding the correct integral equation describing the
\SYM\ cusp anomalous dimension.  Another intriguing result from the
extrapolation to strong coupling is that the next term in the
expansion ($\Ord(\ah^{-1/2})$) should be very small and may even
vanish.  Finally, the remarkably good numerical properties of the
approximate formulas indicate that the transition between weak and
strong coupling is smooth for planar $\NeqFour$ super-Yang-Mills
theory.


\section*{Acknowledgments}

We are grateful to Niklas Beisert, John Joseph Carrasco, Alexander
Gorsky, Henrik Johansson, Igor Klebanov, Radu Roiban, Emery Sokatchev,
Marcus Spradlin,
Matthias Staudacher and Marvin Weinstein for very stimulating
discussions. We also thank Thomas Hahn for communications regarding
{\tt CUBA}. We are particularly indebted to Matthias Staudacher for
repeated encouragement in the course of this work. We thank
Niklas Beisert, Burkhard Eden and Matthias Staudacher for sending us
an advance copy of their new paper~\cite{BESNew} and for discussions.
We also thank Radu Roiban for his useful comments on the manuscript.
We thank Academic Technology Services at UCLA for computer support.
This research was supported by the US Department of Energy under
contracts DE--FG03--91ER40662 and DE--AC02--76SF00515.  MC was
supported by the Sofja Kovalevskaja Award of the Alexander von
Humboldt Foundation sponsored by the German Federal Ministry of
Education and Research.  The work of VAS was supported by the Russian
Foundation for Basic Research through project 05-02-17645.  DAK and
VAS also acknowledge the support of the ECO-NET program of the {\sc
Egide} under grant 12516NC.  The figures were generated using
Jaxodraw~\cite{Jaxo}, based on Axodraw~\cite{Axo}.

\appendix

\section{Evaluating four-loop integrals by MB representation}
\label{MB4loops}

To obtain Laurent expansions in $\ep$ for our integrals we use the
Mellin--Barnes (MB) technique, which has been successfully applied in
numerous calculations (see, e.g.,
refs.~\cite{SmirnovDoubleBox,Tausk,SmirnovTripleBox,%
Iterate3,TwoloopOffandMassive} and chapter~4 of ref.~\cite{Buch}).  It
relies on the identity
\be
\frac{1}{(X+Y)^{\lm}} = \int_{\beta - i  \infty}^{\beta + i
\infty} \frac{Y^z}{X^{\lm+z}} \frac{\Gm(\lm+z) \Gm(-z)}{\Gm(\lm)}
\frac{\dd z}{2\pi  i} \,,
\ee
where $-{\rm Re}\, \lm <\beta <0$.  This identity basically 
replaces a sum over terms raised to some power with a product of factors.

It is convenient to keep arbitrary powers of propagators and numerator
factors in each of the MB representations.  In this way, one can
check the MB representations by setting some of the indices to zero,
in order to obtain a simpler integral whose value is already known.  
Also, we can obtain the values of the two non-rung-rule integrals 
of \fig{nonrrFigure} from two of the rung-rule integrals 
of \fig{rrFigure} by setting some of the indices to zero.

We define a four-loop integral with general indices as, 
\begin{eqnarray}
F^{(x)} (a_1, \ldots, a_{n_i}; s, t; \e) 
&=& (-i e^{\e \gamma} \pi^{-d/2})^4
 \int {\dd^d p\,\dd^d r\,\dd^d u\,\dd^d v \over 
\prod_{j=1}^{n_i} (p_j^2)^{a_j}}\,,
\label{QuadrupleLadderGenIndices}
\end{eqnarray}
where $p,r,u,v$ are the four independent loop, $n_i$ the number of
indices corresponding to the $p_j^2$ propagator and numerator factors
in a given graph.  For each graph $(x)$ the labels correspond to the
propagator and numerator labels of~\fig{rrFigure}.  In this notation a
numerator factor is indicated by negative indices.

Experience shows that a minimal number of MB integrations for planar
diagrams is achieved if one introduces MB integrations loop by loop,
{\it i.e.} one derives a MB representation for a one-loop subintegral, inserts
it into a higher two-loop integral, etc.  This straightforward
strategy provides the following MB representations for the Feynman
integrals corresponding to graphs of \fig{rrFigure}, with general
powers of the propagators and irreducible numerators:
\bea
F^{\rm (a)}(a_1,\ldots,a_{13};s,t;\ep)=
\frac{ e^{4\ep\gamma}\, (-1)^a(-s)^{8-a-4\ep}}{
\prod_{j=2,5,7,9,11,12,13} \Gm(a_j)
\Gm(4 - a_{9,11,12,13}  - 2 \ep)}
&& \nn \\ && \hspace*{-120mm}
\times \frac{1}{(2\pi i)^{11}}
\int_{-i\infty}^{+i\infty}\ldots 
\int_{-i\infty}^{+i\infty}  \prod_{j=1}^{11} \dd z_j
\left(\frac{t}{s}\right)^{z_7}
\frac{
      \Gm(2 - a_{9,12,13} - \ep - z_{1,2})
      \Gm(2 - a_{9,11,12} - \ep - z_{1,3})}
{\Gm(a_{10} - z_2) \Gm(a_8 - z_3)
\Gm(a_6 - z_5) \Gm(a_4 - z_6)}
\nn \\ && \hspace*{-120mm} \times
\frac{      \Gm(a_9 + z_{1,2,3})
\Gm(a_{9,11,12,13}-2 + \ep + z_{1,2,3})
\Gm(z_{10} - z_4) \Gm(z_4-z_1)
}
{\Gm(4 - a_{5,8,10} - 2 \ep + z_{1,2,3})
\Gm(4 - a_{4,6,7} - 2 \ep + z_{4,5,6})
\Gm(4 - a_{1,2,3} - 2 \ep  + z_{8,9,10})}
\nn \\ && \hspace*{-120mm} \times
\frac{\Gm(2 - a_5 - a_8 - \ep + z_1 + z_3 - z_4 - z_6)
\Gm(a_{5,9,10} -2  + \ep - z_{1,2,3} + z_{4,5,6})}{\Gm(a_3 - z_9)\Gm(a_1 - z_8)}
\nn \\ &&  \hspace*{-120mm} \times
\Gm(a_{12} + z_1)\Gm(a_2 + z_7) \Gm(z_7-z_{10})
\Gm(2 - a_{6,7} - \ep - z_{8,10} + z_{4,5})
\Gm(2 - a_{1,2} - \ep + z_{8,10} - z_7)
\nn \\ &&  \hspace*{-120mm} \times
\Gm(2 - a_{4,7} - \ep - z_{9,10} + z_{4,6})
\Gm(a_{1,2,3}-2 + \ep  + z_7 - z_{8,9,10})
\Gm(2 - a_{2,3} - \ep + z_{9,10} - z_7)
\nn \\ &&  \hspace*{-120mm} \times
\Gm(a_7  + z_{8,9,10})\Gm(2 - a_{5,10} - \ep + z_{1,2} - z_{4,5})
\nn \\ &&  \hspace*{-120mm} \times
\Gm(a_5 + z_4 + z_5 + z_6)
\Gm(a_{4,6,7}-2 + \ep  - z_{4,5,6} + z_{8,9,10})
\prod_{j=2,3,5,6,7,8,9} \Gm(-z_j)
\;,
\label{MB-Fa}
\eea
\bea
F^{\rm (b)}(a_1,\ldots,a_{14};s,t;\ep)=
\frac{ e^{4\ep\gamma}\, (-1)^a(-s)^{8-a-4\ep}}{
\prod_{j=2,4,5,6,7,9,12} \Gm(a_j) \Gm(4 - a_{4,5,6,7} - 2 \ep)}
&& \nn \\ && \hspace*{-112mm}
\times \frac{1}{(2\pi i)^{12}}
\int_{-i\infty}^{+i\infty}\ldots \int_{-i\infty}^{+i\infty}
 \prod_{j=1}^{12} \dd z_j
\left(\frac{t}{s}\right)^{z_{9,12}}
\frac{
\Gm(8-a+a_{12}- 4 \ep - z_{9,12})
}{
\Gm(10-a - 5 \ep - z_9)
\Gm(a_{13} - z_{10}) \Gm(a_1 - z_2)
}
\nn \\ && \hspace*{-112mm} \times
\frac{
\Gm(a_9 + z_{9,10})
\Gm(a_{13} - z_{10} + z_{12})
\Gm(2 - a_{5,6,7} - \ep - z_{1,2,4})
\Gm(2 - a_{4,5,7} - \ep - z_{1,3})
}{
\Gm(4 - a_{1,2,3} - 2 \ep + z_{1,2,3})
\Gm(a_{14}  - z_{4,8,11})
\Gm(a_8 - z_6) \Gm(a_3 - z_3)
}
\nn \\ && \hspace*{-112mm} \times
\frac{
\Gm(a_7 + z_{1,2,3})
\Gm(a_{4,5,6,7}-2 + \ep + z_{1,2,3,4})
\Gm(z_{9,11} - z_5 )
\Gm(2 - a_{12,13,14} - \ep + z_{4,8,10,11} - z_{12})
}{\Gm(a_{10} - z_7)\Gm(4 - a_{8,9,10}  - 2 \ep + z_{5,6,7})}
\nn \\ && \hspace*{-112mm} \times
\frac{\Gm(a_5 + z_{1,4})
\Gm(2 - a_{2,3} - \ep + z_{1,3} - z_{5,7})
\Gm(a_2 + z_{5,6,7})
\Gm(a_{14}- z_{4,8,11} + z_{12})}{
\Gm(a_{1,2,3,4,5,6,7,8,9,10,11}-6+ 3 \ep + z_{4,8,9,10,11})}
\nn \\ && \hspace*{-112mm} \times
\Gm(2 - a_{1,2} - \ep + z_{1,2} - z_{5,6,8})
\Gm(z_{5,8}-z_1)
\Gm(a_{1,2,3}-2 + \ep - z_{1,2,3} + z_{5,6,7,8})
\nn \\ && \hspace*{-112mm} \times
\Gm(2 - a_{8,9} - \ep+ z_{5,6} - z_{9,10,11})
\Gm(2 - a_{9,10} - \ep + z_{5,7} - z_9)
\Gm(a-8 + 4 \ep + z_{9,12})
\nn \\ && \hspace*{-112mm} \times
\Gm(a_{8,9,10}-2+ \ep- z_{5,6,7}+ z_{9,10,11})
\prod_{j=2,3,4,6,7,8,9,10,11,12} \Gm(-z_j)
\;,
\label{MB-Fb}
\eea
\bea
F^{\rm (c)}(a_1,\ldots,a_{14};s,t;\ep)=
\frac{ e^{4\ep\gamma}\, (-1)^a(-t)^{8-a-4\ep}}{
\prod_{j=2,4,5,6,7,8,12} \Gm(a_j)
\Gm(4 - a_{4,5,6,7} - 2 \ep)}
&& \nn \\ && \hspace*{-112mm}
\times \frac{1}{(2\pi i)^{11}}
\int_{-i\infty}^{+i\infty}\ldots \int_{-i\infty}^{+i\infty}  \prod_{j=1}^{11} \dd z_j
\left(\frac{s}{t}\right)^{z_{5,8,11}}
\frac{\Gm(a_{14} - z_{10} + z_{11})
\Gm(2 - a_{4,5,7} - \ep - z_{1,3})
 }
{ \Gm(a_{14} - z_{10})
\Gm(a_1 - z_2)
\Gm(4 - a_{1,2,3} - 2 \ep + z_{1,2,3})}
\nn \\ && \hspace*{-112mm} \times
\frac{\Gm(2 - a_{5,6,7} - \ep - z_{1,2,4}) \Gm(a_5 + z_{1,4})
\Gm(a_{4,5,6,7}-2 + \ep + z_{1,2,3,4})
\Gm(2 - a_{2,3} - \ep + z_{1,3} - z_5)
 }
{\Gm(8 - a_{1,2,3,4,5,6,7,8,9,10,11}- 4 \ep - z_5)
\Gm(a_9 - z_6) \Gm(a_{10} - z_{4,7})\Gm(a_3 - z_3) }
\nn \\ && \hspace*{-112mm} \times
\frac{\Gm(a_2 + z_{5,6}) \Gm(2 - a_{1,2} - \ep + z_{1,2} - z_{5,6,7})
\Gm(z_{5,7}-z_1)
\Gm(a_{1,2,3}-2 + \ep - z_{1,2,3} + z_{5,6,7}) }
{ \Gm(a_{1,2,3,4,5,6,7,11}-4+ 2 \ep + z_{4,5,6,7})
\Gm(10 - a- 5 \ep - z_{5,8}) \Gm(a_{13} - z_9)}
\nn \\ &&  \hspace*{-112mm} \times
\Gm(6 - a_{1,2,3,4,5,6,7,9,10,11} - 3 \ep - z_{5,8})
\Gm(8 - a + a_{12} - 4 \ep  - z_{5,8,11})
\Gm(a -8+ 4 \ep + z_{5,8,11})
\nn \\ &&  \hspace*{-112mm} \times
\Gm(a_{10}+ z_{8,10} - z_{4,7}) \Gm(a_{13} + z_{11} - z_9)
\Gm(2  - a_{8,9,10} - \ep  + z_{4,6,7} - z_{8,9,10})\Gm(a_7 + z_{1,2,3})
\nn \\ &&  \hspace*{-112mm} \times
\Gm(2 - a_{12,13,14} - \ep + z_{9,10} - z_{11})
     \Gm(a_9 - z_6 + z_{8,9})
\prod_{j=2}^{11} \Gm(-z_j)
\;,
\label{MB-Fc}
\eea
\bea
F^{\rm (d)}(a_1,\ldots,a_{14};s,t;\ep)=
\frac{  e^{4\ep\gamma}\, (-1)^a(-s)^{8-a-4\ep}}{
\prod_{j=2,4,5,6,7,9,12} \Gm(a_j)
\Gm(4 - a_{4,5,6,7} - 2 \ep)}
&& \nn \\ && \hspace*{-112mm}
\times \frac{1}{(2\pi i)^{14}}
\int_{-i\infty}^{+i\infty}\ldots \int_{-i\infty}^{+i\infty}  
\prod_{j=1}^{14} \dd z_j
\left(\frac{t}{s}\right)^{z_{14}}
\frac{\Gm(a_{12} + z_{11})\Gm(a_9 + z_{14}) \Gm(z_{14}-z_{11})
 }
{\Gm(a_1 - z_2)
\Gm(4 - a_{1,2,3} - 2 \ep + z_{1,2,3})
\Gm(a_{13} - z_{5,10})}
\nn \\ && \hspace*{-112mm} \times
\frac{ \Gm(2 - a_{5,6,7} - \ep - z_{1,2,4})
\Gm(2 - a_{4,5,7} - \ep - z_{1,3,5})  \Gm(a_5 + z_{1,4,5})
\Gm(a_{4,5,6,7}-2 + \ep + z_{1,2,3,4,5}) }
{  \Gm(a_{11} - z_8)
\Gm(a_8 - z_{4,9,12})
\Gm(8 - a_{1,2,3,4,5,6,7,11,12,13,14}  - 4 \ep - z_{4,6,7,9})
 \Gm(a_{10} - z_{7,13}) }
\nn \\ && \hspace*{-112mm} \times
\frac{ \Gm(2 - a_{9,10} - \ep + z_{7,11,13} -z_{14})
\Gm(2 - a_{2,3} - \ep + z_{1,3} - z_{6,8,10})
\Gm(a_2 + z_{6,7,8})\Gm(a_7 + z_{1,2,3})}
{\Gm(4  - a_{8,9,10}- 2 \ep + z_{4,7,9,11,12,13} )
\Gm( a_{1,2,3,4,5,6,7,14} -4 + 2 \ep  + z_{4,5,6,7,8,9,10})\Gm(a_3 - z_3) }
\nn \\ &&  \hspace*{-112mm} \times
\Gm(a_{8,9,10} -2 + \ep - z_{4,7,9,11,12,13} + z_{14} )
\Gm(2 - a_{1,2} - \ep + z_{1,2} - z_{6,7,9})
\nn \\ &&  \hspace*{-112mm} \times
\Gm(6 - a_{1,2,3,4,5,6,7,11,12,14} - 3 \ep  - z_{4,5,6,7,9,10,11,13})
\Gm(6 - a_{1,2,3,4,5,6,7,12,13,14} - 3 \ep - z_{4,6,7,8,9,11,12})
\nn \\ &&  \hspace*{-112mm} \times
\Gm(2 - a_{8,9} - \ep + z_{4,9,11,12} - z_{14} )
\Gm(a_{1,2,3,4,5,6,7,11,12,13,14} -6 + 3 \ep+ z_{4,6,7,9,11,12,13})
\nn \\ &&  \hspace*{-112mm} \times
\Gm(z_{6,9,10}-z_1)
\Gm(a_{1,2,3}-2 + \ep - z_{1,2,3} + z_{6,7,8,9,10})
\nn \\ &&  \hspace*{-112mm} \times
\Gm(a_{1,2,3,4,5,6,7,14}-4 + 2 \ep
 + z_{4,5,6,7,8,9,10,11,12,13})
\prod_{j=2,3,4,5,6,7,8,9,10,12,13,14} \Gm(-z_j)
\;,
\label{MB-Fd}
\eea
\bea
F^{\rm (e)}(a_1,\ldots,a_{15};s,t;\ep)=
\frac{ e^{4\ep\gamma}\, (-1)^a(-s)^{8-a-4\ep}}{
\prod_{j=2,4,5,6,7,9,11} \Gm(a_j)
\Gm(4 - a_{4,5,6,7} - 2 \ep))}
&& \nn \\ && \hspace*{-112mm}
\times \frac{1}{(2\pi i)^{12}}
\int_{-i\infty}^{+i\infty}\ldots \int_{-i\infty}^{+i\infty}  
\prod_{j=1}^{12} \dd z_j
\left(\frac{t}{s}\right)^{z_{12}}
\frac{\Gm(a_9 + z_{12}) \Gm(2 - a_{5,6,7} - \ep - z_{1,2})
\Gm(a_7 + z_{1,2,3})
 }
{  \Gm(a_{10} - z_{11}) \Gm(a_1 - z_2)
 \Gm(4 - a_{1,2,3} - 2 \ep + z_{1,2,3})}
\nn \\ && \hspace*{-112mm} \times
\frac{\Gm(2 - a_{4,5,7} - \ep - z_{1,3,4})\Gm(a_{4,5,6,7} -2+ \ep + z_{1,2,3,4})
\Gm(2 - a_{1,2} - \ep + z_{1,2} - z_{5,6})
 }
{ \Gm(a_8 - z_{6,10} ) \Gm(a_{13} - z_7)\Gm(a_3 - z_3)
\Gm(8-a_{1,2,3,4,5,6,7,11,12,13,14} - 4 \ep - z_{5,6})
\Gm(a_{12} - z_{4,8})}
\nn \\ && \hspace*{-112mm} \times
\frac{\Gm(a_2 + z_{5,6,7})\Gm(2 - a_{2,3} - \ep + z_{1,3} - z_{5,7,8})
\Gm(z_{5,8}-z_1) \Gm(a_{1,2,3}-2 + \ep - z_{1,2,3} + z_{5,6,7,8}) }
{ \Gm(a_{1,2,3,4,5,6,7,14} -4  + 2 \ep + z_{4,5,6,7,8}) \Gm(a_{15} - z_{5,9})
\Gm(4 - a_{8,9,10,15}  - 2 \ep + z_{5,6,9,10,11})}
\nn \\ &&  \hspace*{-112mm} \times
\Gm(6 - a_{1,2,3,4,5,6,7,12,13,14}  - 3 \ep  - z_{5,6,9,10})
\Gm(a_{8,9,10,15} -2 + \ep - z_{5,6,9,10,11} + z_{12} )
\nn \\ &&  \hspace*{-112mm} \times
\Gm(a_{15} + z_{12} - z_{5,9})\Gm(2 - a_{11,12,13} - \ep + z_{4,7,8} - z_{9,11})
\Gm(2 - a_{9,10,15}  - \ep + z_{5,9,11} - z_{12} )
\nn \\ &&  \hspace*{-112mm} \times
\Gm(a_{1,2,3,4,5,6,7,11,12,13,14} -6 + 3 \ep + z_{5,6,9,10,11})
\Gm(2 - a_{8,9,15} - \ep + z_{5,6,9,10} - z_{12} )
\nn \\ &&  \hspace*{-112mm} \times
\Gm(a_5 + z_{1,4})
\Gm(a_{13}- z_7 + z_{9,10,11} ) \Gm(a_{12} - z_{4,8} + z_9)
\prod_{j=2}^{12} \Gm(-z_j)
\;,
\label{MB-Fe}
\eea
\bea
F^{\rm (f)}(a_1,\ldots,a_{15};s,t;\ep)=
\frac{ e^{4\ep\gamma}\, (-1)^a(-s)^{8-a-4\ep}}{
\prod_{j=1}^8 \Gm(a_j)
\Gm(4 - a_{1,2,3,4} - 2 \ep) \Gm(4 - a_{5,6,7,8} - 2 \ep)
}
&& \nn \\ && \hspace*{-128mm}
\times \frac{1}{(2\pi i)^{14}}
\int_{-i\infty}^{+i\infty}\ldots \int_{-i\infty}^{+i\infty}
\prod_{j=1}^{14} \dd z_j
\left(\frac{t}{s}\right)^{z_{5,6}}
\frac{\Gm(2 - a_{6,7,8} - \ep - z_{11,13})\Gm(a_8 + z_{11,14})
 }
{\Gm(a_{13} - z_{10,12}) \Gm(a_{11} - z_{14})
\Gm(a_{10} - z_{3,13}) }
\nn \\ && \hspace*{-128mm} \times
\frac{ \Gm(2 - a_{5,6,8} - \ep - z_{11,12,14}) \Gm(a_6 + z_{11,12,13})
\Gm(a_{13} + z_{1,2,3,4} - z_{10,12})
\Gm(a_{11} + z_{4,5}- z_{14} )
}
{\Gm(a_{14} - z_7) \Gm(a_9 - z_{2,8}) \Gm(a_{12} - z_9)
\Gm(6 - a_{1,2,3,4,9,10,15} - 3 \ep
+ z_{1,2,3,4,11,13}  - z_{7,9,10}) }
\nn \\ && \hspace*{-128mm} \times
\frac{\Gm(a_{14} + z_{2,5} - z_7)
\Gm(4 - a_{5,6,7,8,11,13,14}  - 2 \ep - z_{2,3,4,5,11,13} + z_{7,10})
\Gm(a_9 - z_{2,8} + z_6) }
{ \Gm(a_{15} - z_{4,11})
\Gm(6 - a_{5,6,7,8,11,12,13,14}  - 3 \ep + z_{7,9,10} - z_{11,13} )
 }
\nn \\ &&  \hspace*{-128mm} \times
\frac{\Gm(a_{15} - z_{4,11}  + z_6)\Gm(2 - a_{2,3,4} - \ep  - z_{7,8,10})
\Gm(2- a_{9,10,15}  - \ep + z_{2,3,4,8,11,13}  - z_6)
}{\Gm(a_{1,2,3,4}-2 + \ep - z_1 + z_{7,8,9,10})}
\nn \\ &&  \hspace*{-128mm} \times
 \Gm(2 - a_{1,2,4} - \ep - z_{7,9})
\Gm(a_{5,6,7,8,11,12,13,14} -4 + 2 \ep + z_{1,2,3,4,5,11,13}  - z_{7,9,10})
\nn \\ &&  \hspace*{-128mm} \times
\Gm(4 - a_{1,2,3,4,9,15} - 2 \ep + z_{1,2,4,11} - z_{6,7,9,10})
\Gm(a_4 + z_{7,9,10})\Gm(a_2 + z_{7,8})
\nn \\ &&  \hspace*{-128mm} \times
\Gm(2 - a_{11,12,13,14} - \ep - z_{1,2,4,5}  + z_{7,9,10,12,14})
\Gm(a_{1,2,3,4}-2 + \ep  + z_{7,8,9,10})
\nn \\ &&  \hspace*{-128mm} \times
\Gm(a_{1,2,3,4,9,10,15}-4+ 2 \ep - z_{1,2,3,4,11,13}  + z_{6,7,9,10})
\prod_{j=1}^{14} \Gm(-z_j)
\;,
\label{MB-Ff}
\eea
where
$a_{4,5,6,7}=a_4+a_5+a_6+a_7$,\ $a=\sum a_i$,\ 
$z_{11,12,13}=z_{11}+z_{12}+z_{13}$, etc.

From these general integrals, by choosing appropriate indices,
we may then obtain the eight integrals that appear in the four-loop
amplitude, 
\begin{eqnarray}
\I^{\rm (a)} &= & s^4t \, F^{\rm (a)}(1, 1, 1, 1, 1, 1, 1, 1, 1, 1, 1, 1, 1; 
                                            s, t; \ep)\,, \label{MBIntegralA} \\
\I^{\rm (b)} &= & st^2 \, F^{\rm (b)}(1, 1, 1, 1, 1, 1, 1, 1, 1, 1,
                              -2, 1, 1, 1;  s, t; \ep)\,, \label{MBIntegralB} \\
\I^{\rm (c)} &= & s^3 t \, F^{\rm (c)}(1, 1, 1, 1, 1, 1, 1, 1, 1, 1,
                               -1, 1, 1, 1; s, t; \ep)\,, \label{MBIntegralC} \\
\I^{\rm (d)} &= & s^3 t \, F^{\rm (d)}(1, 1, 1, 1, 1, 1, 1, 1, 1, 1, 1, 
                                  1, 1, -1; s, t; \ep)\,, \label{MBIntegralD} \\
\I^{\rm (e)} &= & s^2 t \,  F^{\rm (e)}(1, 1, 1, 1, 1, 1, 1, 1, 1, 1, 1, 1, 
                              1, -1, -1;  s, t; \ep)\,,  \label{MBIntegralE}  \\
\I^{\rm (f)} &= &  \lim_{\eta\rightarrow 0}
                 s^2 t \, F^{\rm (f)}(1, 1, 1, 1, 1, 1, 1, 1, 1, 1, 1, 
                   1, 1, -1 + \eta, -1; s, t; \ep)\,,  \label{MBIntegralF}  \\
\I^{\rm (d_2)} &=&  s^2 t \, F^{\rm (d)}(1, 1, 1, 1, 1, 1, 1, 1, 1, 0, 
                              1, 1, 0, 0;  s, t; \ep)\,, \label{MBIntegralD2} \\
\I^{\rm (f_2)} &= & s^2 t^2 \, F^{\rm (f)}(1, 1, 1, 1, 1, 1, 1, 1, 1, 1, 1, 
                               1, 0, 0, 0;  s, t; \ep)\,.  \label{MBIntegralF2}
\end{eqnarray}
For the case of integral $\rm(f)$, in order to obtain a valid 
analytic continuation to the region $\ep \rightarrow 0$ 
using {\tt MB}~\cite{CzakonMB}, 
it turns out that we first need to perform an analytic 
continuation in one of the indices, by an auxiliary parameter $\eta$. 
After the analytic continuation we may set $\eta \rightarrow 0$. 

\section{Harmonic polylogarithms}
\label{HarmonicPolyLogAppendix}

We express the leading poles in the integrals and the amplitudes in
terms of harmonic polylogarithms (HPLs)~\cite{HPL}, generalizations of
ordinary polylogarithms~\cite{Lewin}.  Here we briefly summarize their
definitions; see ref.~\cite{HPL} for a more complete treatment.
Routines for numerically evaluating HPLs may be found in
ref.~\cite{HPL2,HPLMaitre}.

The HPLs of weight $n$ are denoted by 
$H_{a_1 a_2 \ldots a_n}(x) \equiv H(a_1,a_2,\ldots,a_n;x)$, 
with $a_i \in \{1,0,-1\}$.  They are defined recursively by,
\begin{equation}
H_{a_1 a_2 \ldots a_n}(x) 
= \int_0^x \dd t \,  f_{a_1}(t) H_{a_2 \ldots a_n}(t)\,,
\end{equation}
where
\begin{eqnarray}
f_{\pm 1}(x) &=& {1 \over 1 \mp x} \,, \qquad f_{0}(x) = {1\over x} \,, \\
H_{\pm 1}(x) &=& \mp \ln(1\mp x) \,, \qquad H_{0}(x)= \ln x \,,
\end{eqnarray}
provided that at least one of the indices $a_i$ is non-zero.
For all $a_i=0$, one defines
\be
H_{0,0,\ldots,0}(x) = \frac{1}{n!}\ln^n x\;.
\ee

For weight less than or equal to four, any HPL having only the 
parameters $a_i=0$ and $1$ can be expressed \cite{HPL} in 
terms of the standard polylogarithms~\cite{Lewin},
\begin{eqnarray}
\Li_n(z) &=& \sum_{j=1}^\infty {z^j \over j^n} =
\int_0^z {\dd t\over t} \Li_{n-1}(t) \,, \nonumber\\
\Li_2(z) &=& -\int_0^z {\dd t\over t} \ln(1-t) \,,
\label{PolyLogDef}
\end{eqnarray}
with $n=2,3,4$, and where $z$ may take the values $x$, $1/(1-x)$, 
or $-x/(1-x)$.  (For $n<4$, polylogarithm identities imply that
not all values are required.)
For the four-loop results presented in \sect{AnalyticSubsection} 
through $\Ord(\e^{-4})$, we indeed need only $a_i \in \{0,1\}$, 
and weights up to four.  In \sect{LowerLoopSymmetricSubsection},
however, we give analytic results for lower-loop amplitudes
at the symmetric point $(s,t)=(-1,-1)$ through $\Ord(\e^{-2})$,
in which we encounter weights up to six.
In the Euclidean region for the planar four-point process,
namely $s<0$, $t<0$, $u>0$, with the identification $x=-t/s$,
the argument $x$ of the harmonic polylogarithms will be negative.

The HPLs obey sets of identities~\cite{HPL} that allow the 
removal of trailing zeroes from the string of parameters 
$a_i$.  The remaining $H_{a_1 a_2 \ldots a_n}(x)$ with $a_n=1$
are well-behaved as $x \to 0$; in fact they all vanish there.
Integrals appear in the \SYM\ amplitudes with both
arguments $(s,t)$ and $(t,s)$, so we also need a set of identities 
relating harmonic polylogarithms with argument $x = -t/s$ to those 
with argument $y = -s/t = 1/x$.  These identities are given in
refs.~\cite{HPL}.  A few directly relevant examples may be found
in the first appendix of ref.~\cite{Iterate3}.



\begin{thebibliography}{99}

\bibitem{Maldacena}
J.~M.~Maldacena,
Adv.\ Theor.\ Math.\ Phys.\ {\bf 2}, 231 (1998)
[Int.\ J.\ Theor.\ Phys.\ {\bf 38}, 1113 (1999)]
[hep-th/9711200];\\
%
S.~S.~Gubser, I.~R.~Klebanov and A.~M.~Polyakov,
Phys.\ Lett.\ B {\bf 428}, 105 (1998)
[hep-th/9802109];\\
%
O.~Aharony, S.~S.~Gubser, J.~M.~Maldacena, H.~Ooguri and Y.~Oz,
Phys.\ Rept.\  {\bf 323}, 183 (2000)
[hep-th/9905111].

\bibitem{BPS}
S.~M.~Lee, S.~Minwalla, M.~Rangamani and N.~Seiberg,
Adv.\ Theor.\ Math.\ Phys.\  {\bf 2}, 697 (1998)
[hep-th/9806074];\\
%
E.~D'Hoker, D.~Z.~Freedman and W.~Skiba,
Phys.\ Rev.\ D {\bf 59}, 045008 (1999)
[hep-th/9807098];\\
%
P.~S.~Howe, E.~Sokatchev and P.~C.~West,
Phys.\ Lett.\ B {\bf 444}, 341 (1998)
[hep-th/9808162];\\
%
S.~Penati, A.~Santambrogio and D.~Zanon,
JHEP {\bf 9912}, 006 (1999)
[hep-th/9910197];\\
%
E.~D'Hoker, D.~Z.~Freedman, S.~D.~Mathur, A.~Matusis and L.~Rastelli,
hep-th/9908160;\\
%
M.~Bianchi and S.~Kovacs,
Phys.\ Lett.\ B {\bf 468}, 102 (1999)
[hep-th/9910016];\\
%
J.~Erdmenger and M.~P\'erez-Victoria,
Phys.\ Rev.\ D {\bf 62}, 045008 (2000)
[hep-th/9912250].

\bibitem{OtherAnomalousDim}
M.~Bianchi, S.~Kovacs, G.~Rossi and Y.~S.~Stanev,
JHEP {\bf 9908}, 020 (1999)
[hep-th/9906188];\\
%
E.~D'Hoker, S.~D.~Mathur, A.~Matusis and L.~Rastelli,
Nucl.\ Phys.\ B {\bf 589}, 38 (2000)
[hep-th/9911222];\\
%
M.~Bianchi, S.~Kovacs, G.~Rossi and Y.~S.~Stanev,
Nucl.\ Phys.\ B {\bf 584}, 216 (2000)
[hep-th/0003203];\\
%
G.~Arutyunov, S.~Penati, A.~C.~Petkou, A.~Santambrogio and E.~Sokatchev,
Nucl.\ Phys.\ B {\bf 643}, 49 (2002)
[hep-th/0206020].

\bibitem{BMN}
D.~Berenstein, J.~M.~Maldacena and H.~Nastase,
JHEP {\bf 0204}, 013 (2002)
[hep-th/0202021].

\bibitem{Nontrivial}
G.~Arutyunov and S.~Frolov,
Phys.\ Rev.\ D {\bf 62}, 064016 (2000)
[hep-th/0002170];\\
%
B.~Eden, A.~C.~Petkou, C.~Schubert and E.~Sokatchev,
Nucl.\ Phys.\ B {\bf 607}, 191 (2001)
[hep-th/0009106].

\bibitem{DhokerTasi}
E.~D'Hoker and D.~Z.~Freedman,
hep-th/0201253.

\bibitem{Iterate2}
C.~Anastasiou, Z.~Bern, L.~J.~Dixon and D.~A.~Kosower,
Phys.\ Rev.\ Lett.\  {\bf 91}, 251602 (2003)
[hep-th/0309040].

\bibitem{Iterate3}
Z.~Bern, L.~J.~Dixon and V.~A.~Smirnov,
Phys.\ Rev.\ D {\bf 72}, 085001 (2005)
[hep-th/0505205].

\bibitem{CSV06}
F.~Cachazo, M.~Spradlin and A.~Volovich,
Phys.\ Rev.\ D {\bf 74}, 045020 (2006)
[hep-th/0602228].

\bibitem{BCKRS}
Z.~Bern, M.~Czakon, D.~A.~Kosower, R.~Roiban and V.~A.~Smirnov,
hep-th/0604074.

\bibitem{Akhoury}
R.~Akhoury,
Phys.\ Rev.\ {\bf D19}, 1250 (1979).

\bibitem{Sudakov}
A.~H.~Mueller,
Phys.\ Rev.\ D {\bf 20}, 2037 (1979);\\
%
J.~C.~Collins,
Phys.\ Rev.\ D {\bf 22}, 1478 (1980);
%
in {\it Perturbative QCD}, ed. 
A.~H. Mueller, Advanced Series on Directions in High Energy 
Physics, Vol. 5 (World Scientific, Singapore, 1989);\\
%
Adv.\ Ser.\ Direct.\ High Energy Phys.\ {\bf 5}, 573 (1989)
[hep-ph/0312336];\\
%
A.~Sen,
Phys.\ Rev.\ D {\bf 24}, 3281 (1981).

\bibitem{Sen83}
A.~Sen,
Phys.\ Rev.\ {\bf D28}, 860 (1983).

\bibitem{SoftGluonSummation}
G.~Sterman,
Nucl.\ Phys.\ B {\bf 281}, 310 (1987);\\
%
S.~Catani and L.~Trentadue,
Nucl.\ Phys.\ B {\bf 327}, 323 (1989);
%
Nucl.\ Phys.\ B {\bf 353}, 183 (1991);\\
%
A.~Vogt,
Phys.\ Lett.\ B {\bf 497}, 228 (2001)
[hep-ph/0010146].

\bibitem{MagneaSterman}
L.~Magnea and G.~Sterman,
Phys.\ Rev.\ D {\bf 42}, 4222 (1990).

\bibitem{IROneLoop}
W.~T.~Giele and E.~W.~N.~Glover,
Phys.\ Rev.\ D {\bf 46}, 1980 (1992);\\
Z.~Kunszt, A.~Signer and Z.~Tr\'ocs\'anyi,
Nucl.\ Phys.\ B {\bf 420}, 550 (1994)
[hep-ph/9401294].

\bibitem{CataniIR}
S.~Catani,
Phys.\ Lett.\ B {\bf 427}, 161 (1998)
[hep-ph/9802439].

\bibitem{TYS}
G.~Sterman and M.~E.~Tejeda-Yeomans,
Phys.\ Lett.\ B {\bf 552}, 48 (2003)
[hep-ph/0210130].

\bibitem{EynckLaenenMagnea}
T.~O.~Eynck, E.~Laenen and L.~Magnea,
JHEP {\bf 0306}, 057 (2003)
[hep-ph/0305179].

\bibitem{QCDIntegrable}
V.~M.~Braun, S.~E.~Derkachov and A.~N.~Manashov,
Phys.\ Rev.\ Lett.\  {\bf 81}, 2020 (1998)
[hep-ph/9805225];\\
%
V.~M.~Braun, S.~E.~Derkachov, G.~P.~Korchemsky and A.~N.~Manashov,
Nucl.\ Phys.\ B {\bf 553}, 355 (1999)
[hep-ph/9902375];\\
%
A.~V.~Belitsky,
Phys.\ Lett.\ B {\bf 453}, 59 (1999)
[hep-ph/9902361].

\bibitem{MinahanZarembo}
J.~A.~Minahan and K.~Zarembo,
JHEP {\bf 0303}, 013 (2003)
[hep-th/0212208].

\bibitem{MoreIntegrable}
N.~Beisert, C.~Kristjansen and M.~Staudacher,
Nucl.\ Phys.\ B {\bf 664}, 131 (2003)
[hep-th/0303060].

\bibitem{BeisertDispersion}
N.~Beisert,
Nucl.\ Phys.\ B {\bf 682}, 487 (2004)
[hep-th/0310252].

\bibitem{Integrable}
A.~V.~Belitsky, A.~S.~Gorsky and G.~P.~Korchemsky,
Nucl.\ Phys.\ B {\bf 667}, 3 (2003)
[hep-th/0304028];\\
N.~Beisert,
Nucl.\ Phys.\ B {\bf 676}, 3 (2004)
[hep-th/0307015];\\
%
N.~Beisert,
JHEP {\bf 0309}, 062 (2003)
[hep-th/0308074];\\
%
N.~Beisert and M.~Staudacher,
Nucl.\ Phys.\ B {\bf 670}, 439 (2003)
[hep-th/0307042];\\
%
L.~Dolan, C.~R.~Nappi and E.~Witten,
JHEP {\bf 0310}, 017 (2003)
[hep-th/0308089];\\
%
G.~Arutyunov and M.~Staudacher,
JHEP {\bf 0403}, 004 (2004)
[hep-th/0310182];\\
%
L.~Dolan, C.~R.~Nappi and E.~Witten, in Proceedings of the 
{\it 3${}^{\rm rd}$ International Symposium on 
Quantum Theory and Symmetries (QTS3)\/}, Cincinnati, Ohio, Sept.~10--14, 2003
[hep-th/0401243];\\
%
A.~V.~Belitsky, S.~E.~Derkachov, G.~P.~Korchemsky and A.~N.~Manashov,
Phys.\ Lett.\ B {\bf 594}, 385 (2004)
[hep-th/0403085];
%
%
Nucl.\ Phys.\ B {\bf 708}, 115 (2005)
[hep-th/0409120];\\
%
A.~V.~Ryzhov and A.~A.~Tseytlin,
Nucl.\ Phys.\ B {\bf 698}, 132 (2004)
[hep-th/0404215];\\
%
S.~A.~Frolov, R.~Roiban and A.~A.~Tseytlin,
hep-th/0503192.

\bibitem{BPR}
I.~Bena, J.~Polchinski and R.~Roiban,
Phys.\ Rev.\ D {\bf 69}, 046002 (2004)
[hep-th/0305116].

\bibitem{Berkovits}
N.~Berkovits,
JHEP {\bf 0503}, 041 (2005)
[hep-th/0411170].

\bibitem{Schubert}
B.~Eden, P.~S.~Howe, C.~Schubert, E.~Sokatchev and P.~C.~West,
Phys.\ Lett.\ B {\bf 466}, 20 (1999)
[hep-th/9906051];\\
%
B.~Eden, C.~Schubert and E.~Sokatchev,
Phys.\ Lett.\ B {\bf 482}, 309 (2000)
[hep-th/0003096];
%
hep-th/0010005.

\bibitem{BGK06}
A.~V.~Belitsky, A.~S.~Gorsky and G.~P.~Korchemsky,
Nucl.\ Phys.\ B {\bf 748}, 24 (2006)
[hep-th/0601112].

\bibitem{Belitsky}
A.~V.~Belitsky,
hep-th/0609068.

\bibitem{AsymptoticBA}
N.~Beisert and M.~Staudacher,
Nucl.\ Phys.\ B {\bf 727}, 1 (2005)
[hep-th/0504190].

\bibitem{EdenStaudacher}
B.~Eden and M.~Staudacher,
hep-th/0603157.

\bibitem{KM}
G.~P.~Korchemsky,
Mod.\ Phys.\ Lett.\ A {\bf 4}, 1257 (1989);\\
G.~P.~Korchemsky and G.~Marchesini,
Nucl.\ Phys.\ B {\bf 406}, 225 (1993)
[hep-ph/9210281].

\bibitem{Makeenko}
Y.~Makeenko,
JHEP {\bf 0301}, 007 (2003)
[hep-th/0210256].

\bibitem{KLV}
A.~V.~Kotikov, L.~N.~Lipatov and V.~N.~Velizhanin,
Phys.\ Lett.\ B {\bf 557}, 114 (2003)
[hep-ph/0301021].

\bibitem{KLOV}
A.~V.~Kotikov, L.~N.~Lipatov, A.~I.~Onishchenko and V.~N.~Velizhanin,
Phys.\ Lett.\ B {\bf 595}, 521 (2004)
[hep-th/0404092].

\bibitem{MOS}
Y.~Makeenko, P.~Olesen and G.~W.~Semenoff,
Nucl.\ Phys.\ B {\bf 748}, 170 (2006)
[hep-th/0602100].

\bibitem{MVV}
S.~Moch, J.~A.~M.~Vermaseren and A.~Vogt,
Nucl.\ Phys.\ B {\bf 688}, 101 (2004)
[hep-ph/0403192];
%
Nucl.\ Phys.\ B {\bf 691}, 129 (2004)
[hep-ph/0404111].

\bibitem{SoftNf}
S.~Moch, J.~A.~M.~Vermaseren and A.~Vogt,
Nucl.\ Phys.\ B {\bf 646}, 181 (2002)
[hep-ph/0209100];\\
%
C.~F.~Berger,
Phys.\ Rev.\ D {\bf 66}, 116002 (2002)
[hep-ph/0209107].
%

\bibitem{KL02}
A.~V.~Kotikov and L.~N.~Lipatov,
Nucl.\ Phys.\ B {\bf 661}, 19 (2003)
[Erratum-ibid.\ B {\bf 685}, 405 (2004)]
[hep-ph/0208220].

\bibitem{Staudacher}
M.~Staudacher,
JHEP {\bf 0505}, 054 (2005)
[hep-th/0412188].

\bibitem{StaudacherPrivate}
M.~Staudacher, private communication.

\bibitem{AFS04}
G.~Arutyunov, S.~Frolov and M.~Staudacher,
JHEP {\bf 0410}, 016 (2004)
[hep-th/0406256].

\bibitem{OtherDressing}
N.~Beisert and A.~A.~Tseytlin,
Phys.\ Lett.\ B {\bf 629}, 102 (2005)
[hep-th/0509084];\\
%
S.~Schafer-Nameki and M.~Zamaklar,
JHEP {\bf 0510}, 044 (2005)
[hep-th/0509096].
%

\bibitem{HernandezLopez}
R.~Hern\'andez and E.~L\'opez,
JHEP {\bf 0607}, 004 (2006)
[hep-th/0603204];\\
%
L.~Freyhult and C.~Kristjansen,
Phys.\ Lett.\ B {\bf 638}, 258 (2006)
[hep-th/0604069].

\bibitem{JanikCrossing}
R.~A.~Janik,
Phys.\ Rev.\ D {\bf 73}, 086006 (2006)
[hep-th/0603038].

\bibitem{BHL}
N.~Beisert, R.~Hern\'andez and E.~L\'opez,
hep-th/0609044.

\bibitem{BESNew}
 N.~Beisert, B.~Eden, and M.~Staudacher, to appear.

\bibitem{NeqFourOneLoop}
Z.~Bern, L.~J.~Dixon, D.~C.~Dunbar and D.~A.~Kosower,
Nucl.\ Phys.\ B {\bf 425}, 217 (1994)
[hep-ph/9403226].

\bibitem{Fusing}
Z.~Bern, L.~J.~Dixon, D.~C.~Dunbar and D.~A.~Kosower,
Nucl.\ Phys.\ B {\bf 435}, 59 (1995)
[hep-ph/9409265].

\bibitem{UnitarityMachinery}
Z.~Bern and A.~G.~Morgan,
Nucl.\ Phys.\ B {\bf 467}, 479 (1996)
[hep-ph/9511336];\\
%
Z.~Bern, L.~J.~Dixon and D.~A.~Kosower,
Nucl.\ Phys.\ Proc.\ Suppl.\  {\bf 51C}, 243 (1996)
[hep-ph/9606378].

\bibitem{OneLoopReview}
Z.~Bern, L.~J.~Dixon and D.~A.~Kosower,
Ann.\ Rev.\ Nucl.\ Part.\ Sci.\  {\bf 46}, 109 (1996)
[hep-ph/9602280].

\bibitem{TwoLoopSplitting}
Z.~Bern, L.~J.~Dixon and D.~A.~Kosower,
JHEP {\bf 0408}, 012 (2004)
[hep-ph/0404293].

\bibitem{BRY}
Z.~Bern, J.~S.~Rozowsky and B.~Yan,
Phys.\ Lett.\ B {\bf 401}, 273 (1997)
[hep-ph/9702424].

\bibitem{DHSS}
J.~M.~Drummond, J.~Henn, V.~A.~Smirnov and E.~Sokatchev,
hep-th/0607160.

\bibitem{SmirnovDoubleBox}
V.~A.~Smirnov,
Phys.\ Lett.\ B {\bf 460}, 397 (1999)
[hep-ph/9905323].

\bibitem{LoopIntegrationAdvance}
S.~Laporta,
Int.\ J.\ Mod.\ Phys.\ A {\bf 15}, 5087 (2000)
[hep-ph/0102033];\\
%
S.~Moch, P.~Uwer and S.~Weinzierl,
J.\ Math.\ Phys.\  {\bf 43}, 3363 (2002)
[hep-ph/0110083];\\
T.~Gehrmann and E.~Remiddi,
Nucl.\ Phys.\ B {\bf 601}, 248 (2001)
[hep-ph/0008287];
%
Nucl.\ Phys.\ B {\bf 601}, 287 (2001)
[hep-ph/0101124].

\bibitem{SmirnovTripleBox}
V.~A.~Smirnov,
Phys.\ Lett.\ B {\bf 567}, 193 (2003)
[hep-ph/0305142].

\bibitem{Tausk}
J.~B.~Tausk,
Phys.\ Lett.\ B {\bf 469}, 225 (1999)
[hep-ph/9909506].

\bibitem{TwoloopOffandMassive}
V.~A.~Smirnov,
Phys.\ Lett.\ B {\bf 491}, 130 (2000)
[hep-ph/0007032];
%
Phys.\ Lett.\ B {\bf 500}, 330 (2001)
[hep-ph/0011056];
%
Phys.\ Lett.\ B {\bf 524}, 129 (2002)
[hep-ph/0111160];
%
Nucl.\ Phys.\ Proc.\ Suppl.\  {\bf 135}, 252 (2004)
[hep-ph/0406052];\\
G.~Heinrich and V.~A.~Smirnov,
Phys.\ Lett.\ B {\bf 598}, 55 (2004)
[hep-ph/0406053];\\
%
M.~Czakon, J.~Gluza and T.~Riemann,
Nucl.\ Phys.\ B {\bf 751}, 1 (2006)
[hep-ph/0604101].

\bibitem{Buch}
V.~A.~Smirnov, {\it Evaluating Feynman integrals}, 
Springer tracts in modern physics, {\bf 211} 
(Springer, Berlin, Heidelberg, 2004).

\bibitem{AnastasiouDaleo}
C.~Anastasiou and A.~Daleo,
hep-ph/0511176.

\bibitem{CzakonMB}
M.~Czakon,
hep-ph/0511200.

\bibitem{HPL}
E.~Remiddi and J.~A.~M.~Vermaseren,
Int.\ J.\ Mod.\ Phys.\ A {\bf 15}, 725 (2000)
[hep-ph/9905237].

\bibitem{HPL2}
T.~Gehrmann and E.~Remiddi,
Comput.\ Phys.\ Commun.\  {\bf 141}, 296 (2001)
[hep-ph/0107173];\\
%
J.~Vollinga and S.~Weinzierl,
hep-ph/0410259.

\bibitem{BeisertKlose}
N.~Beisert and T.~Klose,
J.\ Stat.\ Mech.\  {\bf 0607}, P006 (2006)
[hep-th/0510124].

\bibitem{BeisertInvariant}
N.~Beisert,
Bulg.\ J.\ Phys.\  {\bf 33S1}, 371 (2006)
[hep-th/0511013].

\bibitem{BeisertDynamic}
N.~Beisert,
hep-th/0511082.

\bibitem{BeisertPhase}
N.~Beisert,
hep-th/0606214.

\bibitem{StrongCouplingLeadingGKP}
S.~S.~Gubser, I.~R.~Klebanov and A.~M.~Polyakov,
Nucl.\ Phys.\ B {\bf 636}, 99 (2002)
[hep-th/0204051].

\bibitem{Kruczenski}
M.~Kruczenski,
JHEP {\bf 0212}, 024 (2002)
[hep-th/0210115].

\bibitem{StrongCouplingSubleading}
S.~Frolov and A.~A.~Tseytlin,
JHEP {\bf 0206}, 007 (2002)
[hep-th/0204226].

\bibitem{SWI}
M.~T.~Grisaru, H.~N.~Pendleton and P.~van Nieuwenhuizen,
Phys.\ Rev.\ D {\bf 15}, 996 (1977);\\
M.~T.~Grisaru and H.~N.~Pendleton,
Nucl.\ Phys.\ B {\bf 124}, 81 (1977).

\bibitem{GSWANP}
R.~Grimm, M.~Sohnius and J.~Wess,
Nucl.\ Phys.\ B {\bf 133}, 275 (1978);\\
M.~F.~Sohnius,
Nucl.\ Phys.\ B {\bf 136}, 461 (1978);\\
Y.~Abe, V.~P.~Nair and M.~I.~Park,
Phys.\ Rev.\ D {\bf 71}, 025002 (2005)
[hep-th/0408191].

\bibitem{FDH}
Z.~Bern and D.~A.~Kosower,
Nucl.\ Phys.\ B {\bf 379}, 451 (1992);\\
%
Z.~Bern, A.~De Freitas, L.~Dixon and H.~L.~Wong,
Phys.\ Rev.\ D {\bf 66}, 085002 (2002)
[hep-ph/0202271];\\
%
A.~De Freitas and Z.~Bern,
JHEP {\bf 0409}, 039 (2004)
[hep-ph/0409007].

\bibitem{Siegel}
W.~Siegel,
Phys.\ Lett.\ B {\bf 84}, 193 (1979).

\bibitem{IBP}
F.V.~Tkachov,
Phys.\ Lett.\ B {\bf 100}, 65 (1981); \\
K.G.~Chetyrkin and F.V.~Tkachov,
Nucl.\ Phys.\ B {\bf 192}, 159 (1981).

\bibitem{IBP2loop}
V.A.~Smirnov and O.L.~Veretin,
Nucl.\ Phys.\  {\bf B566}, 469 (2000)
[hep-ph/9907385];\\
T.~Gehrmann and E.~Remiddi,
Nucl.\ Phys.\ B {\bf 580}, 485 (2000)
[hep-ph/9912329];\\
C.~Anastasiou, T.~Gehrmann, C.~Oleari, E.~Remiddi and J.B.~Tausk,
Nucl.\ Phys.\  B {\bf 580}, 577 (2000)
[hep-ph/0003261].

\bibitem{GSB}
M.~B.~Green, J.~H.~Schwarz and L.~Brink,
Nucl.\ Phys.\ B {\bf 198}, 474 (1982).

\bibitem{BDDPR}
Z.~Bern, L.~J.~Dixon, D.~C.~Dunbar, M.~Perelstein and J.~S.~Rozowsky,
Nucl.\ Phys.\ B {\bf 530}, 401 (1998)
[hep-th/9802162].

\bibitem{GravNoTriangle}
Z.~Bern, L.~J.~Dixon, M.~Perelstein and J.~S.~Rozowsky,
Nucl.\ Phys.\ B {\bf 546}, 423 (1999)
[hep-th/9811140];\\
Z.~Bern, N.~E.~J.~Bjerrum-Bohr and D.~C.~Dunbar,
JHEP {\bf 0505}, 056 (2005)
[hep-th/0501137];\\
N.~E.~J.~Bjerrum-Bohr, D.~C.~Dunbar and H.~Ita,
Phys.\ Lett.\ B {\bf 621}, 183 (2005)
[hep-th/0503102];
hep-th/0606268;\\
N.~E.~J.~Bjerrum-Bohr, D.~C.~Dunbar, H.~Ita, W.~B.~Perkins and K.~Risager,
hep-th/0610043.

\bibitem{Finiteness}
S.~Mandelstam,
Nucl.\ Phys.\ B {\bf 213}, 149 (1983);\\
L.~Brink, O.~Lindgren and B.~E.~W.~Nilsson,
Phys.\ Lett.\ B {\bf 123}, 323 (1983);\\
P.~S.~Howe, K.~S.~Stelle and P.~K.~Townsend,
Nucl.\ Phys.\ B {\bf 236}, 125 (1984).

\bibitem{SuperspaceBook}
S.~J.~Gates, M.~T.~Grisaru, M.~Ro\v{c}ek and W.~Siegel,
{\it Superspace}, (Benjamin/Cummings, Reading, Mass., 1983).

\bibitem{GKS}
G.~Cvetic, I.~Kondrashuk and I.~Schmidt,
Mod.\ Phys.\ Lett.\ A {\bf 21}, 1127 (2006)
[hep-th/0407251].

\bibitem{Nakanishi}
N.~Nakanishi, {\it Graph Theory and Feynman Integrals} (Gordon and
Breach, New York, 1971).

\bibitem{CUBA}
T.~Hahn,
Comput.\ Phys.\ Commun.\  {\bf 168}, 78 (2005)
[hep-ph/0404043].

\bibitem{HPLMaitre}
D.~Ma\^{\i}tre,
hep-ph/0507152.

\bibitem{LipatovPotsdam}
 L.~Lipatov, presented at 
Workshop on Integrability in Gauge and String Theory, AEI, Potsdam,
Germany, July 24-28, 2006 [int06.aei.mpg.de/presentations/lipatov.pdf].

\bibitem{DMS}
Yu.~L.~Dokshitzer, G.~Marchesini and G.~P.~Salam,
Phys.\ Lett.\ B {\bf 634}, 504 (2006)
[hep-ph/0511302];\\
G.~Marchesini,
hep-ph/0605262.

\bibitem{RSS}
A.~Rej, D.~Serban and M.~Staudacher,
JHEP {\bf 0603}, 018 (2006)
[hep-th/0512077].

\bibitem{Jaxo}
D.~Binosi and L.~Theussl,
Comput.\ Phys.\ Commun.\ {\bf 161}, 76 (2004)
[hep-ph/0309015].

\bibitem{Axo}
J.~A.~M.~Vermaseren,
Comput.\ Phys.\ Commun.\ {\bf 83}, 45 (1994).

\bibitem{Lewin}
L.~Lewin, {\it Polylogarithms and associated functions}
(North-Holland, New York, 1981).

\end{thebibliography}
\end{document}